\pdfoutput=1
\documentclass[11pt,twoside,a4paper,cmspaper,final,collab]{cms-tdr}
\def\svnVersion{b39adc9}\def\svnDate{2023/06/03}\def\cmsCernNoTag{CERN-EP-2022-080}\def\cmsCernDate{\today}\def\cmsMessage{Published in the Journal of High Energy Physics as \href{http://dx.doi.org/10.1007/JHEP05(2023)228}{\doi{10.1007/JHEP05(2023)228}.}}
\begin{document}\cmsNoteHeader{EXO-21-006}

\newcommand{\pTErr}{\ensuremath{\sigma_{\pt}}\xspace}
\newcommand{\chisq}{\ensuremath{\chi^{2}}\xspace}
\newcommand{\vchisq}{\chisq_\text{vtx}\xspace}
\newcommand{\normchisq}{\ensuremath{\chi^{2}/\mathrm{dof}}\xspace}
\newcommand{\tkiso}{\ensuremath{I^\text{rel}_\text{trk}}\xspace}
\newcommand{\DeltaR}{\ensuremath{\Delta R}\xspace}
\newcommand{\mMuMu}{\ensuremath{m_{\Pgm\Pgm}}\xspace}
\newcommand{\Lxy}{\ensuremath{L_{\mathrm{xy}}}\xspace}
\newcommand{\LxyErr}{\ensuremath{\sigma_{\Lxy}}\xspace}
\newcommand{\LxySig}{\ensuremath{\Lxy/\LxyErr}\xspace}
\newcommand{\dzero}{\ensuremath{d_{\mathrm{0}}}\xspace}
\newcommand{\dzeroErr}{\ensuremath{\sigma_{\dzero}}\xspace}
\newcommand{\dzeroSig}{\ensuremath{\dzero/\dzeroErr}\xspace}
\newcommand{\DeltaPhi}{\ensuremath{\Delta\Phi}\xspace}
\newcommand{\DeltaPhiAbs}{\ensuremath{\abs{\Delta\Phi}}\xspace}
\newcommand{\DeltaPhiAbsStar}{\ensuremath{\abs{\Delta\Phi^{*}}}\xspace}
\newcommand{\DetaMuMu}{\ensuremath{\abs{\Delta \eta _{\Pgm\Pgm}}}\xspace}
\newcommand{\PhiMu}{\ensuremath{\phi_\mu^{\mathrm{TMS}}}\xspace}
\newcommand{\PhiMuAbs}{\ensuremath{\abs{\phi_\mu^{\mathrm{TMS}}}}\xspace}
\newcommand{\PBSMHiggs}{\ensuremath{\mathsf{\Phi}}}

\newcommand{\PSMHiggs}{\PH}
\newcommand{\PDarkHiggs}{\PH\ensuremath{\mathrm{_{D}}}}
\newcommand{\PZD}{\PZ\ensuremath{\mathrm{_{D}}}}
\newcommand{\PLLP}{\PX}
\DeclareRobustCommand{\PAf}{{\HepAntiParticle{\Pf}{}{}}\Xspace}
\newcommand{\htwomu}{\ensuremath{\PBSMHiggs\to\PLLP\PLLP\to\Pgm\Pgm}+anything\xspace}
\newcommand{\fourMu}{4\Pgm}
\newcommand{\twoMu}{2\Pgm}
\newcommand{\mh}{\ensuremath{m(\PSMHiggs)}\xspace}
\newcommand{\mH}{\ensuremath{m(\PBSMHiggs)}\xspace}
\newcommand{\mX}{\ensuremath{m(\PLLP)}\xspace}
\newcommand{\mZD}{\ensuremath{m(\PZD)}\xspace}
\newcommand{\Tau}{\ensuremath{\tau}\xspace}
\newcommand{\cTau}{\ensuremath{c\tau}\xspace}

\renewcommand{\Lone}{L1\xspace}

\newlength\DFigWidth
\setlength\DFigWidth{0.77\textwidth}
\newlength\DSquareWidth
\setlength\DSquareWidth{0.51\textwidth}

\newlength\cmsTabSkip\setlength{\cmsTabSkip}{1ex}

\cmsNoteHeader{EXO-21-006}
\title{Search for long-lived particles decaying to a pair of muons in 
proton-proton collisions at 
\texorpdfstring{$\sqrt{s} = 13\TeV$}{sqrt(s) = 13 TeV}}

\date{\today}

\abstract{An inclusive search for long-lived exotic particles decaying to a pair of muons is presented.  The search uses data collected by the CMS experiment at the CERN LHC in proton-proton collisions at $\sqrt{s} = 13\TeV$ in 2016 and 2018 and corresponding to an integrated luminosity of 97.6\fbinv.  The experimental signature is a pair of oppositely charged muons originating from a common secondary vertex spatially separated from the $\Pp\Pp$ interaction point by distances ranging from several hundred $\mum$ to several meters.  The results are interpreted in the frameworks of the hidden Abelian Higgs model, in which  the Higgs boson decays to a pair of long-lived dark photons $\PZD$, and of a simplified model, in which long-lived particles are produced in decays of an exotic heavy neutral scalar boson.   For the hidden Abelian Higgs model with $\mZD$ greater than 20\GeV and less than half the mass of the Higgs boson, they provide the best limits to date on the branching fraction of the Higgs boson to dark photons for $\cTau(\PZD)$ (varying with $\mZD$) between 0.03 and ${\approx}0.5$\mm, and above ${\approx}0.5$\unit{m}. Our results also yield the best constraints on long-lived particles with masses larger than 10\GeV produced in decays of an exotic scalar boson heavier than the Higgs boson and decaying to a pair of muons.}

\hypersetup{
pdfauthor={CMS Collaboration},
pdftitle={Search for long-lived particles decaying to a pair of muons in proton-proton collisions at sqrt(s) = 13 TeV},
pdfsubject={CMS},
pdfkeywords={CMS, Exotica, long-lived particles, dimuons}}

\maketitle 

\section{Introduction} \label{sec:introduction}
Long-lived particles (LLPs) are predicted by many extensions of the
standard model (SM), in particular by various supersymmetric
scenarios~\cite{Barbier:2004ez, Hewett:2004nw} and ``hidden sector''
models~\cite{Strassler:2006im, Han:2007ae}. 
Such particles could manifest themselves through decays to SM
particles at macroscopic distances from the proton-proton ($\Pp\Pp$)
interaction point (IP).

This paper describes an inclusive search for an exotic massive LLP
decaying to a pair of oppositely charged muons, referred to as a ``displaced dimuon'', that
originates from a common secondary vertex spatially separated from
the IP.  The search is based on an analysis of $\Pp\Pp$ collisions
corresponding to an integrated luminosity of 97.6\fbinv
collected with the CMS detector at $\sqrt{s} = 13\TeV$ during Run~2 of the CERN LHC.
A minimal set of requirements and loose event selection criteria 
allow the search to be sensitive to a wide range of
models predicting LLPs that decay to final states that include a pair of
oppositely charged muons.  We interpret the results of the search in the
framework of two benchmark models: the hidden Abelian Higgs model (HAHM), in
which displaced dimuons arise from decays of hypothetical dark
photons~\cite{Curtin:2014cca}, and a simplified model, in which a
non-SM Higgs boson decays to a pair of long-lived exotic heavy neutral
scalar bosons, at least one of which decays into a pair of
muons~\cite{Strassler:2006ri}.

The present search explores the LLP mass range above 10\GeV accessible
with the Run 2 dimuon triggers and is sensitive to secondary vertex
displacements ranging from several hundred $\mum$ to several meters.
It is a continuation and extension of
two CMS analyses performed using data taken at
$\sqrt{s} = 8\TeV$ during Run~1 of the LHC.  One analysis was
dedicated to a search for LLPs decaying to two electrons or two
muons in the tracker~\cite{EXO-12-037}; the other looked for LLP
decays to final states containing two muons reconstructed only in the
muon system~\cite{CMS-PAS-EXO-14-012}.  The analysis of the Run 2 data
described here contains numerous improvements over these
Run 1 searches, notably in refined event selection and improved
background evaluation procedures. It also benefits from an increase in integrated luminosity by almost a factor of five, collected at a higher $\sqrt{s}$.  A
search for LLPs decaying to displaced dimuons has also been performed by
the ATLAS Collaboration, using 2016 data corresponding to an
integrated luminosity of 32.9\fbinv~\cite{Aaboud:2018jbr}.

This paper is organized as follows.  Section~\ref{sec:detector}
describes the CMS detector.  Section~\ref{sec:samples} presents the
signal models as well as the samples analyzed from data and from the Monte
Carlo simulation.  Section~\ref{sec:selection} describes the analysis
strategy, the triggers, and the offline event selection.  Estimation of
backgrounds and the associated systematic uncertainties are described
in Section~\ref{sec:background}.  Section~\ref{sec:systunc_signal}
summarizes the systematic uncertainties affecting signal
efficiencies.  Section~\ref{sec:results} describes the results
obtained in the individual dimuon categories and their combination.
The analysis summary is presented in Section~\ref{sec:summary}.
Tabulated results and supplementary material for reinterpreting the
results in the framework of models not explicitly considered in this
paper are provided in HEPData~\cite{HEPData}.

\section{The CMS detector} \label{sec:detector}
The central feature of the CMS detector is a superconducting solenoid
of 6\unit{m} internal diameter, providing a magnetic field of
3.8\unit{T}.  Within the solenoid volume are a silicon pixel and strip
tracker extending outwards to a radius of 1.1\unit{m},
a lead tungstate crystal electromagnetic calorimeter,
and a brass and scintillator hadron calorimeter, each composed
of a barrel and two endcap sections.  Forward calorimeters extend the
coverage in pseudorapidity $\eta$ provided by the barrel and endcap
detectors.  Muons are detected in gas-ionization chambers covering the
range $\abs{\eta} < 2.4$ and embedded in the steel flux-return yoke
outside the solenoid.  The muon system is composed of three types of
chambers: drift tubes (DTs) in the barrel, cathode strip chambers
(CSCs) in the endcaps, and resistive-plate chambers in both the
barrel and the endcaps.  The chambers are assembled into four
``stations'' at increasing distance from the IP; each station provides
reconstructed hits in several
detection planes, which are combined into track segments, forming the
basis of muon reconstruction in the muon system~\cite{Sirunyan:2018}.
A more detailed description of the CMS detector, together with a
definition of the coordinate system used and the relevant kinematic
variables, can be found in Ref.~\cite{Chatrchyan:2008zzk}.

Events of interest are selected using a two-tiered trigger system.
The first level (\Lone), composed of custom hardware processors, uses
information from the calorimeters and muon detectors to select events
at a rate of approximately 100\unit{kHz} within a fixed time interval
of less than 4\mus~\cite{Sirunyan:2020zal}.  The second level, known
as the high-level trigger (HLT), consists of a farm of processors
running a version of the full event reconstruction software optimized
for fast processing, and reduces the event rate to about
1\unit{kHz} before data storage~\cite{Khachatryan:2016bia}.

\section{Signal models, data and simulated samples} \label{sec:samples}
The search is performed using $\Pp\Pp$ collision data collected at
$\sqrt{s} = 13\TeV$ in 2016 and 2018 corresponding
to integrated luminosities of $36.3 \pm 0.4$ and $61.3 \pm
1.5$\fbinv, respectively~\cite{CMS:2021xjt, CMS-PAS-LUM-18-002}.
The data collected in 2017 are not used because the triggers
required for the analysis were not included when those data were recorded.

\begin{figure}[htbp]
  \centering
  \includegraphics[width=0.80\DSquareWidth]{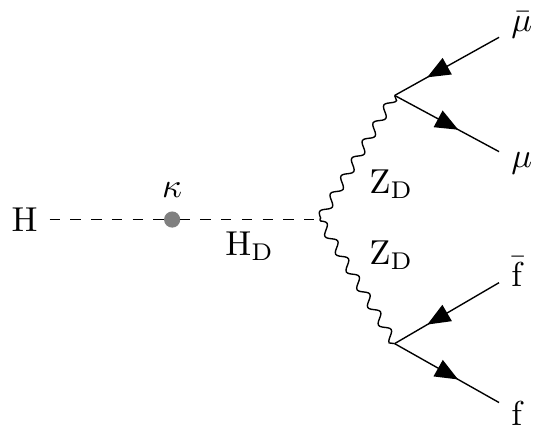}
\hspace*{5em}
  \includegraphics[width=0.58\DSquareWidth]{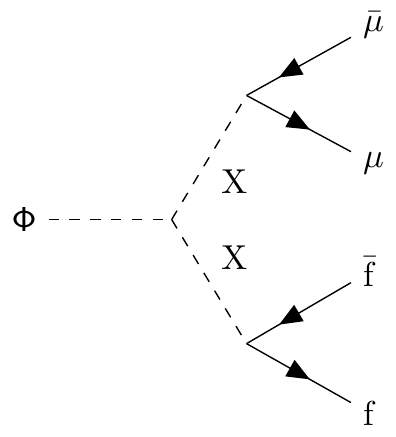}
  \caption{Feynman diagrams for (left) the HAHM model, showing the
    production of long-lived dark photons $\PZD$ via the Higgs portal,
    through $\PSMHiggs$--$\PDarkHiggs$ mixing with the parameter $\kappa$,
    with subsequent decays via the vector portal; and (right) the heavy-scalar
    model with $\PBSMHiggs$ boson decaying to a pair of long-lived bosons \PLLP.  The symbols \Pf and \PAf represent, respectively, fermions
    and antifermions lighter than half the LLP mass.
     \label{fig:SignalDiagram}}
\end{figure}

As mentioned above, two signal models with different final-state topologies and event
kinematics are used in the optimization of event selection criteria
and in the interpretation of results.  The first belongs to a class of
models featuring a ``hidden'' or ``dark'' sector of matter that does
not interact directly with SM particles, but can manifest itself
through mixing effects.  This HAHM benchmark
contains an extra dark $U$(1)$_\text{D}$ gauge
group whose symmetry is broken by a new dark Higgs field~\cite{Wells:2008xg,
  Curtin:2014cca}.  The spin-1 mediator of the $U$(1)$_\text{D}$ group,
known as the dark photon $\PZD$, mixes kinetically with the
hypercharge SM gauge boson (``vector portal''), whereas the dark Higgs
boson $\PDarkHiggs$ mixes with the SM Higgs boson $\PSMHiggs$
(``Higgs portal'') and gives mass $\mZD$ to the dark photon.  If there are no
hidden-sector states with masses smaller than $\mZD$, the mixing
through the vector portal with the SM photon and \cPZ boson causes the
dark photon to decay exclusively to SM particles, with a sizable
branching fraction to leptons.  Pair production of the $\PZD$ via the
Higgs portal with subsequent decays of dark photons
via the vector portal is shown in Fig.~\ref{fig:SignalDiagram} (left).

The present search probes the regime of $\mZD > 10\GeV$ with small values of
the $\cPZ$--$\PZD$ kine\-tic mixing parameter $\epsilon$~\cite{Curtin:2014cca}.
In this regime, the dark photon is long-lived, since its mean proper lifetime
$\tau(\PZD)$ is proportional to $\epsilon^{-2}$.  In particular, the
dark photon with $10\GeV \lesssim \mZD < \mh/2$ is expected
to have macroscopically large mean proper decay lengths $\cTau(\PZD)
\gtrsim \mathcal{O}(100{\micron})$ for $\epsilon < \mathcal{O}(10^{-6})$.  The
$\PZD$ production rate is governed by the branching fraction 
$\mathcal{B}(\PSMHiggs \to \PZD\PZD)$, which does not depend on $\epsilon$ but is
proportional to the square of
$\kappa m^2(\PSMHiggs)/\abs{m^2(\PSMHiggs)-m^2(\PDarkHiggs)}$, where
$\kappa$ is the $\PSMHiggs$--$\PDarkHiggs$ mixing parameter.  Since
$\kappa$ and $m(\PDarkHiggs)$ affect only the overall dark photon
production rate, sampling of $\mZD$ and $\epsilon$ is sufficient to
explore different kinematical and topological scenarios of the model.
We generated a set of 24 HAHM samples with $\mZD$ between 10
and 60\GeV and $\epsilon$ between $10^{-6}$ and $2 \times 10^{-9}$.  In
this mass range, 
the model's prediction for $\mathcal{B}(\PZD \to \Pgm\Pgm)$ varies
between 15.4\% at $\mZD = 10\GeV$ and 10.7\% at $\mZD = 60\GeV$.
The dark Higgs boson is assumed to be heavy enough such that
$\PSMHiggs \to \PDarkHiggs\PDarkHiggs$ decays are kinematically
forbidden. In the sample generation, we specify $m(\PDarkHiggs)=400\GeV$
and $\kappa=0.1$.
The production of dark photons is modeled at leading order
by \MGvATNLO~\cite{Alwall:2014hca} version 2.4.2.
The generation of the samples is done for the dominant gluon-fusion
production mechanism.  The Higgs boson production cross section is
normalized to the most recent theoretical prediction for the sum of
all production modes for $\mh = 125\GeV$,
55.7\unit{pb}~\cite{Cepeda:2019klc}.  The decays of the dark photons
are modeled by \PYTHIA 8.212 and 8.230~\cite{Sjostrand:2014zea} in
samples corresponding to the 2016 and 2018 data sets, respectively.

At the LHC, another way that LLPs might arise is via production of
mediators heavier than the Higgs boson that decay into
LLPs.  To explore ranges of kinematic variables and
event topologies broader than those offered by HAHM,
we also consider a simplified benchmark model~\cite{Strassler:2006ri},
previously used in Run 1 searches for displaced dimuons~\cite{EXO-12-037,
CMS-PAS-EXO-14-012}, in which the LLP is an exotic spin-0
boson \PLLP.  The scalar $\PLLP$ has a non-zero branching fraction to
dimuons and is pair produced in the decay of a new heavier scalar
boson \PBSMHiggs, which is produced in gluon-gluon fusion:
$\Pg\Pg \to \PBSMHiggs \to \PLLP\PLLP$, $\PLLP \to \Pgmp \Pgmm$.  The
Feynman diagram for this process is shown in
Fig.~\ref{fig:SignalDiagram} (right).

The samples for $\PBSMHiggs \to \PLLP\PLLP$ are generated with
\PYTHIA.  Two sets of samples are
produced, depending on whether one or both \PLLP bosons are forced to decay
to dimuons.
Samples in each set are generated with different combinations of
$\PBSMHiggs$ boson masses $\mH$ (ranging from 125\GeV to 1\TeV)
and $\PLLP$ boson masses $\mX$ (ranging from 20 to 350\GeV).  The
width of the $\PBSMHiggs$ boson is assumed to be small for the purpose
of simulation, but the analysis has negligible dependence on this assumption.
Each sample is furthermore produced with three different mean proper
lifetimes $\Tau(\PLLP)$ of
the $\PLLP$ bosons, corresponding to mean transverse
decay lengths of approximately 3, 30, and 250\cm.  Events generated at the
selected values of $\mH$, $\mX$, and $\Tau(\PLLP)$ allow us to study wide ranges of
signal displacements, kinematical variables, and event topologies.

Since the optimization of the event selection criteria and the
evaluation of the residual backgrounds are performed using data, the
simulated background samples are used primarily to
gain a better understanding of the nature and composition of surviving
background events.  Simulated background samples used in the analysis
include Drell--Yan (DY) dilepton production; $\ttbar$, $\PQt\PW$, and
$\PAQt\PW$ events; \PW and \PZ boson
pair production (dibosons); $\PW$+jets; and events comprised
of jets produced through the strong interaction that are
enriched in muons from semileptonic decays of hadrons containing $\PQb$ or $\PQc$
quarks.  

The 2016 simulated signal and background samples are
produced with either the NNPDF2.3 (leading order) or NNPDF3.0 (next-to-leading order) parton
distribution functions (PDFs)~\cite{Butterworth_2016}, using the
CUETP8M1~\cite{Khachatryan:2015pea} tune to model the underlying
event.  All 2018 simulated samples are produced with the NNPDF3.1
PDFs~\cite{Ball:2017nwa} (next-to-next-to-leading order), using the CP5~\cite{Sirunyan:2019dfx} tune,
which is optimized for the NNPDF3.1 PDFs.  Simulation of the
passage of particles through detector material 
is performed by \GEANTfour~\cite{Agostinelli:2002hh}.
Simulated minimum bias events are superimposed on a hard interaction
in simulated events to describe the effect of additional inelastic
$\Pp\Pp$ interactions within the same or neighboring bunch crossings,
known as pileup; the samples are weighted to match the pileup
distribution observed in data.  All simulated events are then
reconstructed with the same algorithms as used for data.

\section{Analysis strategy and event selection} \label{sec:selection}
An LLP produced in the hard interaction of the
colliding protons may travel a significant distance in the detector before
decaying into muons.  
While trajectories of the muons produced well within the silicon
tracker can be reconstructed by both the tracker and the muon system,
tracks of muons produced in the outer tracker layers or beyond
can only be reconstructed by the muon system.  Since the
dimuon vertex resolution and the background composition differ
dramatically depending on whether the muon is reconstructed in the
tracker, we classify all reconstructed dimuon events into three
mutually exclusive categories: a) both muons are reconstructed using
both the tracker and the muon system (TMS-TMS category); b) both muons
are reconstructed using only the muon system, as ``standalone'' muons
(STA-STA category); and c) one muon is reconstructed only in the muon
system, whereas the other muon is reconstructed using both the tracker
and the muon system (STA-TMS category).  These three categories of
events are analyzed separately, each benefiting from dedicated event
selection criteria and background evaluation.  The results in
each category are statistically combined to provide the final results.

The beamspot is identified with the mean
position of the $\Pp\Pp$ interaction vertices.  
The primary vertex (PV) is taken to be the vertex corresponding to the
hardest scattering in the event, evaluated using tracking information
alone, as described in Section 9.4 of Ref.~\cite{CMS-TDR-15-02}.
A pair of
reconstructed muon tracks is fitted to a common vertex (CV), which is
expected to be displaced with respect to the PV.  The transverse decay
length \Lxy is defined as the distance between the PV and the CV in
the plane transverse to the beam direction.  The transverse impact
parameter $\dzero$ is defined as the distance of closest approach (DCA) of
the muon track in the transverse plane with respect to the PV.

Events were collected with dedicated triggers
aimed at recording dimuons produced both within and outside of the
tracker.  Therefore, these triggers require two muons reconstructed in
the muon system alone, without using any information from the tracker,
and do not impose the beamspot constraint in the muon track fit at the
HLT~\cite{CMS:2021yvr}.  Each muon is required to be within the region
$\abs{\eta} < 2.0$ and to have transverse momentum magnitude $\pt > 28 (23)\GeV$ 
in 2016 (2018) data taking.  To reduce the trigger rate caused by
hadron punch-through and poorly measured muons, each muon track is required
to be composed of segments found in two or more muon stations.  To reduce the contribution to the trigger rate from cosmic ray muons
and low-mass dimuon resonances, the trigger used to collect 2016 data
also required that the 3D angle between the muons be less than
2.5\unit{rad}, and that the invariant mass of the two muons be larger than
10\GeV.  The optimization of the online selection prior to the 2018 data
taking made it possible to remove these two requirements from the 2018
trigger, thus providing additional validation regions for background
evaluation and increasing the signal efficiency.  The efficiency of
triggering on signal events in 2018 was further improved by complementing
the above trigger with one very similar to it, but using a
modified version of the initial ``seeding'' stage of the muon trajectory
building at the HLT.  The seed generator used in the new trigger was
specifically designed for muons not pointing to the beamspot and
helped to increase the reconstruction efficiency for displaced muons.

\begin{figure}[htbp]
  \centering
  \includegraphics[width=0.97\DSquareWidth]{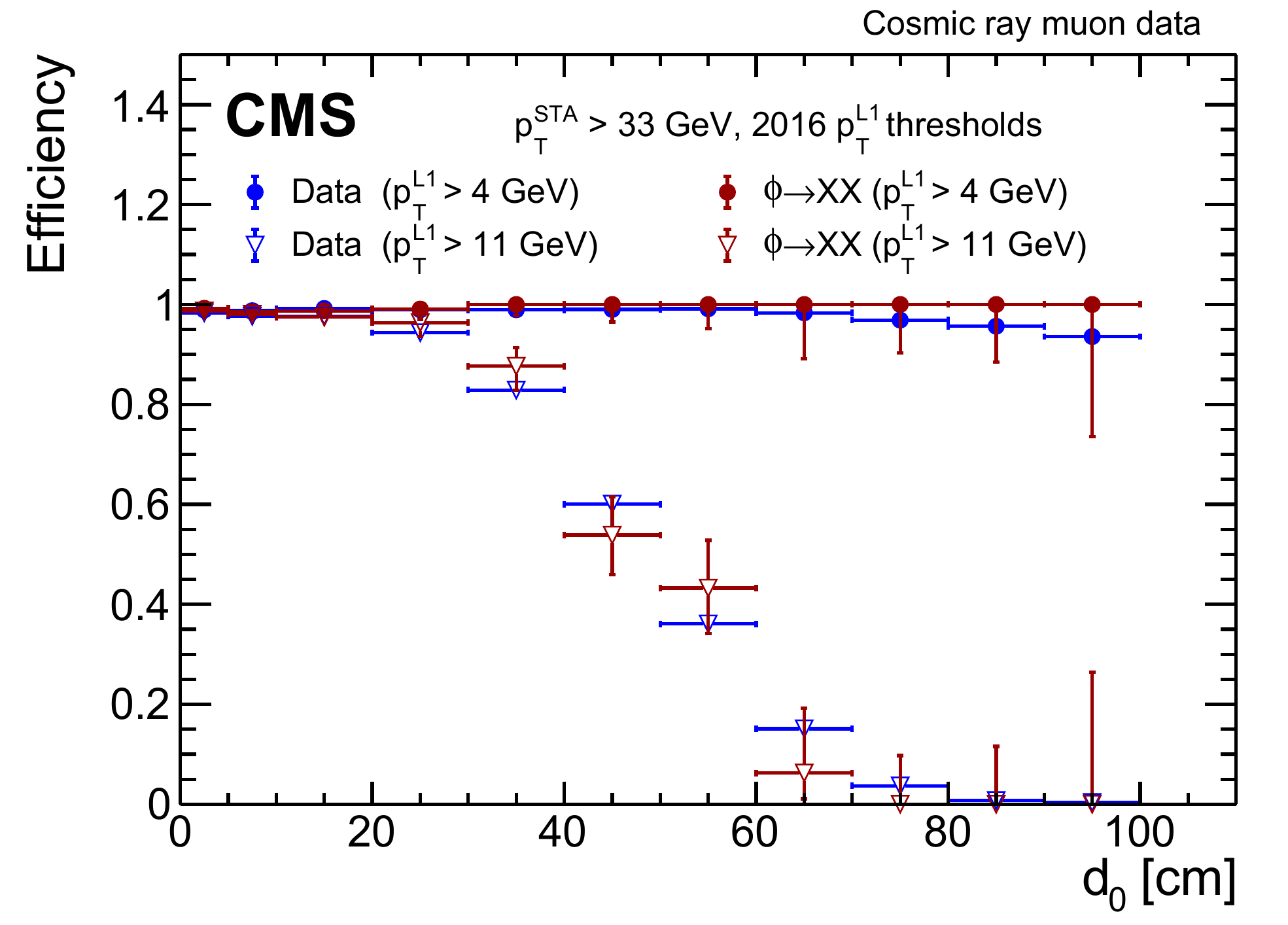}
  \includegraphics[width=0.97\DSquareWidth]{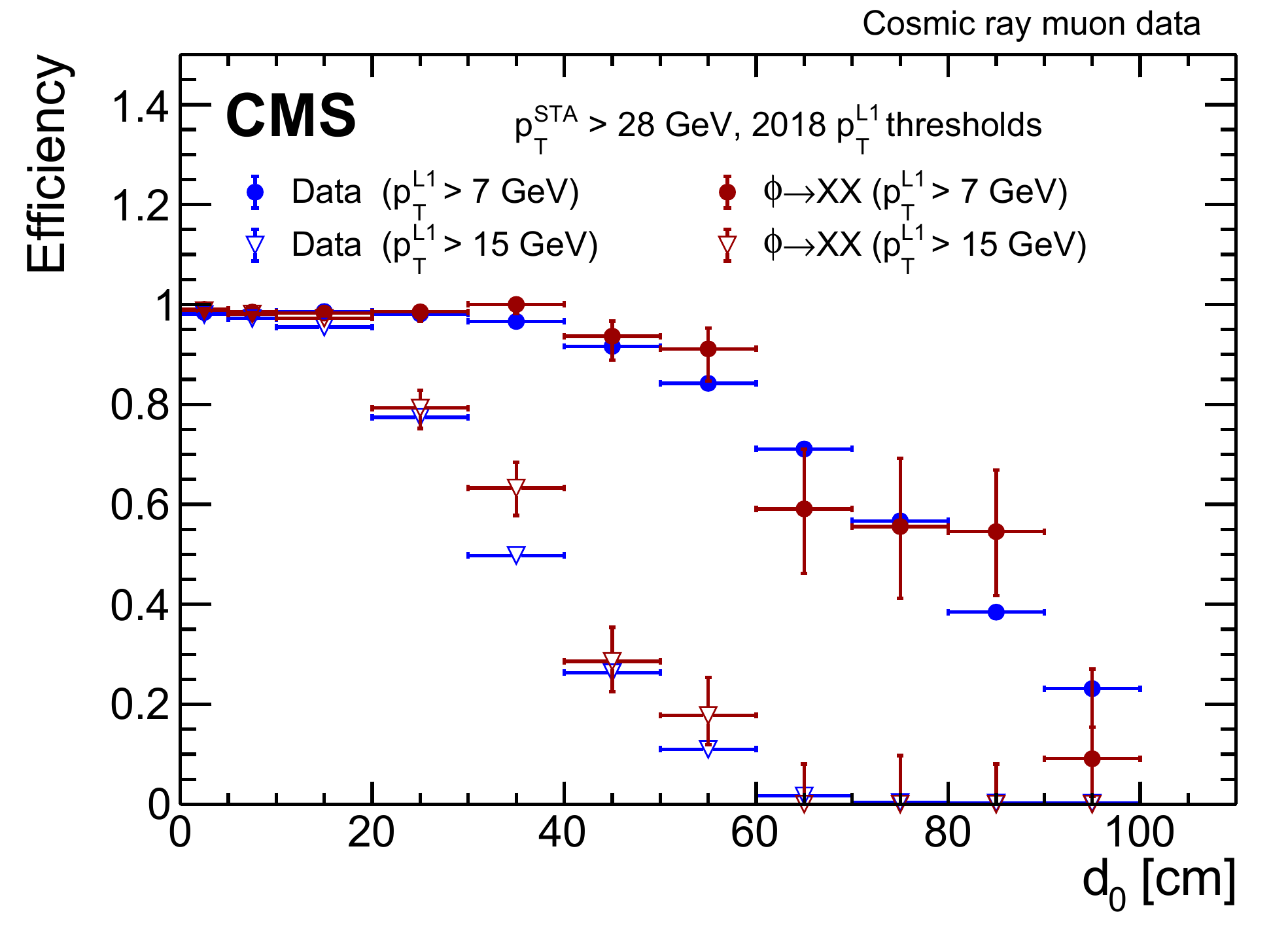}
  \caption{\Lone muon trigger efficiency in cosmic ray muon data
    (blue) and signal simulation (red) as a function of $\dzero$,
    for the \Lone trigger \pt thresholds used in (left) 2016 and
    (right) 2018.  The denominator in the efficiency
    calculations is the number of STA muons with $\abs{\eta} < 1.2$
    and $\pt > 33$ (28)\GeV in 2016 (2018). \label{fig:L1_eff_cosmic}}
\end{figure}

The high-level triggers used in the analysis were seeded by \Lone dimuon triggers that required
the \pt of the muons to be above certain thresholds.  The values of the thresholds
were varied during the data taking, depending on the instantaneous
luminosity, from 11 and 4\GeV (for the leading and subleading \Lone
muons, respectively) during most of 2016, to 15 and 7\GeV at the end of
Run 2.  At \Lone, the \pt assignment for muons was made under the
assumption that the muons originated at the beamspot.  As a result, the
\pt of the displaced muons not pointing to the beamspot were
underestimated and could fall below the \Lone trigger thresholds.
The ensuing signal efficiency loss was larger when higher \Lone
trigger \pt thresholds were used.  Since this
effect is decoupled from the collision environment (\eg, instantaneous
luminosity), it can be studied using cosmic ray muons recorded with
very loose triggers during periods with no beam.
Figure~\ref{fig:L1_eff_cosmic} shows that the decrease in the \Lone
muon trigger efficiency as a function of the impact parameter $\dzero$
for various \Lone trigger \pt thresholds used
in 2016 (left) and 2018 (right) is well
reproduced by the signal simulation in the barrel.

As noted in Section~\ref{sec:introduction}, no single
muon reconstructor provides optimal performance over the wide range of
displacements of secondary vertices considered in the analysis.  
Muons produced near the IP can be accurately reconstructed by using
commonly used algorithms developed for prompt muons and combining
measurements in the tracker and the muon system.  Among them are the
global muon and tracker muon reconstruction
algorithms~\cite{Chatrchyan:2012xi, Sirunyan:2018}.  The first algorithm builds
``global muons'' by using hits in the tracker and segments in the muon
system in a common track fit.  The second constructs ``tracker muons'' by
propagating tracks in the inner tracker to the muon system and requiring loose
geometrical matching to DT or CSC segments.  The efficiency of these
algorithms, however, rapidly decreases as the distance between the IP
and muon origin increases, dropping to zero for muons produced in the
outer tracker layers and beyond.  On the other hand,
such muons can still be efficiently reconstructed by algorithms that
use only information from the muon system.  These STA
algorithms~\cite{Chatrchyan:2012xi, Sirunyan:2018} can reconstruct muons with
displacements of up to a few meters, but they have poorer spatial and
momentum resolution than muons reconstructed using more precise
information from the silicon tracker.

To benefit from the advantages offered by both types of algorithms and
to follow what was done in the trigger, we begin the muon selection with the
most efficient standalone muons and replace them with more
accurately reconstructed global and tracker muons whenever such muons
are found.  We use muons reconstructed by an STA algorithm
with the beamspot constraints removed from all stages of the muon
reconstruction procedure, which yields the highest efficiency and the
best resolution for displaced muons, out of all available STA
algorithms.  The event selection starts with the requirement that the
event is selected by the triggers described above and has at
least two STA muons, each containing more than 12 valid CSC or DT hits.
The requirement of the minimal number of hits suppresses backgrounds
from hadron punch-through and other sources, and ensures that the STA
muons have acceptable \pt resolution and charge assignment.  The STA muons
that satisfy this basic quality requirement form the initial list of
the muon candidates retained for the analysis.

Next, we reject events in which no HLT muon pair that
triggered the event matches two STA muons in the list.  This
requirement suppresses events that triggered on muons not related to
the signal
and facilitates application of trigger
efficiency measurements in the analysis.  We then attempt to
match each STA muon in the list with a TMS muon, \ie, a
global or a tracker muon. 
The STA and TMS muons are
considered to be matched if they share at least two thirds of their
segments or if $\DeltaR_{\mathrm{STA}-\mathrm{TMS}} < 0.1$, 
where $\DeltaR_{\mathrm{STA}-\mathrm{TMS}} =
\sqrt{\smash[b]{(\eta_\mathrm{hit}-\eta_\mathrm{pca})^2 +
    (\phi_\mathrm{hit}-\phi_\mathrm{pca})^2}}$ is the separation 
between $\eta_\mathrm{hit}$ ($\phi_\mathrm{hit}$) of the position of the
innermost hit of the STA muon and $\eta_\mathrm{pca}$ ($\phi_\mathrm{pca}$) of
the point of closest approach of the TMS muon to this hit. 
If an associated TMS muon is found, it replaces the
corresponding STA muon in the list of the muon candidates used
for further analysis; otherwise, the original STA muon is kept.
The matching procedure was optimized using events in the simulated
signal and background samples as well as data in the signal-free
control regions discussed in Section~\ref{sec:background}.
It eliminates most of the $\Pp\Pp$ collision background to
LLP decays outside of the tracker and greatly
increases sensitivity to LLP decays in the tracker, thanks to a far
superior resolution of TMS muons compared to that of STA muons.

\begin{figure}[htb]
  \centering
  \includegraphics[width=\DFigWidth]{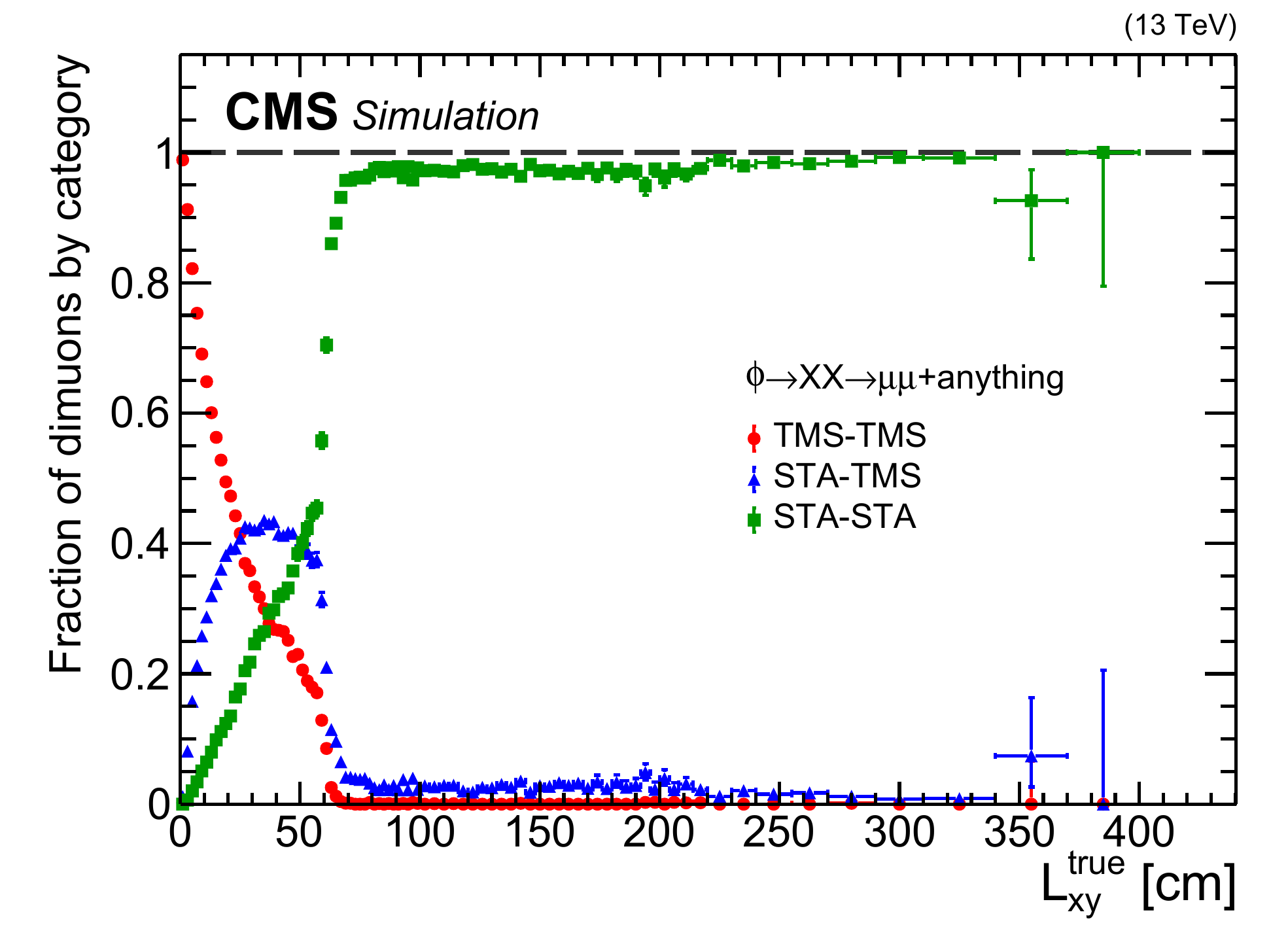}
  \caption{Fractions of signal events with zero (green), one (blue),
    and two (red) STA muons matched to TMS muons by the STA-to-TMS
    muon association procedure, as a function of true \Lxy, in all
    simulated \htwomu signal samples combined.  The fractions are computed
    relative to the number of signal events passing the trigger and
    containing two STA muons with more than 12 muon detector hits and $\pt >
    10\GeV$ matched to generated muons from $\PLLP\to\Pgm\Pgm$
    decays.  \label{fig:replacement_categories}}
\end{figure}

The impact of the STA-to-TMS muon association procedure on the event
selection is further illustrated in
Fig.~\ref{fig:replacement_categories}, which shows the fraction of
simulated \htwomu signal events with zero, one, and two STA muons
matched to TMS muons as a function of $\Lxy^\mathrm{true}$, defined as
the transverse distance between the simulated positions of the
hard-interaction and LLP decay
vertices.  While almost all dimuons produced close to the IP have both
STA muons matched to TMS muons, the fraction of these events rapidly
decreases with $\Lxy^\mathrm{true}$, reflecting the 
dependence on $\Lxy^\mathrm{true}$ of the tracker reconstruction efficiency.  Events with one
STA muon matched to a TMS muon start to dominate at $\Lxy^\mathrm{true} =
25\cm$ and remain the dominant component up to ${\approx}50\cm$, where
events with no STA-to-TMS matches take over.
When LLPs decay in the outer tracker layers or beyond the tracker, all
STA-to-TMS associations are purely accidental and occur for fewer than
5\% of the simulated signal muons.  Therefore, the association
procedure gives rise to three categories of dimuons, each dominating
in a certain $\Lxy^\mathrm{true}$ range: TMS-TMS at small
$\Lxy^\mathrm{true}$, STA-TMS at intermediate $\Lxy^\mathrm{true}$,
and STA-STA at large $\Lxy^\mathrm{true}$.  The number of STA-STA
dimuons beyond the solenoid, at $\Lxy^\mathrm{true} > 3.2\unit{m}$, is
low because of the low trigger efficiency.

The STA and TMS muons are then subjected to additional selection criteria
optimized using simulated signal and background samples, and samples of
dimuons misreconstructed as displaced in the signal-free regions in data.
The STA muons are required to have $\pt > 10\GeV$ and to satisfy the
following criteria:
\begin{itemize}
  \item relative \pt uncertainty $\pTErr/\pt < 1.0$, where $\pTErr$ is the internal uncertainty from
    the muon track fit;
  \item \normchisq of the muon track fit less than 2.5;
  \item more than 18 DT hits for muons reconstructed only in the
    barrel;
  \item time difference $\abs{\Delta t}$ with respect to the current bunch
    crossing less than 12\unit{ns};
  \item muon identified as traveling outwards based on the timing
    measurements~\cite{Sirunyan:2018}.
\end{itemize}
The $\pTErr/\pt$, \normchisq, and DT hits requirements
suppress background events arising from poorly measured prompt muons
and ensure acceptable \pt
resolution and charge assignment for signal muons.  The muon timing
requirement serves several purposes: it rejects events in which the
trigger timing is early, causing the tracker hits that are read out to be from unrelated
$\Pp\Pp$ collisions 25\unit{ns} earlier and not to include any tracker hits of the
triggering STA muons; it rejects out-of-time collision muons; and it helps suppress
background arising from cosmic ray muons crossing the detector outside of
the tracker acceptance, which are often reconstructed as two STA muons
with no associated TMS muons.  The $\abs{\Delta t} < 12\unit{ns}$
requirement does not impact signal efficiency in
the studied mass range.  The muon direction, determined using
the muon timing measurements, offers another handle for suppressing
cosmic ray muon background by identifying cosmic ray muons in the upper part
of the detector, which travel inwards.  We do not impose any explicit
requirements on the isolation of the STA muons, thus making the
STA-STA analysis sensitive to models predicting highly displaced \PQb
quarks, such as LLPs decaying to $\bbbar$.

The initial list of TMS muon candidates consists of the global and
tracker muons with $\pt > 10\GeV$ that are matched to STA muons by the
STA-to-TMS muon association procedure.  These TMS muon candidates are
further required to satisfy the following criteria:
\begin{itemize}
  \item $\pTErr/\pt<1.0$;
  \item be identified as tracker muons with two or more matched CSC or
    DT segments;
  \item be isolated from other activity in the event.
\end{itemize}
The first two requirements reject
poorly measured muons and misidentified hadrons.  The isolation requirement
suppresses copious background from dijet and multijet events.
Specifically, we calculate the sum of the \pt of other
tracks reconstructed within a cone of radius $\DeltaR =
\sqrt{\smash[b]{\left(\Delta\eta\right)^2 + \left(\Delta\phi\right)^2}}$ centered
on the direction of the TMS muon track.  Both the \pt of the muon
track itself and the \pt of the track of any other muon forming a
dimuon (if it lies within the cone) are not included in the sum.  For
the muon to be considered isolated, the relative tracker isolation
$\tkiso$, defined as the ratio of the \pt sum in a cone
of $\DeltaR < 0.3$ to the muon track \pt, is required to be smaller
than 0.075.  The cone size, isolation threshold, and other
parameters of the algorithm such as the minimum $\abs{\Delta z}$ between
the reference point of the track and that of the muon, are the result
of the optimization procedure that uses displaced muons in the
simulated signal samples and background muons in the signal-free
control regions in data.

Dimuon candidates are then formed
from pairs of STA and TMS muons passing the above muon selection
criteria.  All possible combinations of muons (STA-STA, STA-TMS, and
TMS-TMS) are considered.  Each pair of selected muons with the
distance of closest approach of the muon tracks smaller than
15\unit{cm} is fit to a common vertex by means of the
Kalman vertex fitter algorithm~\cite{Fruhwirth:1987fm,
  Speer:927395} and forms a dimuon.  
If more than two muons pass the selection
criteria, multiple dimuons can be formed, including dimuons sharing
one of the two muons; this situation would be particularly common in
signal events with two LLPs decaying to two muons each.  Using
simulated signal samples, we developed a set of muon pairing
criteria that efficiently choose up to two dimuons among the formed
dimuons.  The selection is first done separately in each of the three
dimuon categories.  In events with three selected muons, we choose
a dimuon with the lowest value of \chisq of the vertex fit, $\vchisq$;
in events with four or more selected
muons, we choose the pair of dimuons whose $\vchisq$ sum is the smallest
among all distinct pairs of dimuons formed from the four highest \pt
muons in the event.  In rare cases, when more than two dimuons are found
in all categories combined, up to two dimuons with the smallest $\vchisq$ are kept for further analysis; we also ensure that there are no
STA or TMS muons included in more than one dimuon.  
The efficiency of choosing the right dimuon is 85--95\% for
\fourMu signal samples and close to 100\% for \twoMu samples.

To suppress background from dimuons formed from unrelated
muons, the dimuons chosen by the muon pairing criteria are required to
have \chisq of the common vertex fit below a certain threshold.  Since
\chisq distributions in data are difficult to reproduce
accurately in the simulation, the requirements on $\vchisq$
are chosen to be loose, leaving the signal efficiency
almost intact while rejecting a sizable fraction of background events
in the control regions in data.  The values of the thresholds
are determined separately in each of the dimuon categories and are
$\vchisq < 10$ in the STA-STA and TMS-TMS
categories, and $\vchisq < 20$ in the STA-TMS category.
To further ensure that the muon tracks are consistent with the vertex,
we require that the number of tracker hits assigned to the track of
the TMS muon upstream of the vertex position does not exceed 2 in the
TMS-TMS category and 5 in the STA-TMS category.  In the TMS-TMS
category, background from dimuons formed from unrelated or mismeasured
muons is further suppressed by requiring TMS muons forming the dimuon
to have a similar number of pixel hits, namely that the difference
between the number of pixel hits on two TMS muons, $\Delta
N(\text{pixel hits})$, is smaller than 3 hits.

A cosmic ray muon crossing the detector within the acceptance of the muon
system is often reconstructed as two back-to-back muons, one in the
upper half and one in the lower half of the detector.  This background
is very efficiently suppressed by requiring that the 3D opening angle $\alpha$ 
between the two muons be less than a certain threshold value.
In the analysis of 2016 data, we require $\alpha< 2.5\unit{rad}$ 
(or, equivalently, $\cos{\alpha} > -0.8$), consistent with the
presence of the $\cos{\alpha} > -0.8$ requirement in the 2016 version
of the trigger.  Since the 2018 version of the trigger has no
$\cos{\alpha}$ requirement, the offline requirements on $\alpha$ in
the analysis of 2018 data are re-optimized in each of the dimuon
categories separately, taking into account the angular resolution of
STA and TMS muons.  As the result of this optimization, the 2018
selection criteria are loosened to $\alpha < 2.7\unit{rad}$
($\cos{\alpha} > -0.9$) for STA-STA and STA-TMS dimuons and to $\alpha
< 3.0\unit{rad}$ ($\cos{\alpha} > -0.99$) for TMS-TMS dimuons.

The $\cos{\alpha}$ requirements fail to reject one particular type of
cosmic ray muon background occurring when a cosmic ray muon with a large
incoming angle crosses the muon system diagonally in the $r$-$z$ plane.  A cosmic ray muon of
this type can lead to multiple STA muons, each reconstructed from a
small number of segments.  These poorly measured STA muons are not
necessarily back-to-back, and can give rise to a mistakenly formed
displaced dimuon passing the full event selection.  To suppress such
events, we reject STA-STA dimuons
if the sum of the numbers of segments belonging to the two muons,
$N$(dimuon segments), is smaller than 5.

Furthermore, there are events with
multiple cosmic ray muons produced in an atmospheric shower.  Such events
typically result in a large number of nearly parallel STA muons
reconstructed in both upper and lower halves of the detector.  While
dimuons formed from muons in different hemispheres are rejected by the
$\cos{\alpha}$ requirement, dimuons formed from two muons in the same
hemisphere are not.  To reject the background from cosmic ray
muon showers, we require that the event contain fewer than 4 nearly
parallel STA muons and have at least one reconstructed $\Pp\Pp$
interaction vertex with more than 3 associated tracks and with
transverse (longitudinal) coordinates within 2 (24) cm of the IP.  In
the STA-STA dimuon category, in which the contribution from the cosmic ray
muon showers is the largest, we also require that neither of the two
muons forming a dimuon be back-to-back ($\cos{\alpha} < -0.9$) with
another STA muon with $\pt > 10\GeV$.
The dimuon is rejected only if the time difference $\abs{\Delta t_\text{b2b}}$
between the muon in the dimuon and its back-to-back
muon is larger than 20\unit{ns}, \ie, consistent with the time
difference between the two reconstructed parts of a cosmic ray muon.  This
requirement also helps to suppress dimuons reconstructed from
a cosmic ray muon and a muon from an overlapping $\Pp\Pp$ collision.

Displaced dimuons produced in decays of nonprompt \JPsi mesons and other
low-mass SM resonances, or formed from the products of the $\PQb$ hadron 
cascade decays ($\PQb \to \PQc\Pgm_{1}X$ followed by $\PQc
\to \Pgm_{2}X$), are suppressed by requiring that the
reconstructed dimuon invariant mass \mMuMu be larger than 10\GeV.
This requirement, applied in all three dimuon categories, also
suppresses
dimuons from decays of promptly produced low-mass SM resonances.
However, low-\pt muons
can appear as muons with higher \pt, with straighter tracks, when
reconstructed from a small number of measurements.
This gives rise to dimuons with an overestimated $\mMuMu$ (above the
10\GeV threshold) and a mistakenly formed displaced vertex.  To
suppress such events, we reject STA-STA dimuons whose separation
in $\eta$ is small ($\DetaMuMu < 0.1$)
if one of the muons is reconstructed in the barrel from fewer than 25
DT hits or if $N\text{(dimuon segments)} < 6$.  In the other two
dimuon categories, this background is suppressed by requiring that the
hits associated with each TMS muon populate a minimum number of
tracker layers, $N$(tracker layers).  This minimum number depends on
$\Lxy$ and decreases from 6 for $\Lxy < 15\cm$ to 3 for $\Lxy >
45\cm$.

Another source of SM background events is prompt high-mass dimuons
that are reconstructed as displaced due to instrumental or
reconstruction failures.  Such dimuons mostly arise from DY
dimuon production; contributions from processes such as $\ttbar$ and
diboson production are relatively small.  Events from DY $\Pgt\Pgt$ production with both
$\Pgt$ leptons decaying to muons lead to a background with
characteristics similar to those of the mismeasured DY
$\Pgm\Pgm$ events.  An important discriminating variable between
signal and these backgrounds is the transverse collinearity angle
$\DeltaPhi$ between $\vec{\pt}^{\Pgm\Pgm}$ of the dimuon system and
$\vec{\Lxy}$.
When a pair of muons is produced in the decay of an LLP originating at the PV, both the resulting
$\vec{\pt}^{\Pgm\Pgm}$ and $\vec{\Lxy}$ point away from the PV, and
\DeltaPhiAbs is small.  On the other hand, mismeasured prompt dimuons,
in particular those arising from the DY process, are expected
to have a $\DeltaPhiAbs$ distribution symmetric around $\pi/2$,
because the directions of $\vec{\pt}^{\Pgm\Pgm}$ and $\vec{\Lxy}$ are independent.  We require $\DeltaPhiAbs < \pi/4$ in all three dimuon
categories.  The requirement is kept loose in order to preserve
sensitivity to signal models featuring LLPs that decay into a dimuon
and a particle (or particles) escaping detection, such as neutrinos,
dark matter particles, or lightest supersymmetric particles~\cite{Allanach:2006st, 
 Deppisch:2019kvs}.  We use the symmetric region $\DeltaPhiAbs
> 3\pi/4$ as a control region for evaluating the contribution from
DY and other prompt backgrounds, and the regions with $\pi/4 <
\DeltaPhiAbs < \pi/2$ and $\pi/2 < \DeltaPhiAbs < 3\pi/4$ for
validating background predictions.

The last important source of SM backgrounds is dijet and multijet
events yielding dimuons that are formed from particles arising from
different jets, either genuine muons or particles misidentified as
muons.  Such events contribute mainly to the TMS-TMS and STA-TMS
categories and are strongly suppressed by the isolation criteria
applied to TMS muons, as well as by the requirement that at least one
of the muons that form the TMS-TMS dimuon has $\pt > 25\GeV$.
Furthermore, such events are expected to result in dimuons with both opposite-sign
(OS) and same-sign (SS) electric charges, each with a symmetric $\DeltaPhiAbs$
distribution.
Therefore, in addition to binning in $\DeltaPhiAbs$, we
classify selected dimuons as OS or SS, based on the observed muon charges.   The signal selection requires that dimuons be OS, while SS dimuons constitute a control region
used to evaluate backgrounds arising from dijet and multijet events,
and from $\PQb$ hadron cascade decays.

A class of background events peculiar to the STA-TMS category
consists of dimuons with their common vertex reconstructed on the wrong side of
the PV relative to the direction of the TMS
momentum vector. Since the TMS muon is usually reconstructed much more
precisely than the STA muon, the reconstructed CV in
STA-TMS events is usually located at or near the trajectory of the TMS
muon.  With $\PhiMu$ defined as the
angle between the TMS muon $\vec{\pt}$ and $\vec{\Lxy}$, the location of the CV along the TMS muon trajectory
depends on the STA muon and can be situated on either the correct side ($\PhiMuAbs<\pi/2$) or the wrong side ($\PhiMuAbs>\pi/2$) of the PV.  We reject events on the extreme wrong side ($\PhiMuAbs \sim \pi$) by
requiring $\PhiMuAbs < 2.9$.

Next, we require that the dimuons be displaced with respect to the
primary vertex.  This is achieved by imposing requirements on the \Lxy
significance \LxySig and the muon \dzero significance \dzeroSig.  The
\Lxy uncertainty $\LxyErr$ is calculated by combining uncertainties in
the transverse positions of the CV and the PV;
the $\dzero$
uncertainty $\dzeroErr$ includes both
track and PV uncertainties.  The optimal values of
requirements on \LxySig and \dzeroSig are chosen separately, for each
of the dimuon categories, by maximizing an approximate figure of merit
for the expected statistical
significance of a discovery~\cite{Cousins:ZBi2008}.  The expected
number of signal events used for the optimization is obtained from
simulation, while the number of background events is estimated from
events in control regions in data using the procedures described in
the next section.  The criteria applied to \LxySig and \dzeroSig are:
\begin{itemize}
\item $\LxySig > 6$ in the STA-STA dimuon category;
\item $\LxySig > 6$ and $\dzeroSig > 6$ in the TMS-TMS dimuon category; and
\item $\LxySig > 3$ and $\dzeroSig > 6$ in the STA-TMS dimuon category.
\end{itemize}

In the STA-STA category, the analysis sensitivity does not increase
appreciably for any requirement on the \dzero significance once a
requirement on \LxySig is in place; therefore, we do not apply any selection
on \dzeroSig.  In the TMS-TMS category, a requirement on \dzeroSig is
applied to both muons and, since the residual background has a falling
\dzeroSig distribution, the signal region is divided into three
bins in the minimum of the two
$\dzeroSig$ values, min($\dzeroSig$):  6--10, 10--20, and ${>}20$.  In the
STA-TMS category, a requirement on \dzeroSig is applied only to the
TMS muon, with no selection on \dzeroSig of the STA muon.  Since the
\Lxy resolution in the STA-TMS category
is of the order of a few cm, a requirement on \LxySig is kept loose to
preserve good efficiency for events with $10 \lesssim \Lxy^\mathrm{gen}
\lesssim 60\cm$, many of which are found in this category.

Table~\ref{tab:fullsel} summarizes the event, muon, and dimuon
selection criteria used in the analysis.

\begin{table}
  \centering
  \topcaption{Summary of the selection criteria used in the analysis,
    grouped into event, muon, and dimuon requirements.}
  \label{tab:fullsel}
  \begin{tabular}{lccc} 
  \hline
    {\textit{Event selection}}                   &          & & \\ [\cmsTabSkip]
    $N$(PV)                           & ${\geq}1$ & & \\
    HLT-STA muon matching             & yes      & & \\
    $N$(nearly parallel STA muons)    & ${<}4$    & & \\
    & & & \\ [\cmsTabSkip]
    {\textit{Muon selection}}                   & \multicolumn{2}{c}{\textit{Muon type}} & \\ [\cmsTabSkip]
                                     & STA             & TMS         & \\ [\cmsTabSkip]
    STA-to-TMS muon association      & not matched to TMS $\mu$ & matched to STA $\mu$ & \\
    $N$(CSC+DT hits)                 & ${>}12$          & \NA         & \\
     \hspace*{1cm} - associated STA muon & \NA         & ${>}12$      & \\
    $N$(DT hits) for muons in barrel & ${>}18$          & \NA         & \\
    tracker muon                     & \NA             & yes         & \\
    $N$(matched muon segments)       & \NA             & ${>}1$       & \\
    \pt                              & ${>}10\GeV$      & ${>}10\GeV$  & \\
    $\pTErr/\pt$                     & ${<}1.0$         & ${<}1.0$     & \\
    $\chisq_\text{trk}/\text{dof}$    & ${<}2.5$         & \NA         & \\
    $\tkiso$                         & \NA             & ${<}0.075$   & \\
    $\abs{\Delta t}$                     & ${<}12\unit{ns}$ & \NA         & \\
    muon direction                   & inside-out      & \NA         & \\
    \dzeroSig                        & \NA             & ${>}6$       & \\
    & & & \\ [\cmsTabSkip]
    {\textit{Dimuon selection}}                 & \multicolumn{3}{c}{\textit{Dimuon category}}  \\ [\cmsTabSkip]
                                     & STA-STA    & STA-TMS    & TMS-TMS    \\ [\cmsTabSkip]
    DCA                              & ${<}15\cm$  & ${<}15\cm$  & ${<}15\cm$  \\
    pairing criteria                 & \multicolumn{3}{c}{best 1--2 ranked dimuons selected} \\
    $\vchisq$           & ${<}10$     & ${<}20$     & ${<}10$     \\
    $\Delta N(\text{pixel hits})$    & \NA        & \NA        & ${<}3$      \\
    $N$(hits before vertex)          & \NA        & ${<}6$      & ${<}3$      \\
    $N$(tracker layers) $+$ floor(\Lxy [cm]/15) & \NA & ${>}5$  & ${>}5$      \\
    $\PhiMuAbs$                      & \NA        & ${<}2.9$    & \NA        \\
    $\cos{\alpha}$                   &            &            &            \\
      \hspace*{1cm} - 2016 data analysis  & ${>}{-}0.8$   & ${>}{-}0.8$   & ${>}{-}0.8$   \\
      \hspace*{1cm} - 2018 data analysis  & ${>}{-}0.9$   & ${>}{-}0.9$   & ${>}{-}0.99$  \\
    $N$(dimuon segments)             & ${>}4$       & \NA        & \NA        \\
    if $\DetaMuMu < 0.1$             &            &            &            \\
      \hspace*{1cm} - $N$(dimuon segments)              & ${>}5$    & \NA    & \NA \\
      \hspace*{1cm} - $N$(DT hits) for muons in barrel  & ${>}24$   & \NA    & \NA \\
    no back-to-back muon             &            &            &            \\
      \hspace*{0.3cm} with $\abs{\Delta t_\text{b2b}} > 20\unit{ns}$ & yes & \NA & \NA \\
    $\mMuMu$                         & ${>}10\GeV$ & ${>}10\GeV$ & ${>}10\GeV$ \\
    \pt of the leading muon          & \NA        & \NA        & ${>}25\GeV$ \\
    \LxySig                          & ${>}6$      & ${>}3$      & ${>}6$      \\
    \DeltaPhiAbs                     & ${<}\pi/4$  & ${<}\pi/4$  & ${<}\pi/4$  \\
    opposite-sign muons              & yes        & yes        & yes        \\
  \hline
  \end{tabular}
\end{table}

Finally, to test for the existence of an LLP with a given mass,
dimuons satisfying the selection criteria are required to have \mMuMu 
within a specified interval 
containing the probed LLP mass.  The width of each interval is chosen
according to the mass resolution and the expected background.  The
resulting intervals typically contain a large fraction (90--99\%) of
putative signal with the probed mass.  Since the mass resolution in
the TMS-TMS category is far superior to that in the other two
categories (1--3\% compared to 10--25\%, for LLP masses between 20 and
350\GeV), the minimum width of mass intervals varies from 3\GeV in the
TMS-TMS category to ${\approx}20\GeV$ in the STA-STA category.

The signal efficiency is defined as the ratio of the number of
simulated signal events in which at least one dimuon candidate of any
type passes all selection criteria (including the trigger) to the
total number of simulated signal events.  The efficiencies are estimated for
LLP lifetimes corresponding to mean proper decay lengths in the range of
$50\micron$--$1\unit{km}$ by reweighting events in the available simulated
signal samples.  The maximum efficiency, which is attained in the
heavy-scalar model with $\mH = 1\TeV$, $\mX = 150\GeV$, and $\cTau =
1\cm$, is approximately 50\% for events that have only one LLP ($\PZD$
or $\PLLP$) decaying to muons.  The efficiency becomes significantly
smaller at low masses or at longer and shorter lifetimes, mostly
because of a lower trigger and geometric acceptance and insufficient
displacement.  For example, the efficiency at
the same set of masses decreases to approximately 15\% at $\cTau =
100\cm$, while the efficiency at $\cTau = 100\cm$ drops to
below 5\% when $\mH = 125\GeV$ and $\mX = 20\GeV$.  In accordance
with Fig.~\ref{fig:replacement_categories}, the vast majority
of the signal with the shortest (longest) lifetimes is found in the
TMS-TMS (STA-STA) dimuon category, whereas all three categories
contain important fractions of the signal with the intermediate
lifetimes.  The efficiency in 2018 is higher than that in 2016, thanks
mostly to the improved trigger and its relaxed requirements.  The gain
in efficiency is especially large, about a factor of two, at $\mH =
125\GeV$ and long lifetimes.

\section{Background estimation and associated systematic uncertainties} \label{sec:background}
Since the background events passing the event selection criteria arise
mostly from misreconstructed prompt muons and muons in jets, their
yield cannot be reliably ascertained from simulation.  Therefore, we
evaluate the expected background using events in data.  The control
regions used to estimate contributions from various types of
background processes are chosen by inverting one or more selection
criteria in order to obtain a region populated mostly by a given type
of background and containing a negligible contribution from the signal
processes.  The definitions of the control regions and the details of
the background evaluation procedure differ for different dimuon
categories and are described in the rest of this section.  To avoid
potential bias in the event selection, the events passing the full
selection (\ie, those in the signal region) were ``blinded'' until the
last steps of the analysis.

The contribution from cosmic ray muons is evaluated separately for
each dimuon category from the number of dimuons satisfying all
selection criteria but failing the $\cos{\alpha}$ requirements.  The
evaluation procedure makes use of the efficiency of the $\cos{\alpha}$
requirements measured from a sample of cosmic ray muons collected
during periods with no beam.  In all dimuon categories, the
residual background arising from cosmic ray muons is estimated to be
smaller than 0.1 events in all mass intervals combined.

\subsection{Estimation of Drell--Yan and other prompt backgrounds} \label{subsec:bkg_evaluation_DY}
In all three dimuon categories, the contribution from prompt
misreconstructed dimuons, collectively referred to as DY-like
events, is evaluated from events in the signal-free
$\DeltaPhiAbs$-symmetric control region, $\DeltaPhiAbs > 3\pi/4$:
\begin{equation}
   N_{\text{DY}}^{i}(\text{OS}; \DeltaPhiAbs < \pi/4) =
   N_{\text{DY}}^{i}(\text{OS}; \DeltaPhiAbs > 3\pi/4) \,
   R_\text{DY}^{i}\,,
  \label{eq:DYestimate}
\end{equation}
where $N_{\text{DY}}^{\text{i}}(\text{OS}; \DeltaPhiAbs < \pi/4)$ and
$N_{\text{DY}}^{\text{i}}(\text{OS}; \DeltaPhiAbs > 3\pi/4)$ are,
respectively, the numbers of DY background events in the
signal and its $\DeltaPhiAbs$-symmetric control region;
$R_\text{DY}^{i}$ is the transfer factor accounting for the residual
asymmetry in the population of events in the two $\DeltaPhiAbs$ regions
and obtained from auxiliary measurements; and
the index $i$ denotes the dimuon category (STA-STA, STA-TMS, or
TMS-TMS).  The number of DY dimuons in the $\DeltaPhiAbs >
3\pi/4$ region is taken to be the total number of events in that
region minus the expected contribution from other types of background
events estimated as discussed in
Section~\ref{subsec:bkg_evaluation_QCD}.

The symmetry of the \DeltaPhiAbs distributions in this class of
background events is studied using data and simulated events.  In
the STA-STA and STA-TMS categories, we use events in the control regions
obtained by reversing the STA-to-TMS association. Specifically, we select
events that consist of STA-STA or, alternatively, STA-TMS dimuons passing
all selection criteria, but in which each of the constituent STA muons
is associated with a TMS muon.  To ensure that such STA-STA
and STA-TMS dimuons are promptly produced (and thus are not signal),
we require that the associated TMS-TMS dimuons, which have a far
superior spatial resolution, are prompt.  This is achieved by
requiring $\LxySig<1.0$ for the associated TMS-TMS dimuon
in the STA-STA category, and $\dzeroSig<1.5$ for the TMS muon associated
with the STA muon in the STA-TMS category.  To minimize contamination from muons from jets, which we discuss
separately in what follows, each TMS muon in the associated TMS-TMS
dimuon is required to satisfy the isolation requirement $\tkiso<0.05$.

\begin{figure}[htbp]
  \centering
  \includegraphics[width=0.97\DSquareWidth]{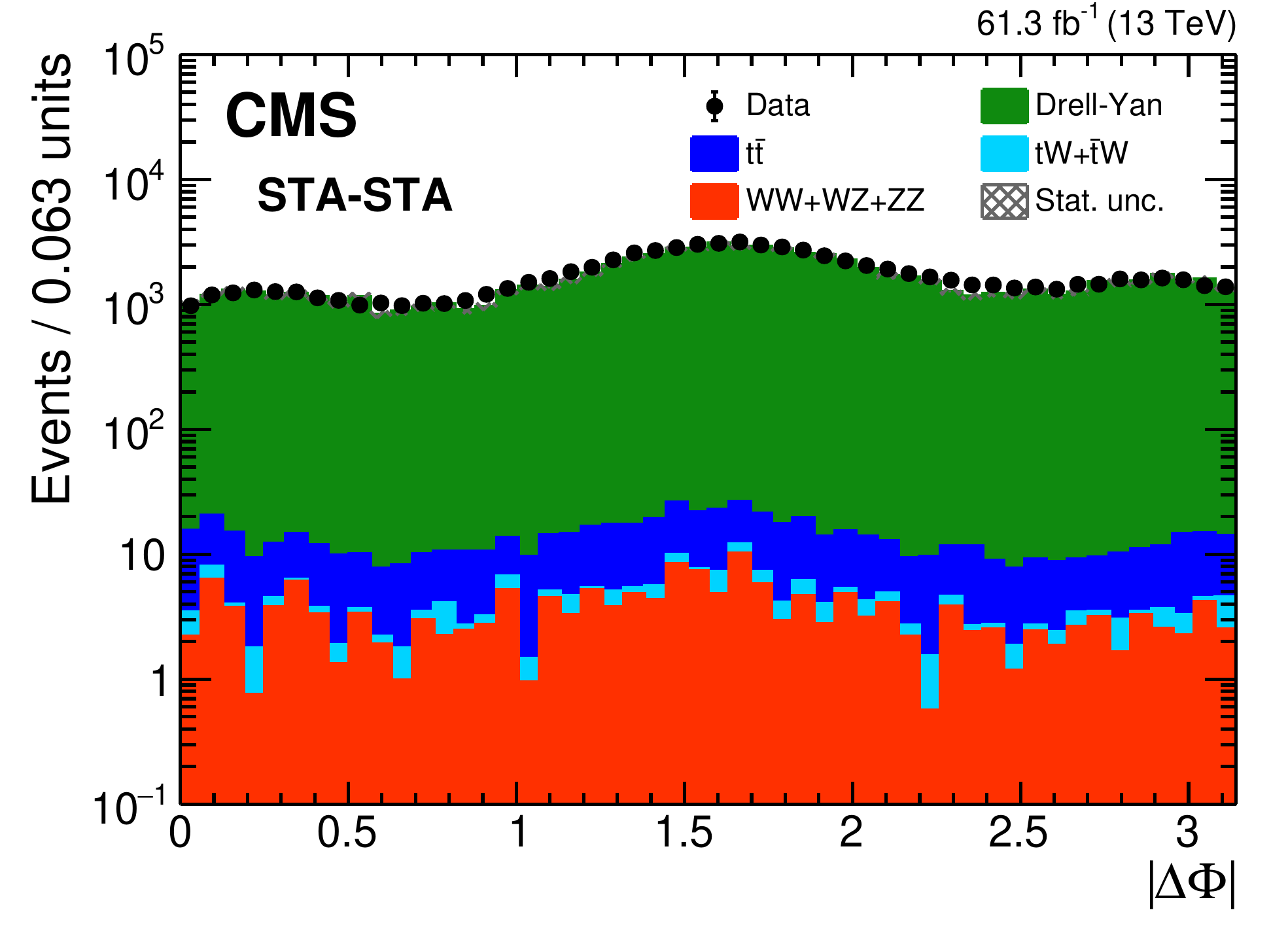}
  \includegraphics[width=0.97\DSquareWidth]{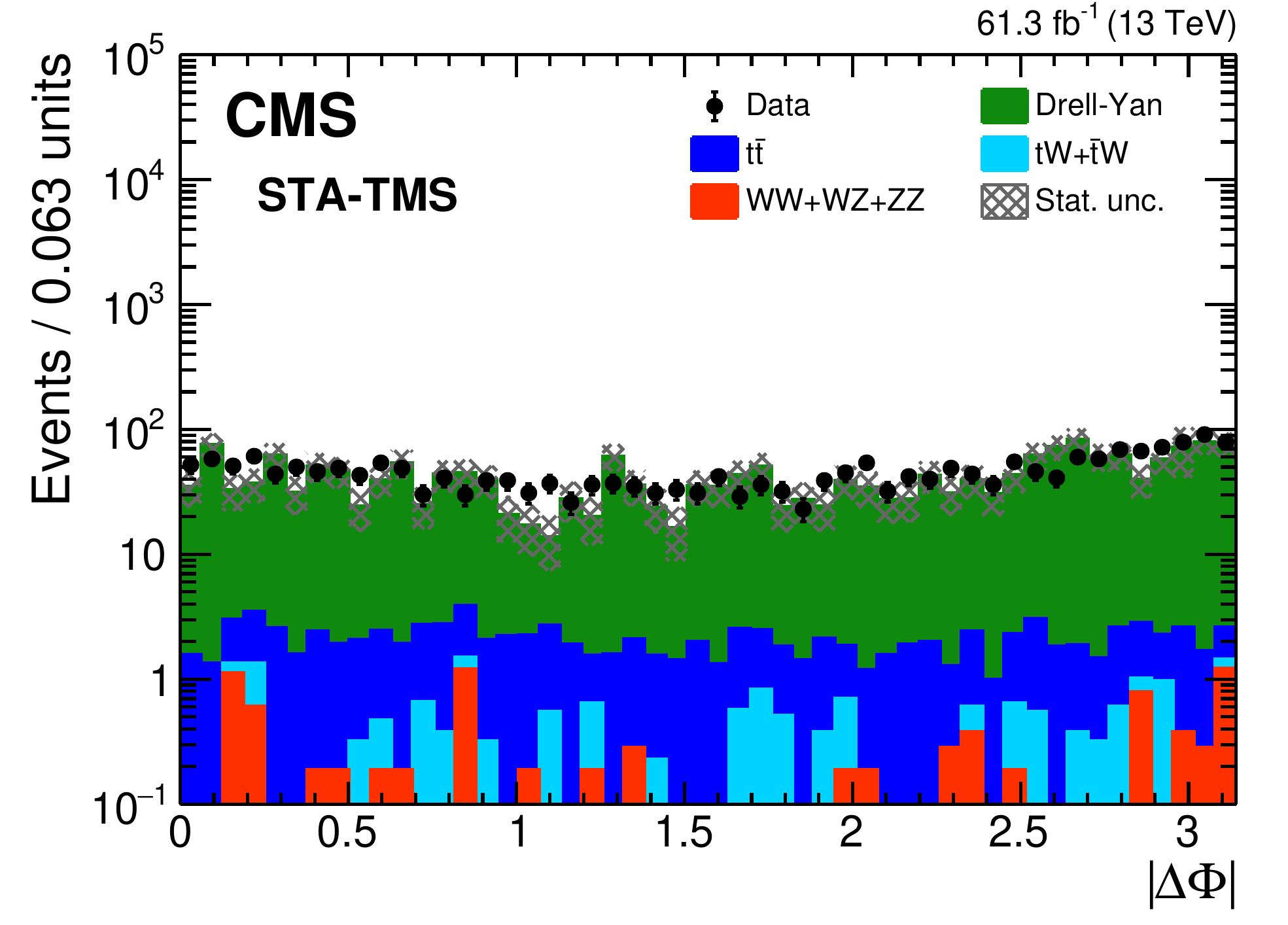}
  \caption{Distributions of $\DeltaPhiAbs$ for (left) STA-STA and
    (right) STA-TMS dimuons in 2018 data (black dots) and simulated
    background processes (stacked histograms), for events in the
    control regions with the STA-to-TMS association of the STA muons
    reversed, as described in the text.  All nominal selection
    requirements, including dimuon $\LxySig$ and TMS muon $\dzeroSig$,
    are applied to the STA-STA and STA-TMS dimuons.  The simulated
    processes are scaled to correspond to the integrated luminosity of the
    data.  The shaded area shows the statistical uncertainty in the
    simulated background yield.\label{fig:deltaphi_DY}}
\end{figure}

Since the TMS and STA muons are predominantly reconstructed from
information in different detectors (the tracker and the muon system,
respectively), a genuine prompt muon giving rise to a displaced STA
muon is usually accurately reconstructed as prompt by the TMS muon
reconstruction.  As a result,
the aforementioned control regions contain genuine prompt dimuons that are
reconstructed as displaced STA-STA or STA-TMS dimuons because of reconstruction
failures or vertex fit anomalies in these categories,
\ie, exactly the type of background events that we wish to
study.  The $\DeltaPhiAbs$ distributions of STA-STA and STA-TMS
dimuons in these control regions, in 2018 data and simulated background
samples, are shown in Fig.~\ref{fig:deltaphi_DY}.  (The distributions in
2016 data are very similar.)  The observed
distributions, sculpted by the interplay between the geometric effects
and the trigger and offline selection requirements, are well reproduced
by the simulation.  The distributions are still fairly symmetric around
$\pi/2$, but there is a small asymmetry caused by the event selection
criteria.  Corrections accounting for this
asymmetry are obtained from the ratio of events with $\DeltaPhiAbs <
\pi/4$ and $\DeltaPhiAbs > 3\pi/4$ in the aforementioned control
regions,
\begin{equation}
R_\text{DY}^{i} = \frac{N^{\text{rev},i}_{\text{DY}}(\text{OS};
  \DeltaPhiAbs < \pi/4)}{N^{\text{rev},i}_{\text{DY}}(\text{OS};
  \DeltaPhiAbs > 3\pi/4)}\,.
  \label{eq:DYtransferfactor}
\end{equation}

In both STA-STA and STA-TMS categories, no dependence of the transfer
factor $R_\text{DY}^{i}$ on mass is observed, and a
single value is used for all signal mass intervals.  
The resulting $R_\text{DY}^{i}$ values only weakly depend on the
dimuon category and the data-taking year, and are in the range of 0.8--0.9.
The statistical uncertainties of the measurements do not exceed
a few per cent.  The
systematic uncertainties in $R_\text{DY}^{i}$ are assessed by comparing
$R_\text{DY}^{i}$ measured in individual mass intervals with the
result of the inclusive measurement and by varying
the boundaries and definitions of the auxiliary control regions.
The latter includes
repeating the measurements of $R_\text{DY}^{\text{STA-STA}}$ in the
region with only one STA-to-TMS muon association and of
$R_\text{DY}^{\text{STA-TMS}}$ in the region obtained by requiring
$\LxySig < 1.5$ for the associated TMS-TMS dimuon.
Based on these studies, we assign systematic uncertainties of 15\%  in
$R_\text{DY}^{\text{STA-STA}}$ and 40\% in $R_\text{DY}^{\text{STA-TMS}}$.

In the TMS-TMS dimuon category, the symmetry of the \DeltaPhiAbs
distribution in DY-like backgrounds
is assessed from events in the control region obtained by reversal of the
requirement on $\vchisq$.  We observe a strong correlation between
vertex $\chisq$ and DCA in both data and simulated DY
events, which suggests that $\vchisq$ is effectively a measure of
the distance between the TMS muons forming the dimuon.  Further
studies of TMS-TMS dimuons in DY events passing and failing
the $\vchisq$ requirement confirm that they differ only in how far
away the two muons are reconstructed from each other, and have very
similar properties otherwise.

We use events in the inverted vertex $\chisq$ control region to
evaluate the transfer factor $R_\text{DY}^{\text{TMS-TMS}}$ following
Eq.~(\ref{eq:DYtransferfactor}).  The measured value of
$R_\text{DY}^{\text{TMS-TMS}}$ agrees with unity within the
statistical uncertainties in most $\mMuMu$ intervals and $\dzeroSig$
bins.  Since no systematic trends are observed, we use the value of
$R_\text{DY}^{\text{TMS-TMS}} = 1$ at all masses and in all
min($\dzeroSig$) bins, and assign a 15\% systematic uncertainty to
account for the largest
deviations of $R_\text{DY}^{\text{TMS-TMS}}$ from unity.

\subsection{Estimation of nonprompt backgrounds} \label{subsec:bkg_evaluation_QCD}
The background evaluation method described above is based on the
symmetry of the $\DeltaPhiAbs$ distribution and does not account for the
contributions from background sources that yield di\-muons exclusively
or predominantly at small $\DeltaPhiAbs$.  Such background sources
include: dimuon decays of nonprompt low-mass resonances such as a
$\JPsi$ meson from $\PQb$ hadron decay; cascade decays of $\PQb$ hadrons; 
and dimuons formed from a
pair of unrelated nonprompt muons in the same jet.  If well
reconstructed, most such background events have \mMuMu
not exceeding a few GeV and are rejected by the $\mMuMu > 10\GeV$
requirement.  However, a small fraction of them with mismeasured
\mMuMu can satisfy this requirement and pass
the event selection.  Such dimuons mostly have small
\DeltaPhiAbs values (because the $\pt^{\Pgm\Pgm}$ and \Lxy vectors are
collinear) and may have large, signal-like \LxySig and $\dzeroSig$
values.  They are also likely to have invariant masses close to the
10\GeV threshold.  The other source of nonprompt background consists
of dimuons formed from muons embedded in different jets.  Since all
these nonprompt background events arise from jets produced through the
strong interaction, we collectively refer to them as quantum
chromodynamics (QCD) events.

\begin{figure}[htbp]
  \centering
  \includegraphics[width=0.97\DSquareWidth]{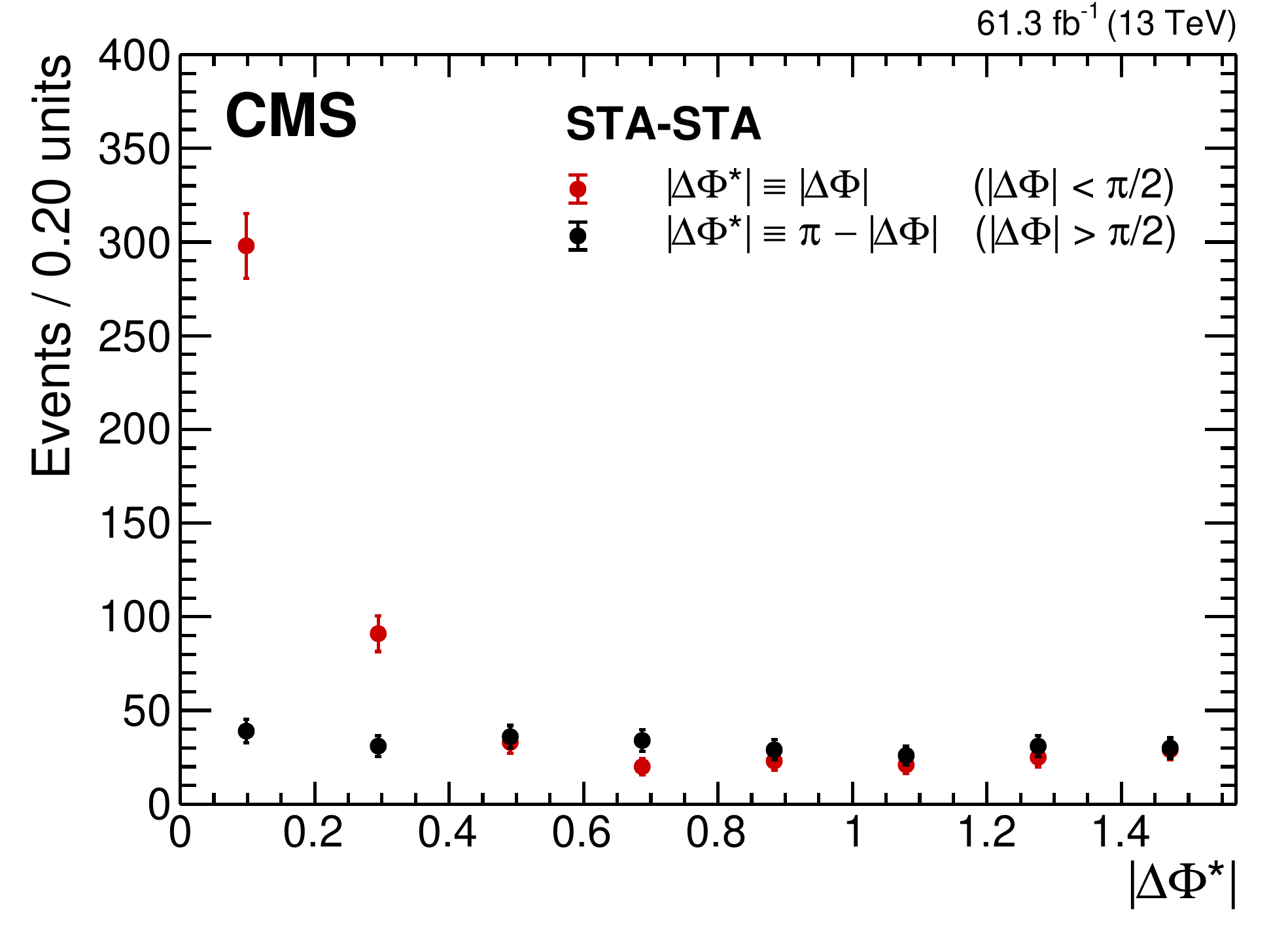}
  \includegraphics[width=0.97\DSquareWidth]{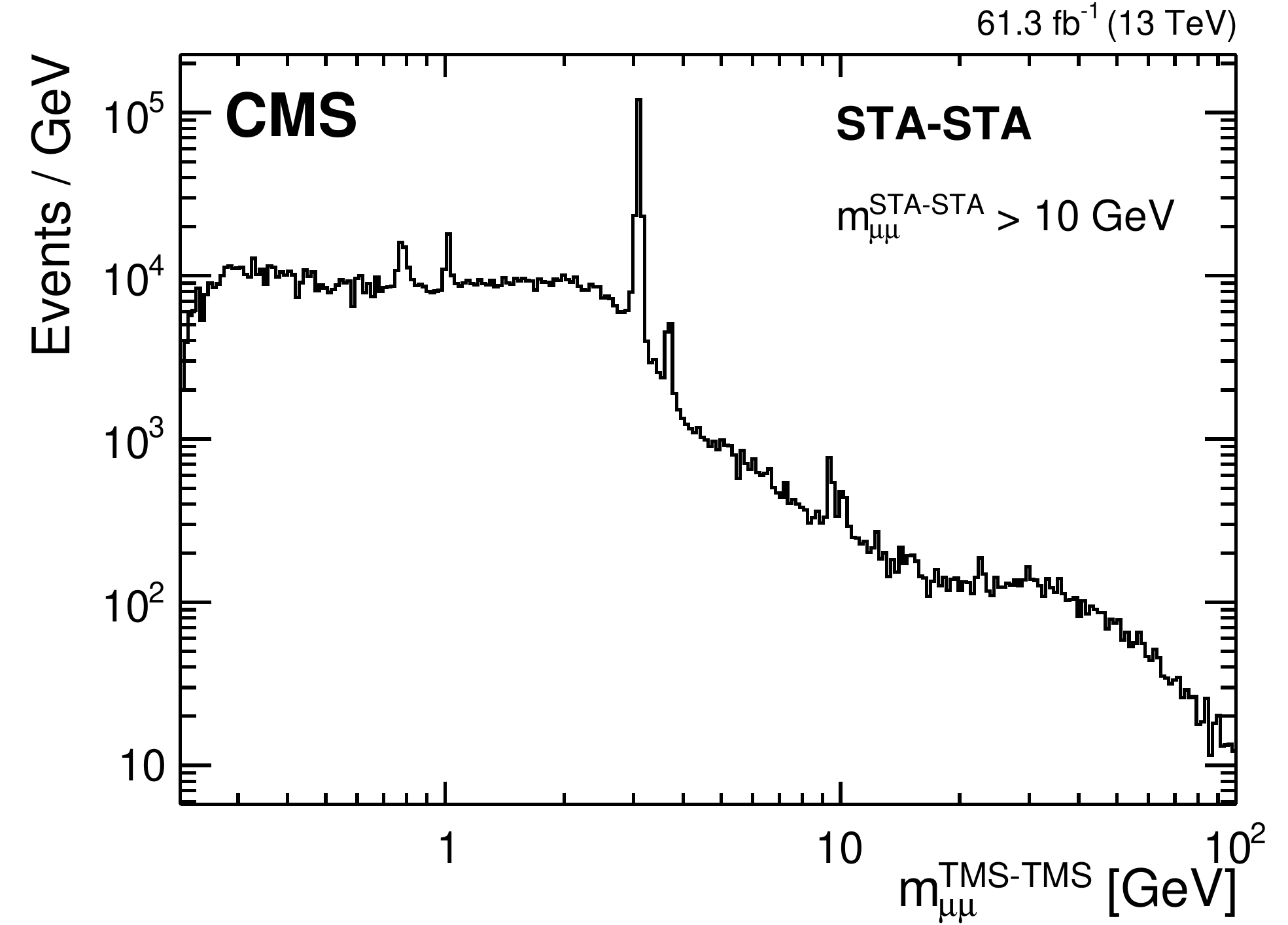}
  \caption{Distributions of (left) $\DeltaPhiAbsStar$ (defined in the
    legend) of STA-STA dimuons and (right) \mMuMu of TMS-TMS dimuons
    associated with STA-STA dimuons.  Both distributions show STA-STA
    dimuons in 2018 data in the control region enriched in QCD events,
    as described in the text.   (The distributions in 2016 data are very
    similar.) In the left plot, exactly one STA muon is associated with
    a TMS muon, while in the right plot, both are.  All nominal
    selection requirements, including $\mMuMu > 10\GeV$ and $\LxySig >
    6$, are applied to the STA-STA dimuons.
  \label{fig:deltaphi_QCD}}
\end{figure}

To gain insight into a contribution from this class of background
events to the STA-STA and STA-TMS categories, we study events in
control regions similar to those described above, but tailored to
select events with muons embedded in jets.  Once again, we invert the
STA-to-TMS association and select STA-STA or, alternatively, STA-TMS
dimuons passing all selection criteria, except that at least one of the
constituent STA muons is associated with a TMS muon.  To
suppress $\DeltaPhiAbs$-symmetric background events such as DY
as well as potential contributions from signal processes, we require
each TMS muon to be nonisolated, defined as $\tkiso > 0.1$ in the
STA-STA and $\tkiso > 0.125$ in the STA-TMS category.  According to
the simulation, this requirement selects a subset of events almost
entirely composed of QCD events.  Figure~\ref{fig:deltaphi_QCD} (left)
shows the \DeltaPhiAbs distribution of OS STA-STA
dimuons in 2018 data in the samples thus obtained.  Unlike the DY
events in Fig.~\ref{fig:deltaphi_DY}, which are approximately
symmetric around $\pi/2$, the QCD events have a signal-like peak at
$\DeltaPhiAbs = 0$.  Most of these events are genuine low-mass dimuons
that are reconstructed at higher \mMuMu because of poor \pt
resolution of STA muons, and hence pass the $\mMuMu > 10\GeV$ requirement.
This is demonstrated by Fig.~\ref{fig:deltaphi_QCD} (right), which
shows the distribution of well-measured \mMuMu of TMS-TMS dimuons
associated with STA-STA dimuons with $\mMuMu > 10\GeV$.

Many of the background processes yielding small-$\DeltaPhiAbs$
OS dimuons also give rise to small-$\DeltaPhiAbs$ SS
dimuons, either because these processes are charge symmetric or via
the muon charge misassignment.  Thus, we evaluate the contribution
from the QCD background to the signal region,
$N_{\text{QCD}}^{i}(\text{OS}; \DeltaPhiAbs < \pi/4)$, from the number
of small-$\DeltaPhiAbs$ SS dimuons, $N^{i}(\text{SS}; \DeltaPhiAbs <
\pi/4)$:
\begin{equation}
   N_{\text{QCD}}^{i}(\text{OS}; \DeltaPhiAbs < \pi/4) =
   N^{i}(\text{SS}; \DeltaPhiAbs < \pi/4) \, R_\text{QCD}^{i}\,.
  \label{eq:QCDestimate}
\end{equation}
The transfer factor $R_{\text{QCD}}^{i}$ between the numbers of QCD
events in these two regions is obtained from the ratio of OS to SS
dimuons in the aforementioned control region with the STA-to-TMS
association reversed and TMS muons not isolated:
\begin{equation}
R_\text{QCD}^{i} = \frac{N^{\text{rev},i}_{\text{QCD}}(\text{OS};
  \DeltaPhiAbs < \pi/4)}{N^{\text{rev},i}_{\text{QCD}}(\text{SS};
  \DeltaPhiAbs < \pi/4)}\,.
  \label{eq:QCDtransferfactor}
\end{equation}
Because the charge misassignment and the probability of finding the
STA-to-TMS association were found to be anticorrelated, the measurement of
$R_{\text{QCD}}^{i}$ in the STA-STA category is performed using
dimuons with exactly one of the constituent STA muons associated with
a TMS muon, providing events that are more representative of those in the
signal region.
Since the composition of the QCD background varies as a function
of \mMuMu, the evaluation of $R_\text{QCD}^{i}$ is
performed separately in the individual mass intervals, with the
exception of the STA-STA category, where a common value is used for
$\mMuMu > 35$\GeV to avoid large statistical fluctuations of
$R_\text{QCD}^{\text{STA-STA}}$.  The measured values of
$R_\text{QCD}^{i}$ in these two categories vary between 1.1 and 2.3
depending on the mass interval and year, with statistical
uncertainties in the range of 10--30\% in the STA-STA and 2--20\% in
the STA-TMS category.  The systematic uncertainties in
$R_\text{QCD}^\text{STA-STA}$ are assessed by evaluating the potential
impact on $R_\text{QCD}^\text{STA-STA}$ of the correlation between the
success rate of the STA-to-TMS association and STA muon charge
misassignment.  The systematic uncertainties in
$R_\text{QCD}^{\text{STA-TMS}}$ are evaluated by varying the
definitions of the auxiliary control regions, \eg, performing the
measurement of $R_\text{QCD}^{\text{STA-TMS}}$ in the region obtained
by inverting the isolation requirement applied to the TMS muon.  Based
on these studies, we assign a systematic uncertainty in the range of
10--30\% in $R_\text{QCD}^{\text{STA-STA}}$, depending on the mass
interval and the year, and a fixed 30\% systematic uncertainty in
$R_\text{QCD}^{\text{STA-TMS}}$.

A priori, we do not expect a large contribution from
$\DeltaPhiAbs$-asymmetric low-mass dimuons in the TMS-TMS category,
because of a far superior dimuon invariant mass resolution.  Indeed,
the study of simulated QCD events shows that a vast majority of both
OS and SS TMS-TMS dimuons passing all selection
criteria arise from a pair of unrelated nonprompt muons in two
different jets.
Such events do not contain genuine displaced dimuons and are expected
to have a symmetric $\DeltaPhiAbs$ distribution.  Nevertheless, since
some contribution from $\DeltaPhiAbs$-asymmetric dimuons may still be
present in the background events in data, we prefer not to rely on the
$\DeltaPhiAbs$ symmetry in the evaluation of nonprompt backgrounds.
Instead, similarly to the STA-STA and STA-TMS categories, we use
the fact that dijet and multijet events give rise to both OS and
SS dimuons, and base our estimate of the QCD background on
the number of SS dimuons following Eq.~(\ref{eq:QCDestimate}).

The transfer factor $R_{\text{QCD}}^{\text{TMS-TMS}}$ is obtained from
the ratio of OS to SS dimuons in the control region with the muon
isolation requirement reversed, which comprises dimuons passing the
nominal event selection but with at least one muon with $\tkiso >
0.075$ and both with $\tkiso < 0.5$.  We have verified that these
events, as well as SS dimuons passing isolation requirements, contain
negligible contributions from signal and DY events.
As the signal region is divided into several min($\dzeroSig$) bins, the
evaluation of $R_{\text{QCD}}^{\text{TMS-TMS}}$ is performed
separately in each min($\dzeroSig$) bin.  Since no dependence of the
value of $R_{\text{QCD}}^{\text{TMS-TMS}}$ on $\mMuMu$ is observed,
$R_{\text{QCD}}^{\text{TMS-TMS}}$ in each min($\dzeroSig$) bin is
calculated by integrating events in the entire invariant mass spectrum.
The measured values of $R_{\text{QCD}}^{\text{TMS-TMS}}$ decrease from
${\approx}2$ to ${\approx}1$ as min($\dzeroSig$) increases, with statistical
uncertainties in the range of 5--20\%.
A systematic uncertainty of 15\% is assigned to account for variations of
$R_{\text{QCD}}^{\text{TMS-TMS}}$ as a function of the invariant mass
and as the result of changing the definition and boundaries of the
auxiliary control region.

To avoid overestimating the DY
background, the same QCD background evaluation method is applied to
dimuons in the $\DeltaPhiAbs > 3\pi/4$ control region.  The obtained
estimate of the QCD background is then subtracted from the total
observed number of OS dimuons with $\DeltaPhiAbs > 3\pi/4$ to obtain
the number of DY dimuons in this $\DeltaPhiAbs$ region,
$N_{\text{DY}}^{i}(\text{OS}; \DeltaPhiAbs > 3\pi/4)$, used for the
evaluation of the DY backgrounds in the signal $\DeltaPhiAbs <
\pi/4$ region according to Eq.~(\ref{eq:DYestimate}).  This procedure
is not applied in the STA-STA category, where the
$\DeltaPhiAbs$-symmetric QCD background is negligible.  The sum of the
QCD and DY background estimates constitute the total predicted
background in the signal region.  According to the background
evaluation method, the DY backgrounds are
expected to dominate at small $\dzeroSig$ and $\LxySig$ values,
whereas the relative QCD contribution becomes larger as $\dzeroSig$
and $\LxySig$ increase.  The uncertainty in the background
predictions is dominated by the statistical uncertainty in the numbers
of events in the $\DeltaPhiAbs > 3\pi/4$ and SS control
regions.

\subsection{Validation of background predictions} \label{subsec:bkg_validation}
\begin{figure}[htbp]
  \centering
  \includegraphics[width=0.97\DSquareWidth]{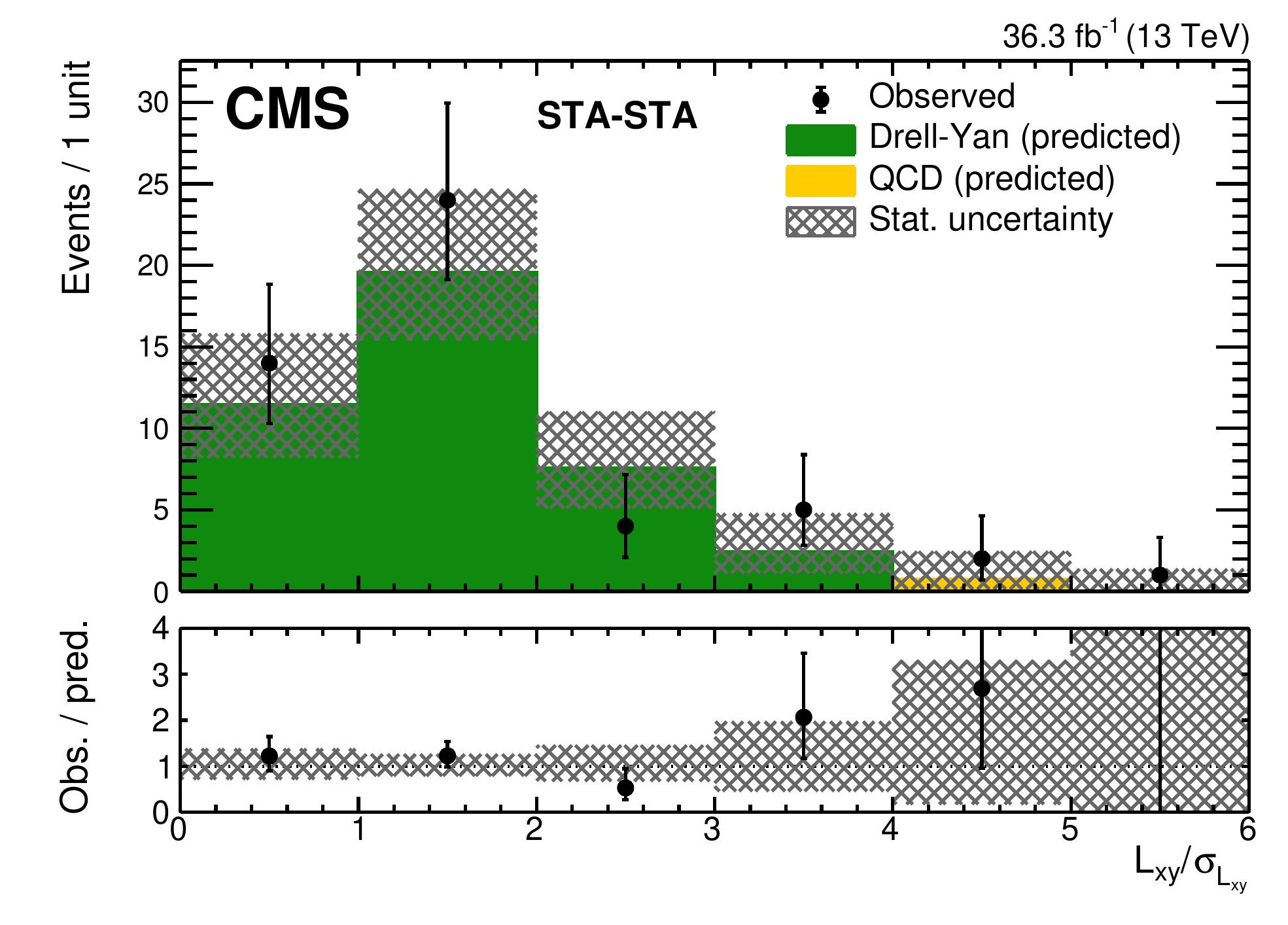}
  \includegraphics[width=0.97\DSquareWidth]{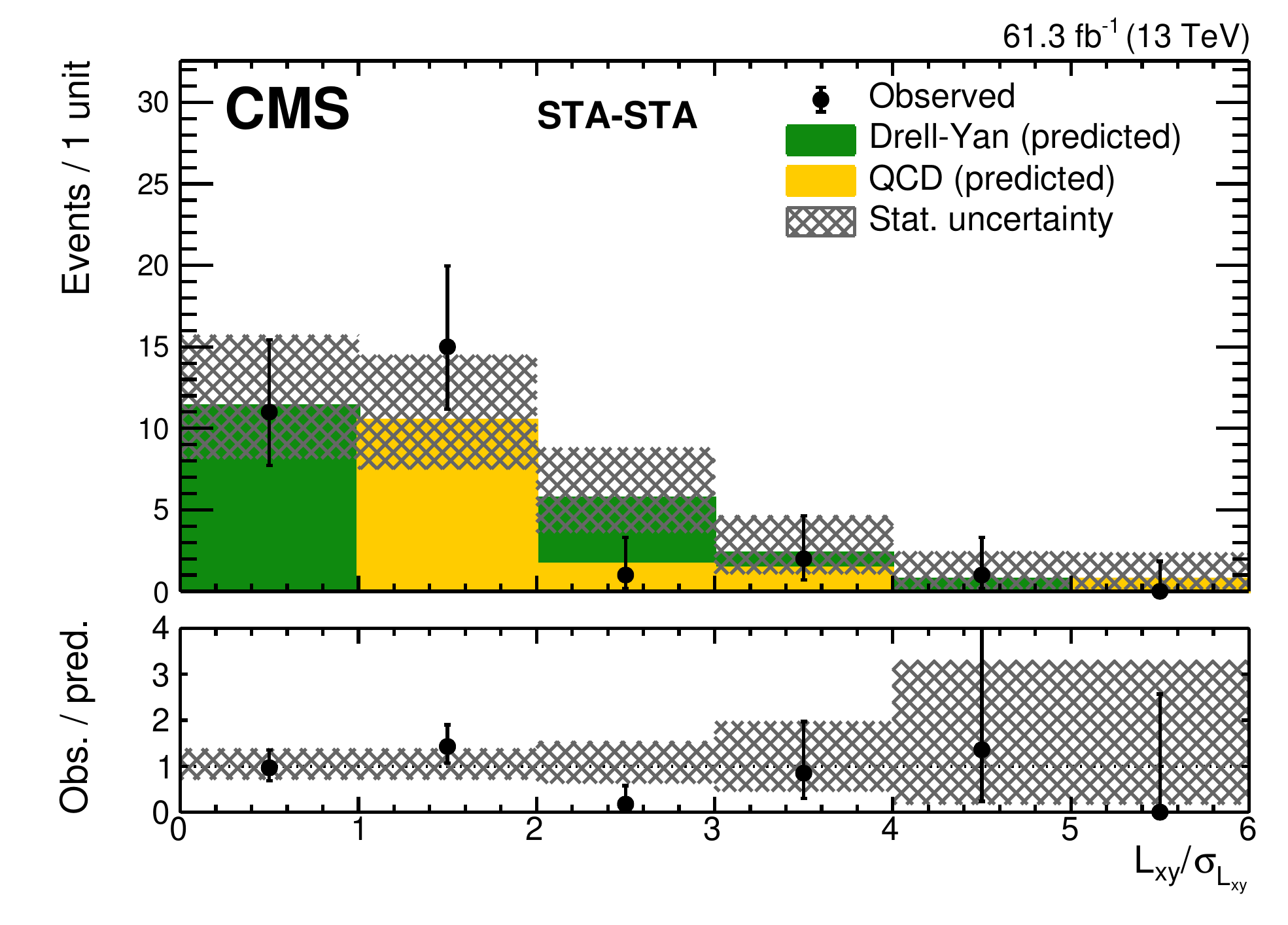}
  \caption{Distributions of $\LxySig$ of STA-STA dimuons in the
    $\LxySig < 6$ VR, in (left) 2016 and (right) 2018
    data, compared to the background predictions.  The observed
    distributions (black points with error bars) are overlayed on stacked histograms
    containing the expected numbers of DY (green) and QCD (yellow)
    background events.  The lower panels show the ratio of the
    observed to predicted numbers of events.  The shaded area shows
    the statistical uncertainty in the total background prediction;
    the admixture of the QCD background in this validation region is
    estimated separately and has a larger statistical uncertainty than
    the total background.
    \label{fig:STASTA_LxySig_closure}}
\end{figure}

The background evaluation method described in
Sections~\ref{subsec:bkg_evaluation_DY} and
\ref{subsec:bkg_evaluation_QCD} is tested in several validation
regions (VRs) that are expected to contain negligible contribution from
signal.  The evaluation of DY backgrounds is
examined in the VRs obtained by inverting the $\LxySig$ and
$\dzeroSig$ requirements and thereby enriched in this class of events.  One
example of such studies is shown in
Fig.~\ref{fig:STASTA_LxySig_closure}, which compares the background
predictions to the observed distributions in the $\LxySig < 6$
VR in the STA-STA category.  The yields in data are
consistent with predictions of the method, which also correctly
predicts a larger STA-STA background in 2016 compared to 2018 due to a
lower tracking efficiency in a part of 2016
data~\cite{Sirunyan:2018njd}.

\begin{figure}[thbp]
  \centering
  \includegraphics[width=0.97\DSquareWidth]{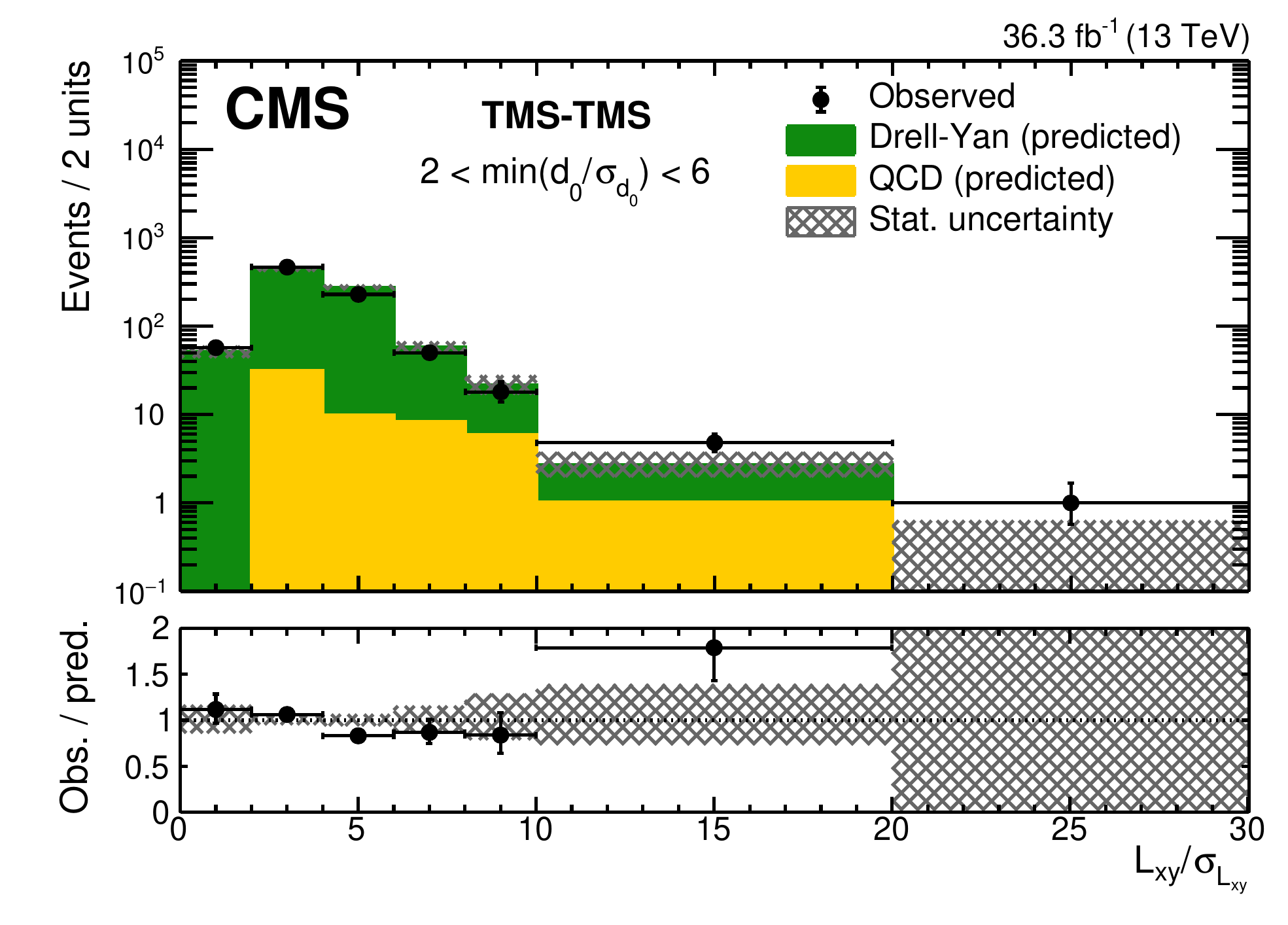}
  \includegraphics[width=0.97\DSquareWidth]{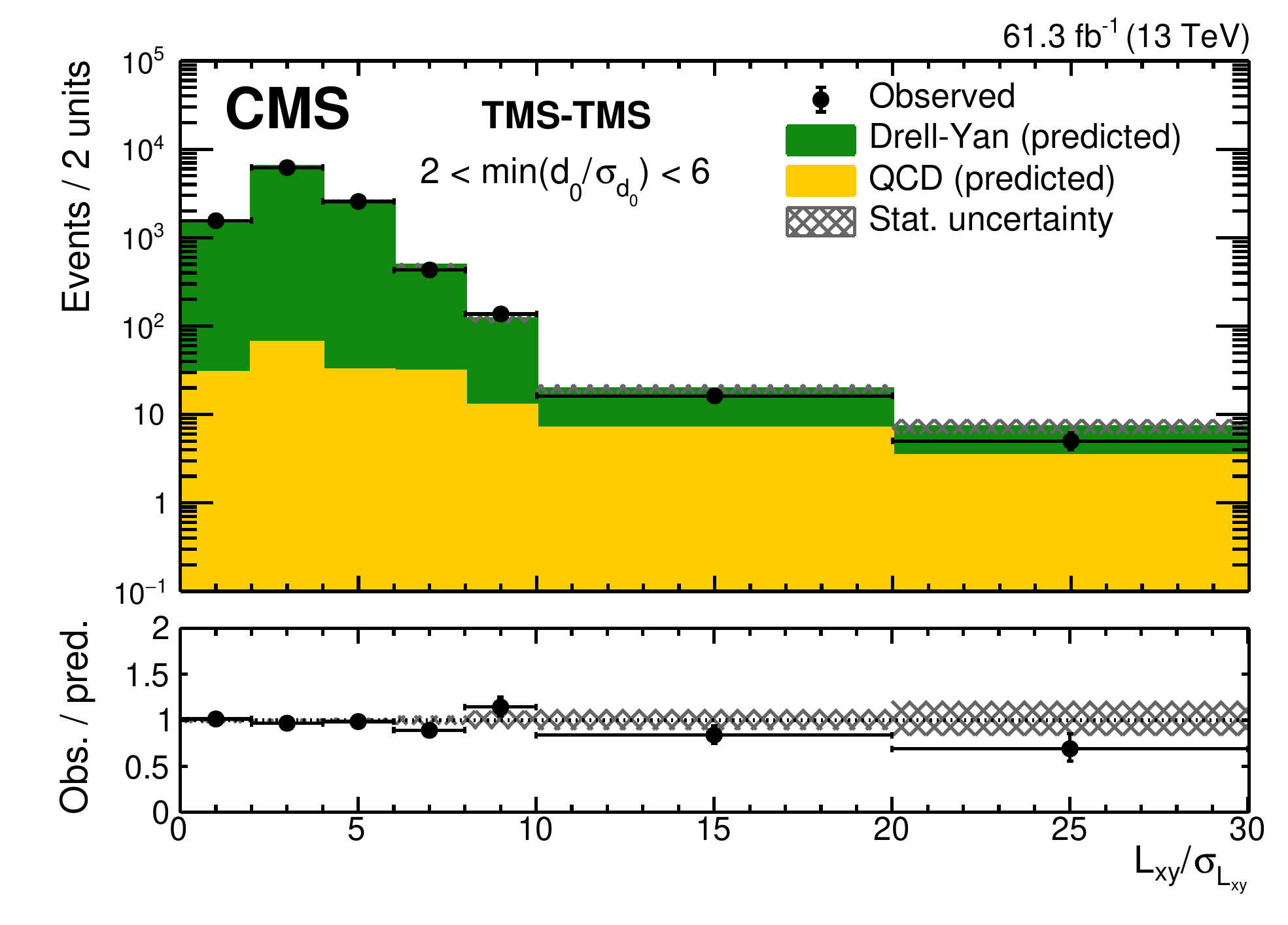}
  \caption{Distributions of $\LxySig$ of TMS-TMS dimuons in the $2 <
    \text{min}(\dzeroSig) < 6$ VR, in (left) 2016 and
    (right) 2018 data, compared to the background predictions.
    The observed distributions (black points with error bars) are
    overlayed on stacked histograms containing the expected numbers of
    DY (green) and QCD (yellow) background events.  The last bin
    includes events in the overflow.  The lower panels show the ratio of the
    observed to predicted numbers of events.  The shaded area shows
    the statistical uncertainty in the background prediction.
    \label{fig:TMSTMS_LxySig_DPHI1_closure}}
\end{figure}

In another check, we apply the background evaluation procedure to the
TMS-TMS dimuons in the $2 < \text{min}(\dzeroSig) < 6$ sideband.  The
comparison of the predicted background and data in bins of $\LxySig$
is shown in Fig.~\ref{fig:TMSTMS_LxySig_DPHI1_closure}.  The expected
and observed numbers of events are in agreement in the entire probed
$\LxySig$ range.  There are more background events in 2018 data than
in 2016 data because of looser trigger requirements and larger
integrated luminosity.

\begin{figure}[htbp]
 \centering
  \includegraphics[width=\DFigWidth]{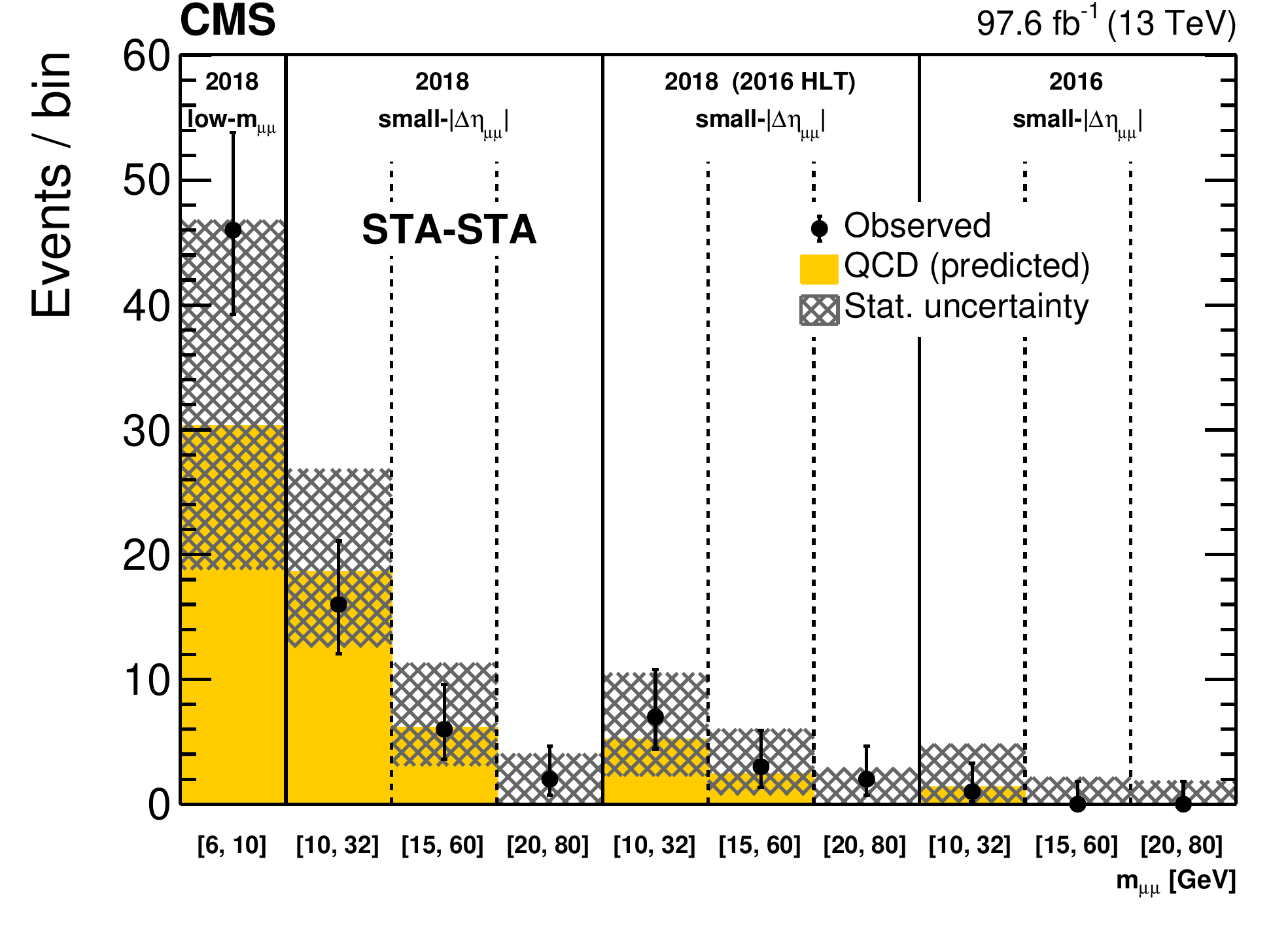}
  \caption{Comparison of observed (black points with error bars) and predicted
    (histograms) yields of STA-STA dimuons in the validation regions
    enriched in QCD background events.  The first bin shows the yields
    in the low-mass VR in 2018 data.  The other three groups of bins,
    separated by solid lines, show the yields in the small-\DetaMuMu
    VR, in (from left to right) the entire 2018 data set, a subset of
    2018 data
    enriched in events collected in 2016, and the 2016 data set.
    Each of these three VRs is further subdivided into
    three \mMuMu intervals, 10--32, 15--60, and 20--80\GeV.  The expected
    number of background events is computed according to
    Eqs.~(\ref{eq:QCDestimate}) and (\ref{eq:QCDtransferfactor}),
    separately in each $\mMuMu$ bin.  The shaded area shows the
    statistical uncertainty in the background
    prediction.  \label{fig:STASTA_lowM_closure}}
\end{figure}

The evaluation of the $\DeltaPhiAbs$-asymmetric component of QCD
backgrounds, which is particularly important in the STA-STA
category, is tested in the low-mass ($6 < \mMuMu <
10\GeV$) VR, as well as in the region obtained by
inverting the requirement on the minimum number of DT hits and muon
segments applied to dimuons with $\DetaMuMu < 0.1$, referred to as the
small-\DetaMuMu VR.  Using dimuons with STA muons associated with TMS muons
and taking well-measured \mMuMu and $\DeltaPhiAbs$ values of corresponding
TMS-TMS dimuons as proxies for true values of these quantities, we
have verified that
the samples of events in these VRs predominantly consist of
small-$\DeltaPhiAbs$ dimuons with mismeasured \mMuMu.
Figure~\ref{fig:STASTA_lowM_closure} shows the comparison of the
predicted background yields and data in these two VRs.  The low-mass VR is
only available in 2018 data because the trigger used to
collect 2016 data for this analysis included the $\mMuMu > 10\GeV$ requirement.  The
small-\DetaMuMu VR is available in both data sets, but since the
number of events in this VR in 2016 data is small, an additional test
is performed on a subset of 2018 data collected using the 2016 trigger,
and therefore enriched in events similar to those recorded in 2016.
The \mMuMu intervals of 10--32, 15--60, and 20--80\GeV shown for the
small-\DetaMuMu VRs are the intervals chosen to probe LLP masses of
20, 30, and 50\GeV, respectively.
The yields in data are found to be consistent with background
predictions in all tests and \mMuMu intervals.

\begin{figure}[htbp]
  \centering
  \includegraphics[width=0.97\DSquareWidth]{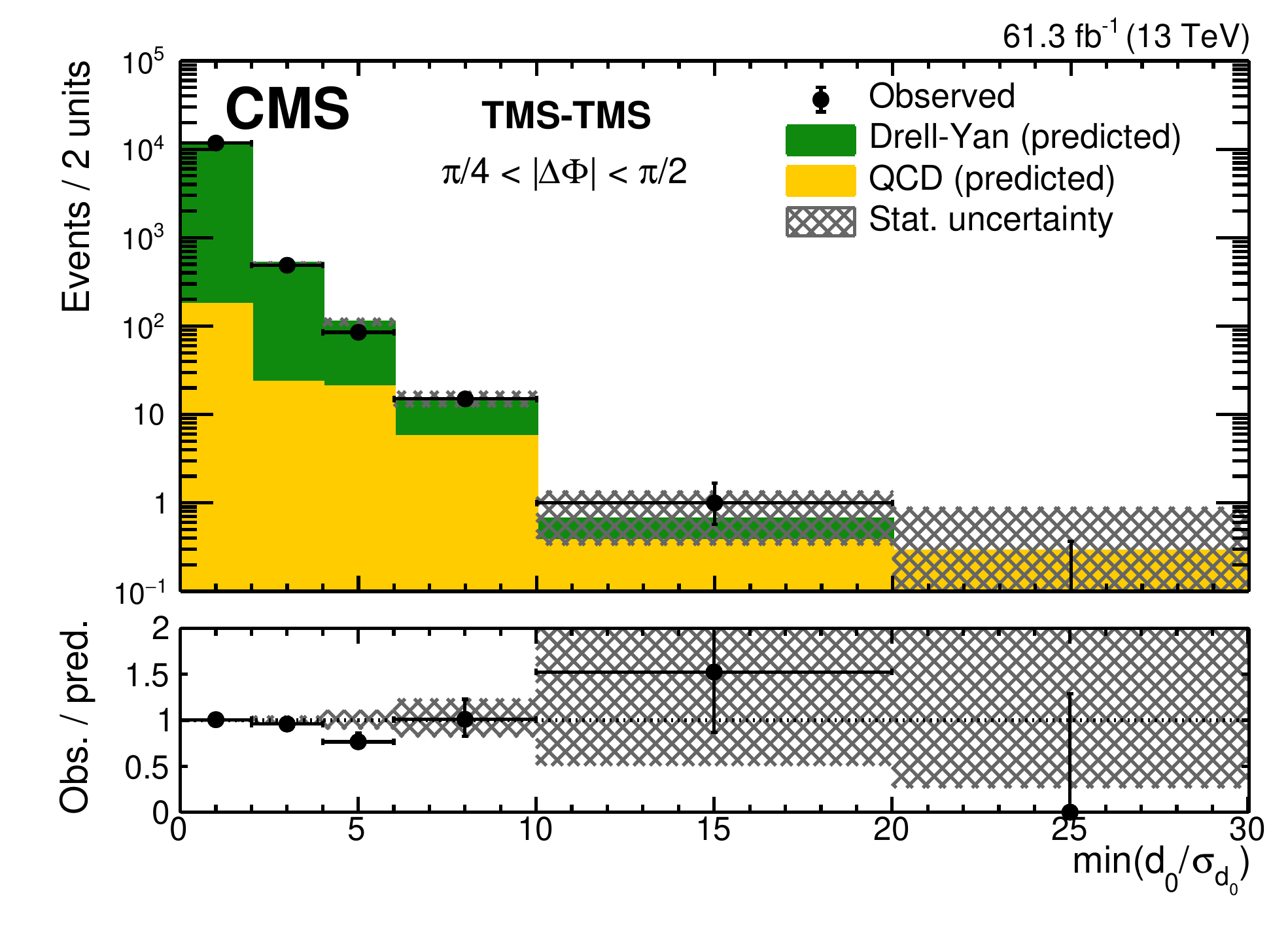}
  \includegraphics[width=0.97\DSquareWidth]{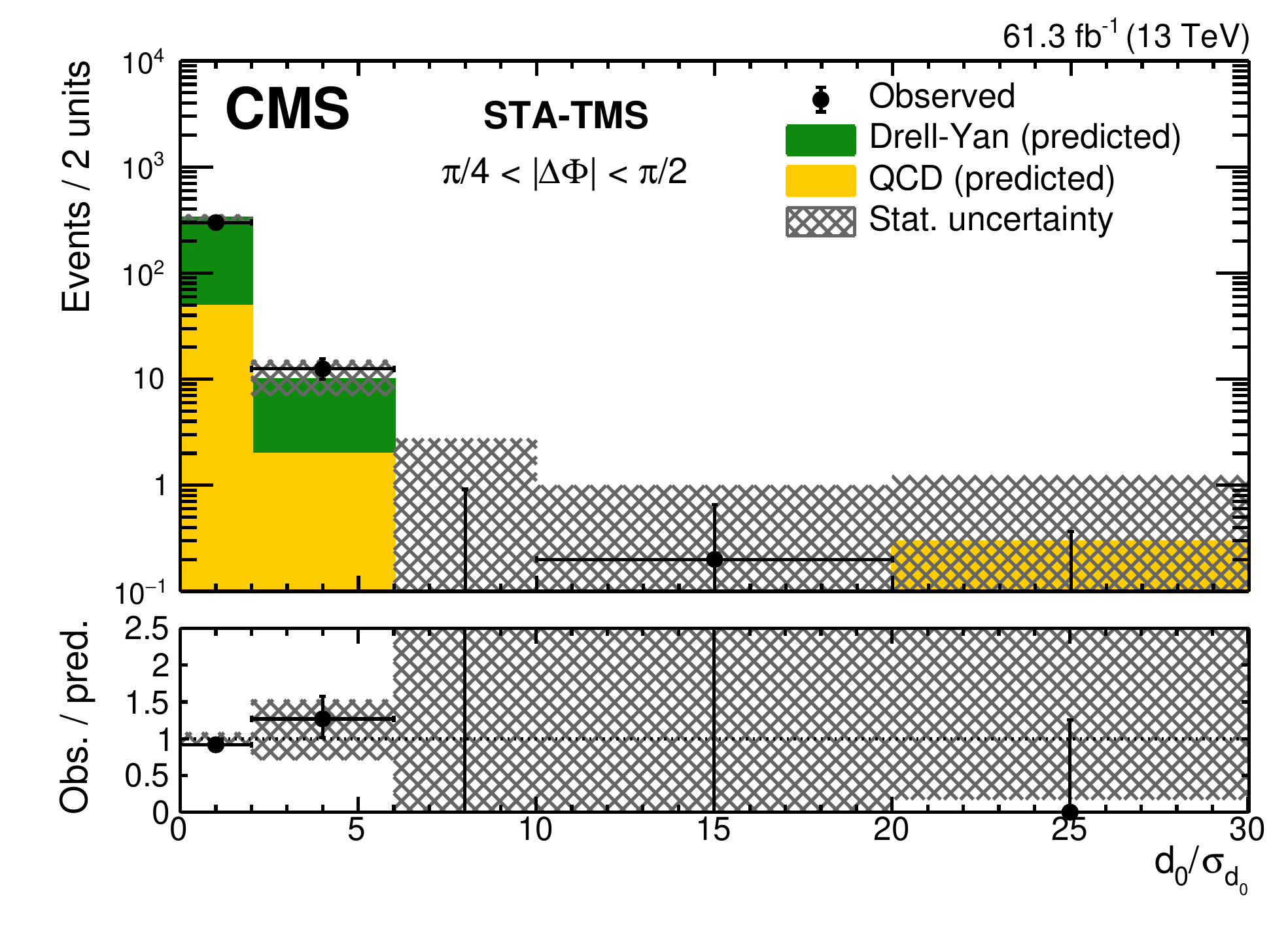}
  \caption{Distributions in the $\pi/4 < \DeltaPhiAbs < \pi/2$
    VR in 2018 data: (left) the smaller of the two
    $\dzeroSig$ values for the TMS-TMS dimuon; (right) $\dzeroSig$ of
    the TMS muon in the STA-TMS dimuon.  The observed distributions
    (black points with error bars) are compared to the results of the background
    prediction method applied to events with $\pi/2 < \DeltaPhiAbs <
    3\pi/4$.  The stacked histograms show the expected numbers of
    DY (green) and QCD (yellow) background events.  The last bin
    includes events in the overflow. The lower panels show the ratios
    of the observed to predicted numbers of events.  The shaded area
    shows the statistical uncertainty in the background prediction.
    \label{fig:d0sigpv_DPHI2_closure}}
\end{figure}

Finally, to ensure the validity of the method at different
values of the main discriminating variable in the TMS-TMS and STA-TMS
categories, the validation checks are performed in bins of $\dzeroSig$
of the TMS muon.  Such checks include comparisons in the $\dzeroSig$
sideband ($\dzeroSig < 6$) in the signal \DeltaPhiAbs region, as well
as those in the entire $\dzeroSig$ range in the $\DeltaPhiAbs$ sideband,
$\pi/4 < \DeltaPhiAbs < \pi/2$.  In the latter, the region with
$\pi/4 < \DeltaPhiAbs < \pi/2$ is used as a signal-free proxy for the
$\DeltaPhiAbs < \pi/4$ signal region.  The background evaluation
procedure is applied to the OS and SS dimuons in the
$\DeltaPhiAbs$-symmetric region, $\pi/2 < \DeltaPhiAbs < 3\pi/4$, as
well as SS dimuons with $\pi/4 < \DeltaPhiAbs < \pi/2$.  The
comparisons of the predicted background and 2018 data in the TMS-TMS
and STA-TMS categories in this VR are shown in
Fig.~\ref{fig:d0sigpv_DPHI2_closure}.  The observed and expected
numbers of events are consistent within statistical uncertainties.

\section{Systematic uncertainties affecting the signal} \label{sec:systunc_signal}
Most of the systematic uncertainties affecting the signal efficiencies
are evaluated separately in each dimuon category and for each
data-taking year.  Unless stated otherwise, we consider sources of
uncertainties to be uncorrelated among different dimuon categories and
years.

In the STA-STA and STA-TMS categories, the dominant systematic
uncertainties come from the STA muon identification and trigger
efficiencies.  At small displacements, both efficiencies are
accurately measured as a function of muon \pt and $\eta$ by applying
the ``tag-and-probe method''~\cite{Chatrchyan:2012xi} to muons from $\JPsi$ meson and $\PZ$ boson decays.  The
differences in the identification and trigger efficiencies between
data and simulation are used to correct the signal simulation yields.
In the STA-STA category, these corrections range from 0.78 to 1.13,
depending on the signal sample.  
The evolution of efficiencies with displacement is studied using a
sample of cosmic ray muons collected during periods with no beam,
and additional $\dzero$-dependent corrections and systematic
uncertainties are derived.  At $\dzero = 10$ (100)\cm, the correction
amounts to 0.99 (0.96) per muon,
whereas the uncertainty is on the order of 10 (35)\%.  Since the
$\dzero$-dependent uncertainty is dominated by the accuracy of the
\Lone trigger efficiency measurements, it is taken to be correlated
among all dimuon categories.

The dominant systematic uncertainties in the TMS-TMS category come
from the $\dzero$ dependence of the \Lone trigger efficiency discussed
above and the efficiency to
reconstruct displaced muons in the tracker.  The evolution of the
tracking efficiency with $\dzero$ is measured using a sample of cosmic
ray muons and compared to the tracking efficiency predicted by
simulation.  Based on the results of the comparison, we assign a 5\%
systematic uncertainty per muon for muons with $\dzero > 1\cm$.  The
overall efficiency corrections applied to the simulated signal yields
range from 0.74 to 1.08, depending on the signal sample, and arise
mostly from imperfect modeling of the HLT efficiencies at small
displacements.

The remaining systematic uncertainties related to the signal
efficiency are much smaller.  The impact of mismodeling of the muon
\pt resolution on the signal yield is evaluated by smearing the muon
\pt in simulated signal events according to the measurements
performed using cosmic ray muons and muons from $\PZ$ boson decays.  This
leads to variations that are less than 2\% at all signal masses
except for $\mZD = 10\GeV$.  Corrections of up to 2\% are applied 
to the TMS muon efficiency to account for the difference in
efficiency of isolation requirements measured using muons from $\PZ$ boson
decays, and an additional systematic uncertainty of 2\% is assigned.
A systematic uncertainty ranging from 1 to 8\%, depending on the
signal sample, is assigned to account for mismodeling of the DCA
requirement in the STA-STA category. 
The efficiency of the vertex $\chisq$
requirement as a function of displacement is studied using cosmic ray
muons and muons from $\PZ$ boson decays in the STA-STA category, and muons
from decays of nonprompt $\JPsi$ mesons in the TMS-TMS category.    The
differences between data and simulation contribute an uncertainty of
2\% in each category.  The efficiencies of several other selection
criteria, such as requirements on the number of tracker hits upstream
of the vertex position and the difference between the number of pixel
hits on two TMS muons, are found to be well modeled by simulation, and
no additional uncertainty is assigned.

The uncertainty in the integrated luminosity, partially correlated
between the years, is 1.2\% in
2016~\cite{CMS:2021xjt} and 2.5\% in
2018~\cite{CMS-PAS-LUM-18-002}.  The uncertainty in the signal
efficiency due to pileup is 2\%.  Both uncertainties are correlated
among dimuon categories.

\section{Results} \label{sec:results}
\begin{figure}[htbp]
  \centering
  \includegraphics[width=0.97\DSquareWidth]{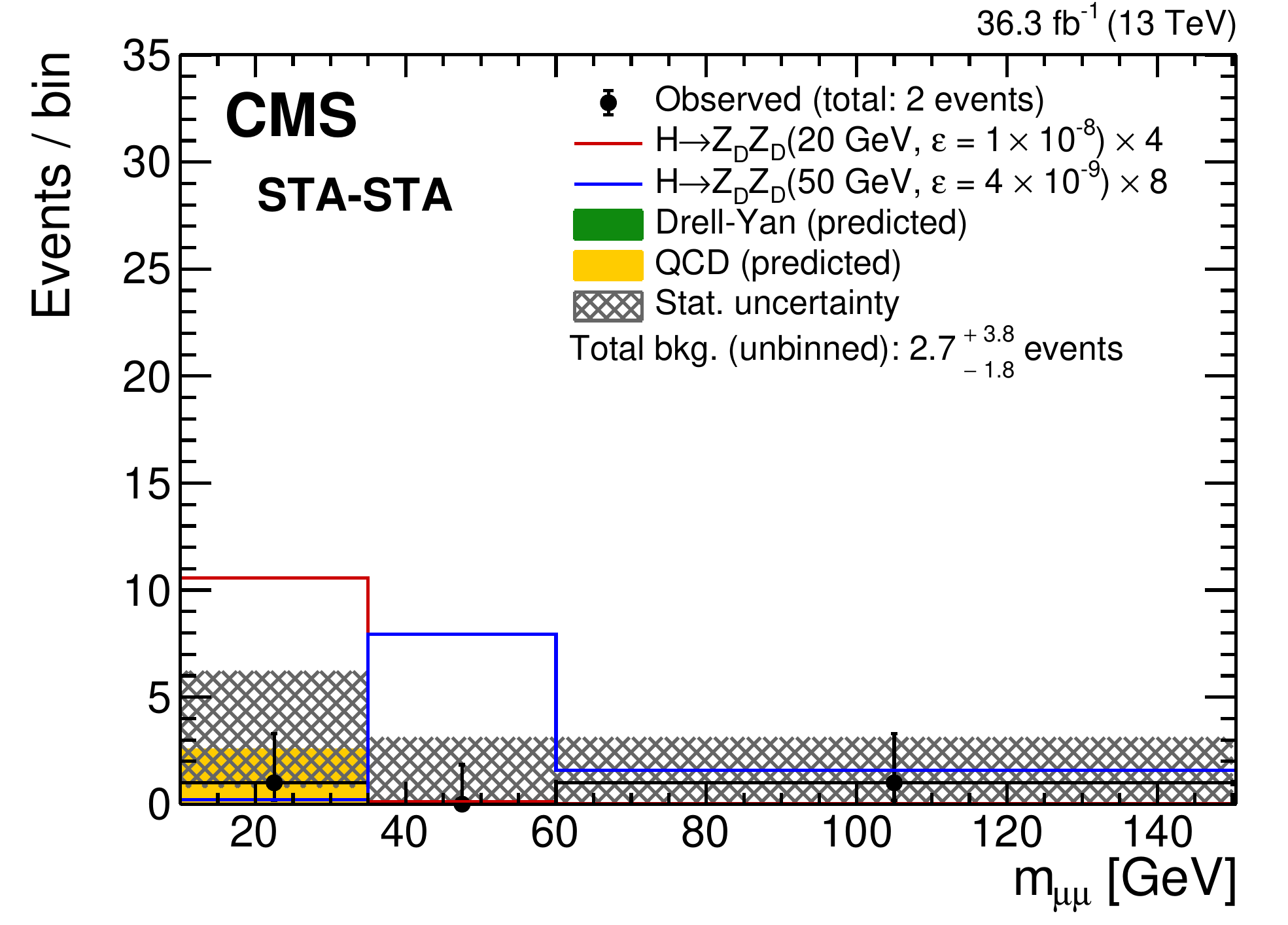}
  \includegraphics[width=0.97\DSquareWidth]{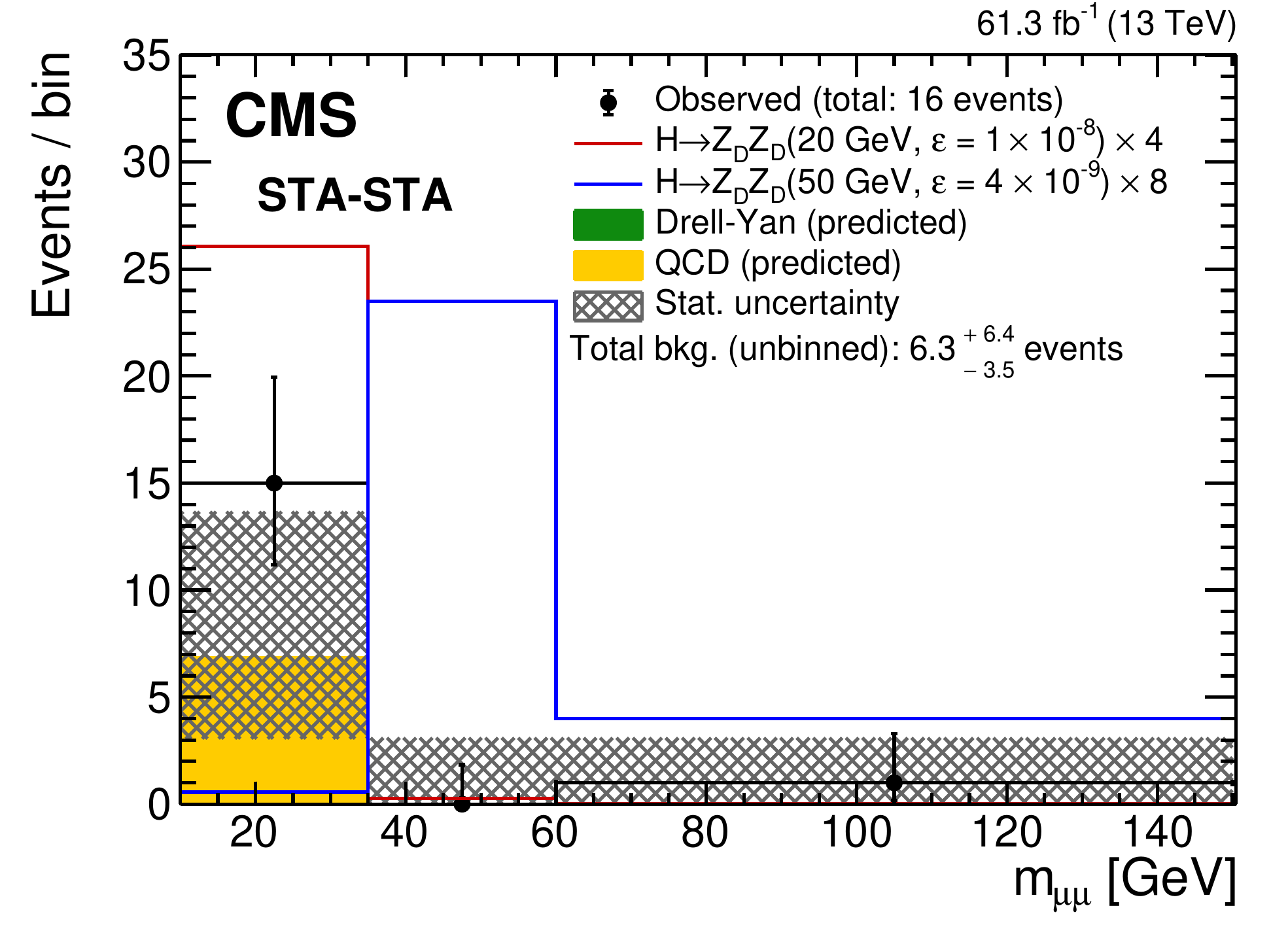}
  \caption{Comparison of the number of events observed in (left)
    2016 and (right) 2018 data in the STA-STA dimuon category with the
    expected number of background events, in representative \mMuMu
    intervals.  The black points with error bars show the number of observed events;
    the green and yellow components of the stacked histograms
    represent the estimated numbers of DY and QCD events,
    respectively.  The last bin includes events in the overflow.  The
    uncertainties in the total expected background (shaded area) are
    statistical only.  Signal contributions expected from simulated
    $\PSMHiggs \to \PZD\PZD$ with $\mZD$ of 20 and 50\GeV are shown in
    red and blue, respectively. Their yields are set to the corresponding 
    combined median expected exclusion limits at 95\% \CL,
    scaled up as indicated in the legend to improve visibility.
    The legends also include the total number of
    observed events as well as the number of
    expected background events obtained inclusively, by applying the
    background evaluation method to the events in all $\mMuMu$ intervals
    combined.  \label{fig:obs_vs_exp_invmass_STASTA}}
\end{figure}

\begin{figure}[htbp]
  \centering
  \includegraphics[width=0.97\DSquareWidth]{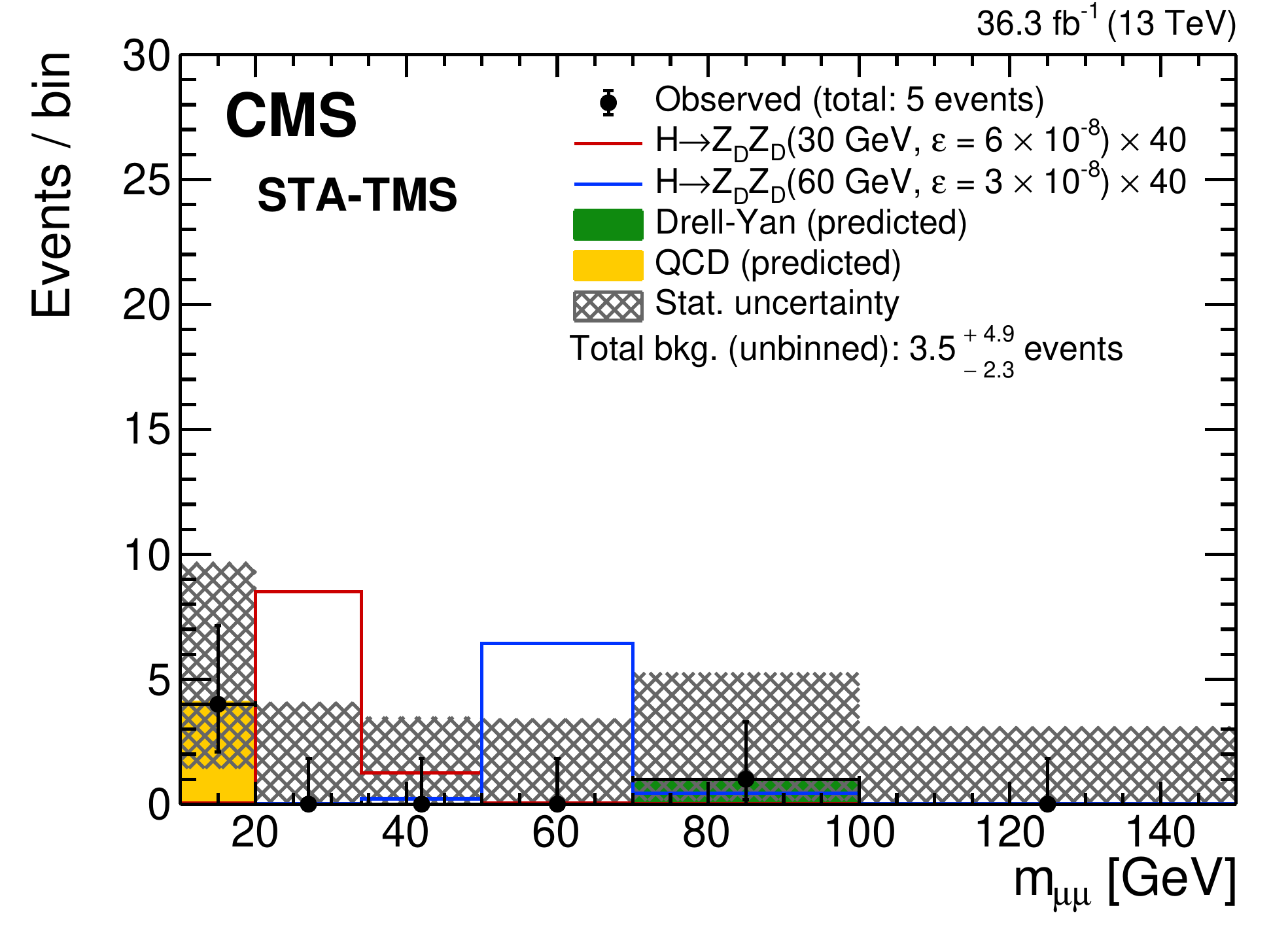}
  \includegraphics[width=0.97\DSquareWidth]{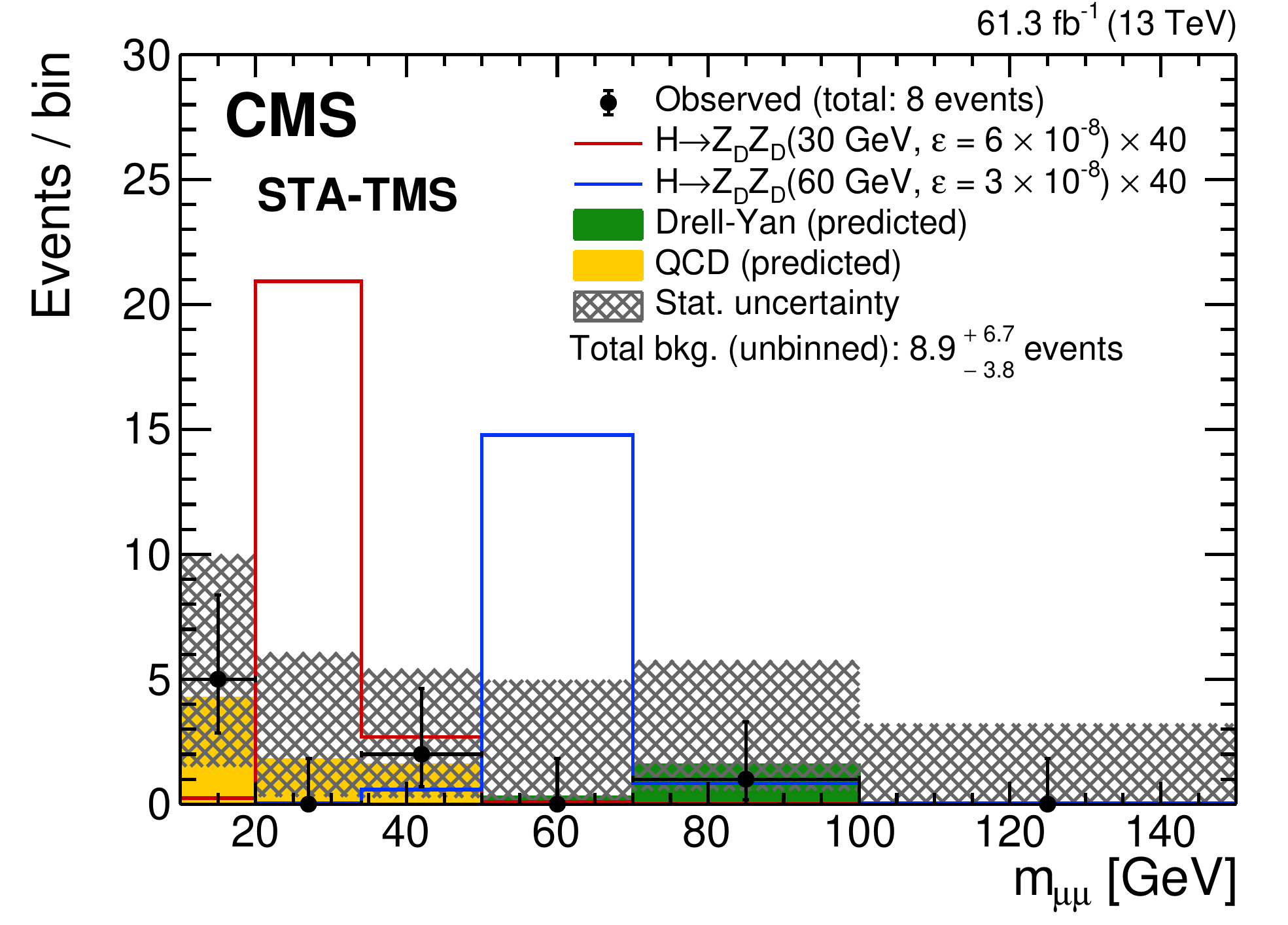}
  \caption{Comparison of the number of events observed in (left)
    2016 and (right) 2018 data in the STA-TMS dimuon category with the
    expected number of background events, in representative \mMuMu
    intervals.  The black points with error bars show the number of observed events; the
    green and yellow components of the stacked histograms represent
    the estimated numbers of DY and QCD events, respectively.  The
    last bin includes events in the overflow.  The uncertainties in
    the total expected background (shaded area) are statistical only.
    Signal contributions expected from simulated
    $\PSMHiggs \to \PZD\PZD$ with $\mZD$ of 30 and 60\GeV are shown in
    red and blue, respectively. Their yields are set to the corresponding 
    combined median expected exclusion limits at 95\% \CL,
    scaled up as indicated in the legend to improve visibility.
    The legends also include the total number of observed events as well as
    the number of expected background events obtained inclusively,
    by applying the background evaluation method to
    the events in all $\mMuMu$ intervals
    combined.  \label{fig:obs_vs_exp_invmass_STATMS}}
\end{figure}

\begin{figure}[htbp]
  \centering
  \includegraphics[width=2.\DSquareWidth]{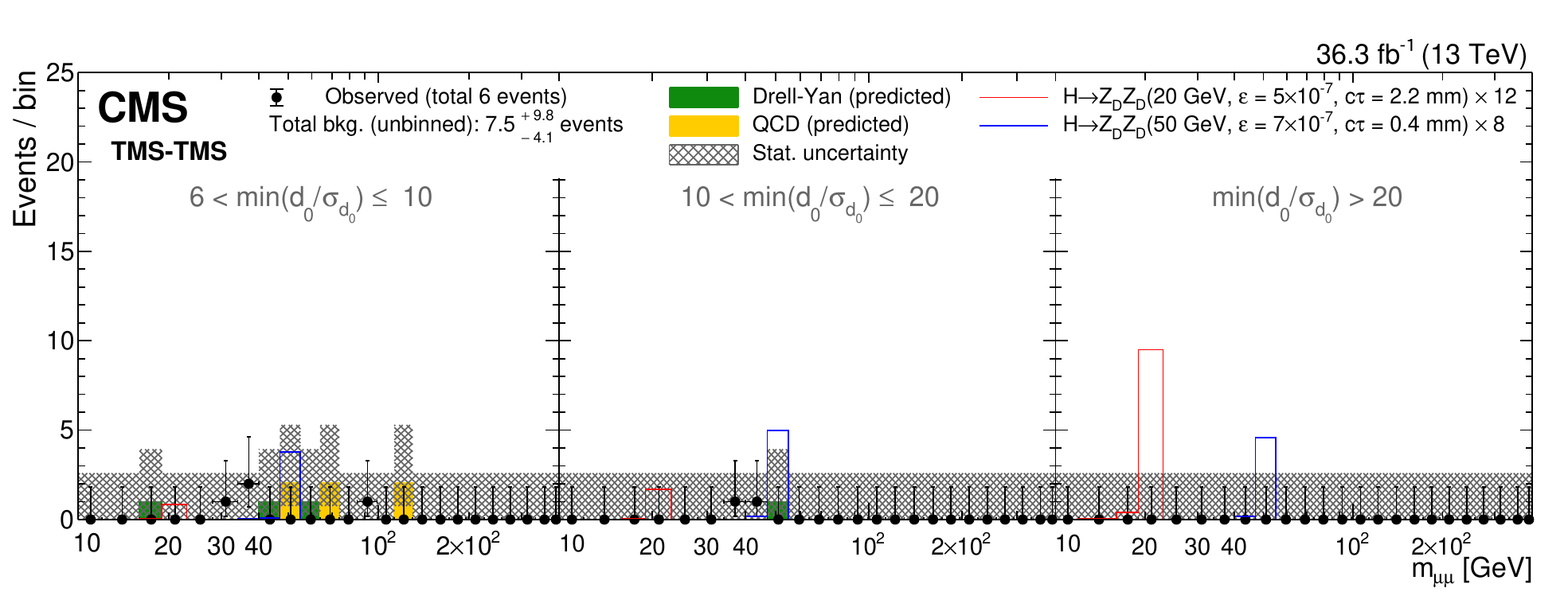}
  \includegraphics[width=2.\DSquareWidth]{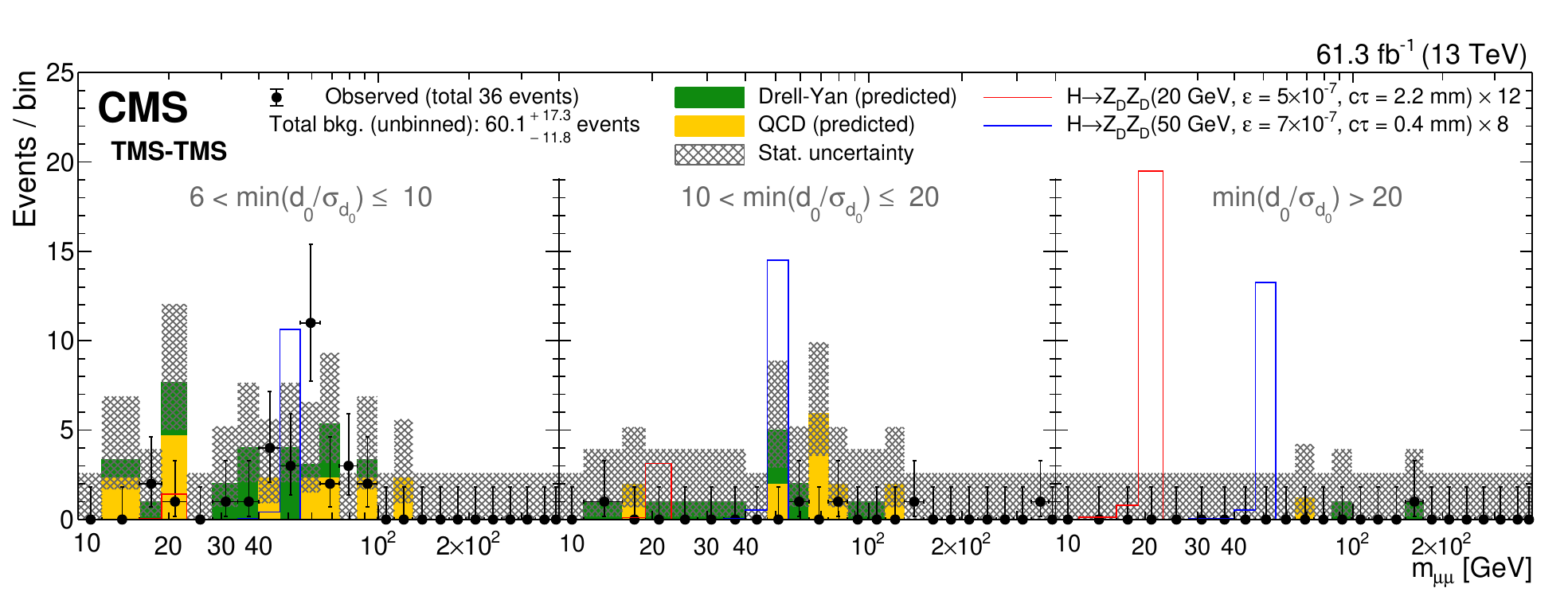}
  \caption{Comparison of the number of events observed in (upper)
    2016 and (lower) 2018 data in the TMS-TMS dimuon category with the
    expected number of background events, in representative \mMuMu
    intervals in each of the three min($\dzeroSig$) bins.
    The black points with error bars show the number of observed events;
    the green and yellow components of the stacked histograms
    represent the estimated numbers of DY and QCD events, respectively.  The
    last bin in each min($\dzeroSig$) interval includes events in the overflow.
    The uncertainties in the total expected background (shaded area) are
    statistical only.  Signal contributions expected from simulated
    $\PSMHiggs \to \PZD\PZD$ with $\mZD$ of 20 and 50\GeV are shown in
    red and blue, respectively. Their yields are set to the corresponding 
    combined median expected exclusion limits at 95\% \CL,
    scaled up as indicated in the legend to improve visibility.
    The legends also include the total number of observed events as
    well as the number of expected background events obtained inclusively,
    by applying the background evaluation method to the
    events in all $\mMuMu$ and min($\dzeroSig$) intervals
    combined.  \label{fig:obs_vs_exp_invmass_TMSTMS}}
\end{figure}

\begin{figure}[htbp]
  \centering
  \includegraphics[width=0.97\DSquareWidth]{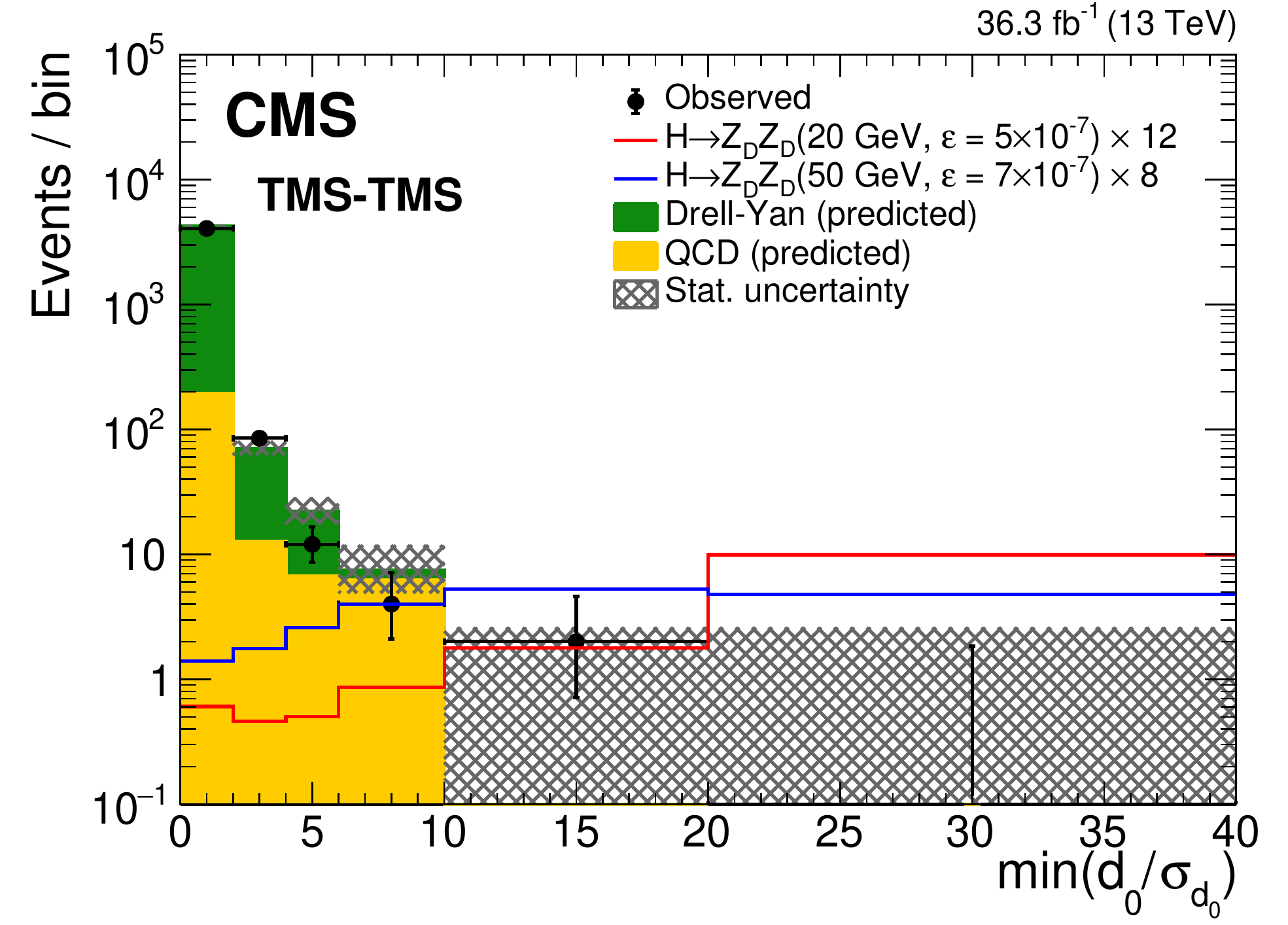}
  \includegraphics[width=0.97\DSquareWidth]{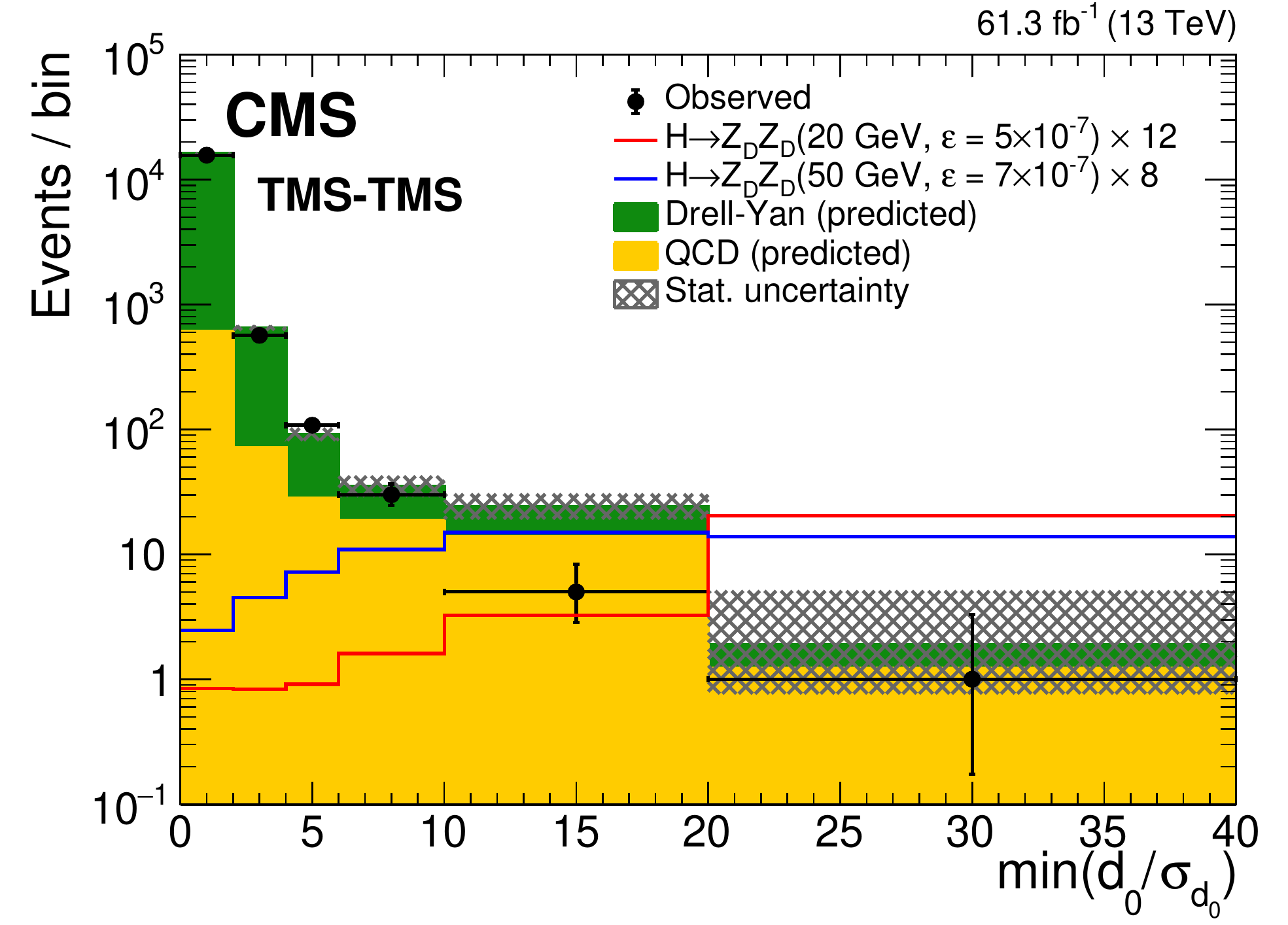}
  \caption{Comparison of the number of events observed in (left)
    2016 and (right) 2018 data in the TMS-TMS dimuon category with the
    expected number of background events, as a function of the smaller
    of the two $\dzeroSig$ values for the TMS-TMS dimuon.  The black
    points with error bars show the number of observed events; the green and yellow
    components of the stacked histograms represent the estimated
    numbers of DY and QCD events, respectively.  The last bin includes
    events in the overflow.  The uncertainties in the total expected
    background (shaded area) are statistical
    only.  Signal contributions expected from simulated
    $\PSMHiggs \to \PZD\PZD$ with $\mZD$ of 20 and 50\GeV are shown in
    red and blue, respectively. Their yields are set to the corresponding 
    combined median expected exclusion limits at 95\% \CL, scaled up as indicated
    in the legend to improve visibility. \label{fig:obs_vs_exp_d0sig_TMSTMS}}
\end{figure}

The predicted background yields in the representative \mMuMu intervals
and the corresponding numbers of observed events are shown in
Fig.~\ref{fig:obs_vs_exp_invmass_STASTA} for the STA-STA category and
Fig.~\ref{fig:obs_vs_exp_invmass_STATMS} for the STA-TMS category. 
For illustration, signals at the level of the median expected exclusion
limits at 95\% confidence level (\CL) in the absence of signal are also shown.
As expected, events observed in the STA-STA and
STA-TMS categories are predominantly at low masses---14 out of 18
STA-STA and 9 out of 13 STA-TMS events have $\mMuMu < 20\GeV$---and
have characteristics typical of those for QCD background events.  The
numbers of observed events and the predicted background and signal yields in the
TMS-TMS category are shown in Fig.~\ref{fig:obs_vs_exp_invmass_TMSTMS}
as functions of \mMuMu in each of the three min($\dzeroSig$) bins
and in Fig.~\ref{fig:obs_vs_exp_d0sig_TMSTMS} as a function of min($\dzeroSig$).
The observed TMS-TMS events have a steeply falling $\text{min}(\dzeroSig)$
distribution and cluster at \mMuMu values of a few tens of GeV, which
are both consistent with the characteristics of the expected background.
The numbers of observed events are consistent with the predicted
background yields in all dimuon categories and \mMuMu intervals, in
both data sets.  No significant excess of events above the SM background is
observed.

For each of the two benchmark models, we compute upper limits on the
product of the signal production cross section $\sigma$ and the branching
fraction $\mathcal{B}$ to two muons as a function of mass and mean proper decay
length.  The limit extraction is based on a modified frequentist
approach~\cite{CLS1, CLS2} and uses the CMS \textsc{Combine}
package developed for statistically combining the
results of Higgs boson searches~\cite{CMS-NOTE-2011-005}.  The method 
yielding background predictions in the signal region is implemented
using a multibin likelihood, which is a product of Poisson
distributions corresponding to the signal region and the control
regions.  The systematic uncertainties affecting the signal yield are
incorporated as nuisance parameters using log-normal distributions.
The expected and observed upper limits are evaluated through the use
of simulated pseudo-experiments.
For each signal model, the limits are first computed separately in
each dimuon category and for each data-taking year.  The individual
likelihoods are then combined to obtain the combined limits.

The signal efficiencies are obtained from simulation and further
corrected by the data-to-simu\-lation scale factors described in
Section~\ref{sec:systunc_signal}; they are computed separately for
each year, signal model, dimuon category, and mass interval.
A reweighting procedure is employed to calculate an estimated number of
signal events for lifetimes other than the lifetimes used to produce
the available simulated signal samples.
Given the smallness of the expected
background
and taking into account the
selection efficiencies discussed in Section~\ref{sec:selection}, an
introduction of a separate category for events with two dimuons does
not increase the sensitivity of the analysis significantly even in the
most favorable case for the 4$\Pgm$ signal events, namely
$\mathcal{B}(\PLLP\to\Pgm\Pgm) = 1$.  The gain is negligible for
smaller $\mathcal{B}(\PLLP\to\Pgm\Pgm)$ values such as those predicted
by the HAHM model.  Therefore, no distinction is made between events
with one and two dimuons of the same type.  Events with two TMS-TMS
dimuons are assigned to the min($\dzeroSig$) bin encompassing the
larger of the two min($\dzeroSig$) values.

\begin{figure}[htbp]
  \centering
  \includegraphics[width=\DSquareWidth]{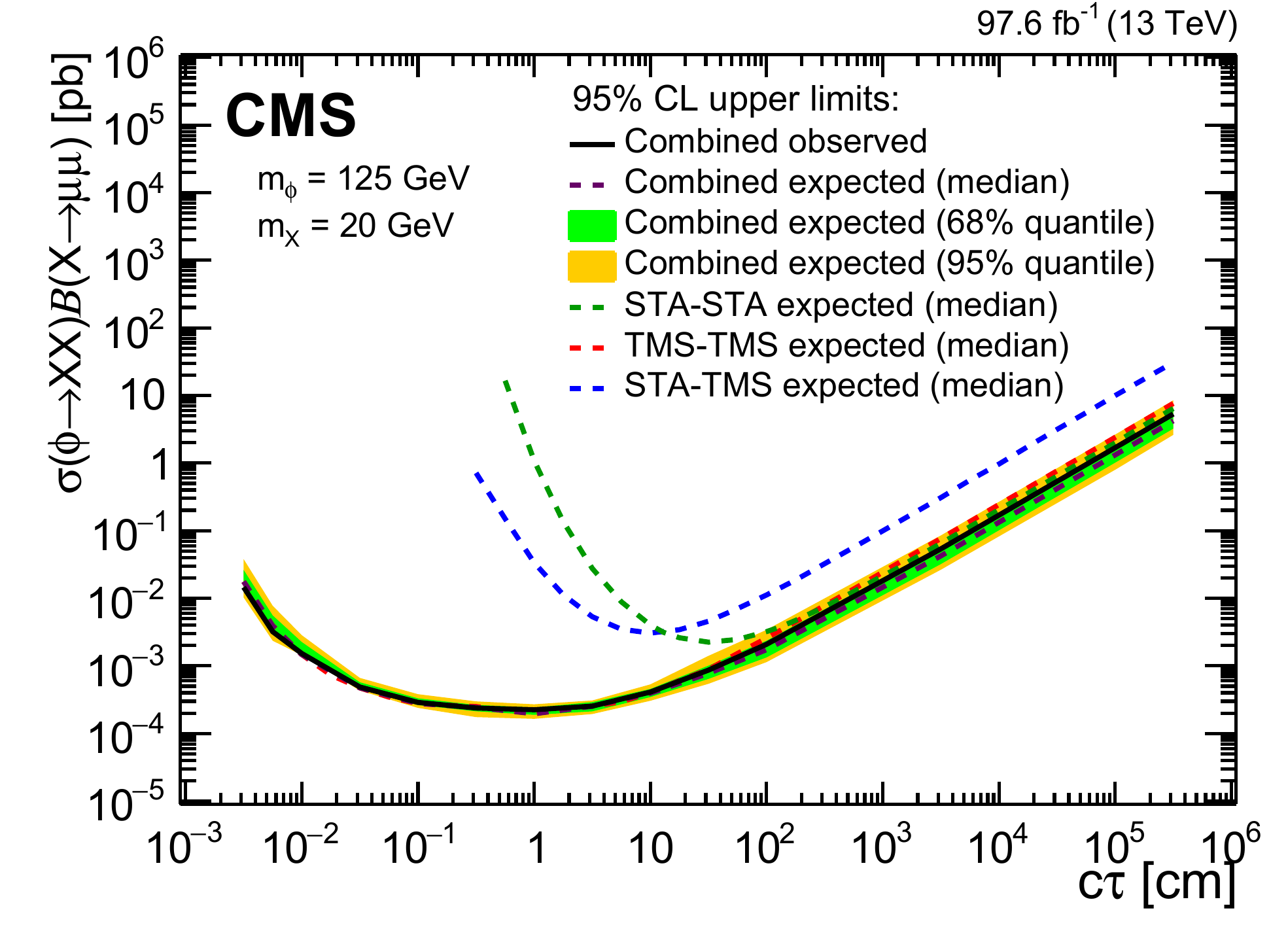}
  \hspace*{-1.3em}
  \includegraphics[width=\DSquareWidth]{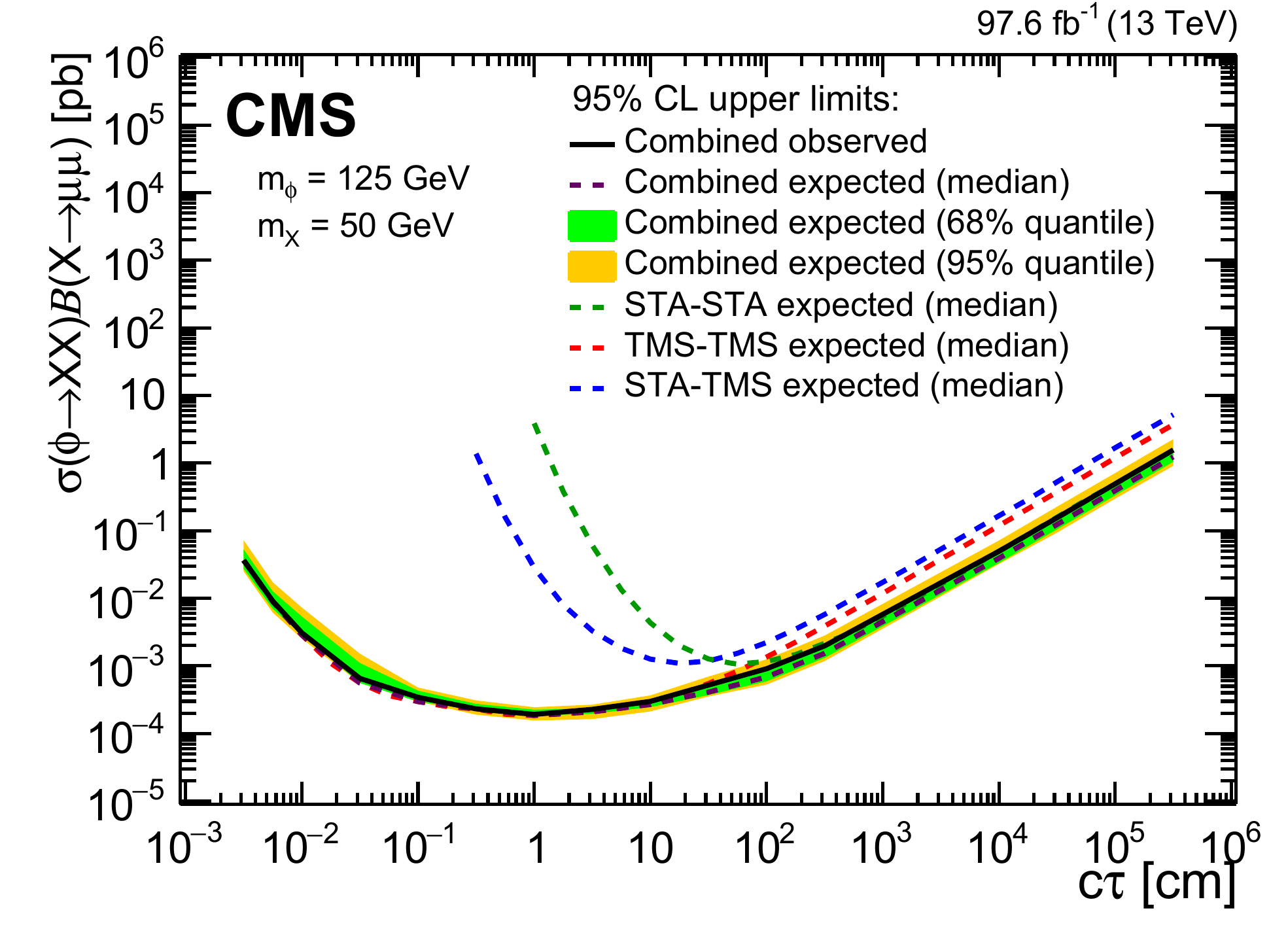}
  \caption{The 95\% \CL upper limits on $\sigma(\PBSMHiggs \to \PLLP
    \PLLP)\mathcal{B}(\PLLP\to\Pgm\Pgm)$ as a function of $\cTau(\PLLP)$
    in the heavy-scalar model, for $\mH=125\GeV$ and (left) $\mX=20\GeV$
    and (right) $\mX=50\GeV$.  The median expected limits
    obtained from the STA-STA, STA-TMS, and TMS-TMS dimuon categories
    are shown as dashed green, blue, and red curves, respectively; the
    combined median expected limits are shown as dashed black curves; and
    the combined observed limits are shown as solid black curves.
    The green and yellow bands correspond, respectively, to the
    68 and 95\% quantiles for the combined expected
    limits.  \label{fig:benchmark_limits_125}}
\end{figure}

\begin{figure}[htbp]
  \centering
  \includegraphics[width=\DSquareWidth]{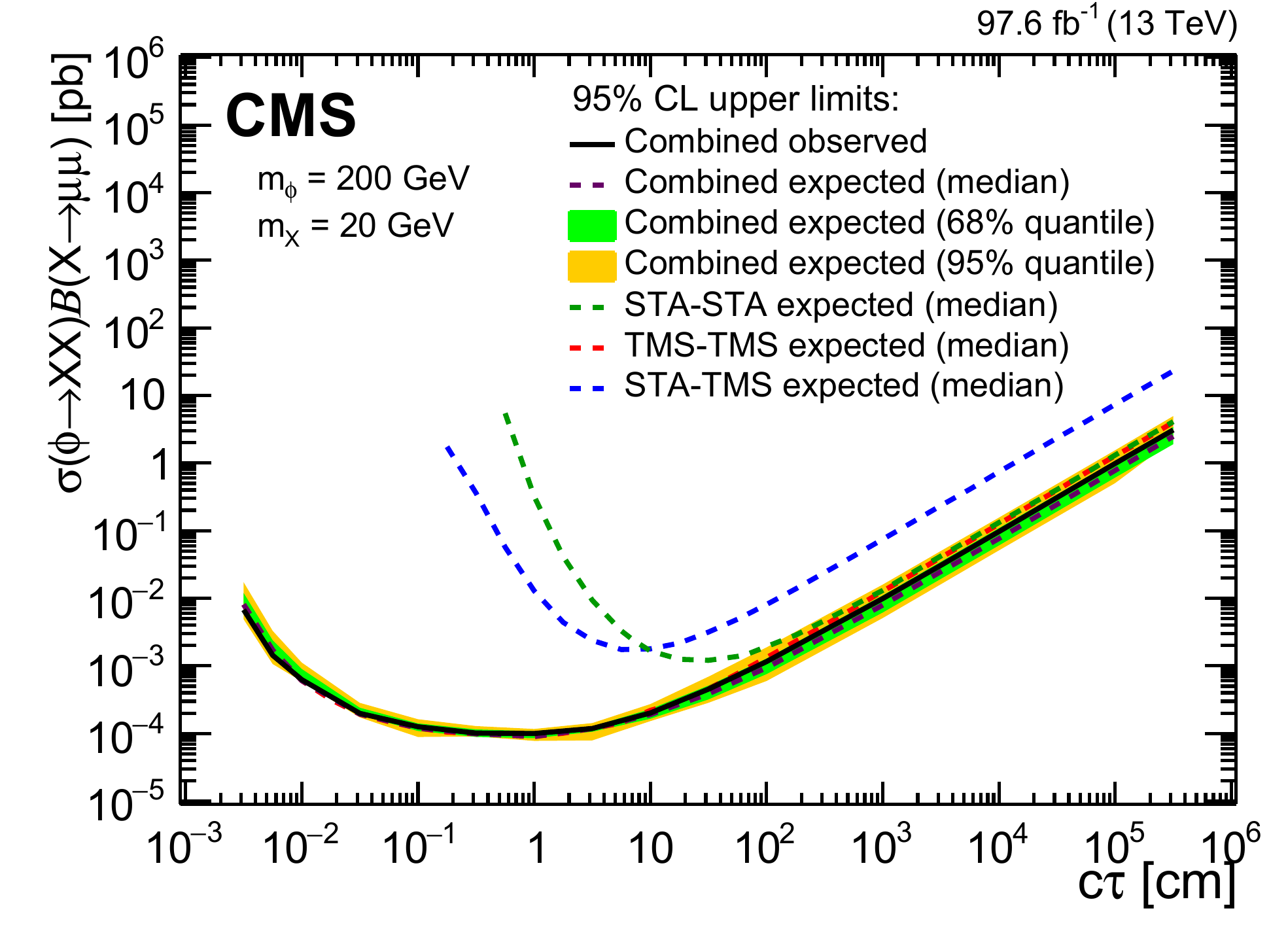}
  \hspace*{-1.3em}
  \includegraphics[width=\DSquareWidth]{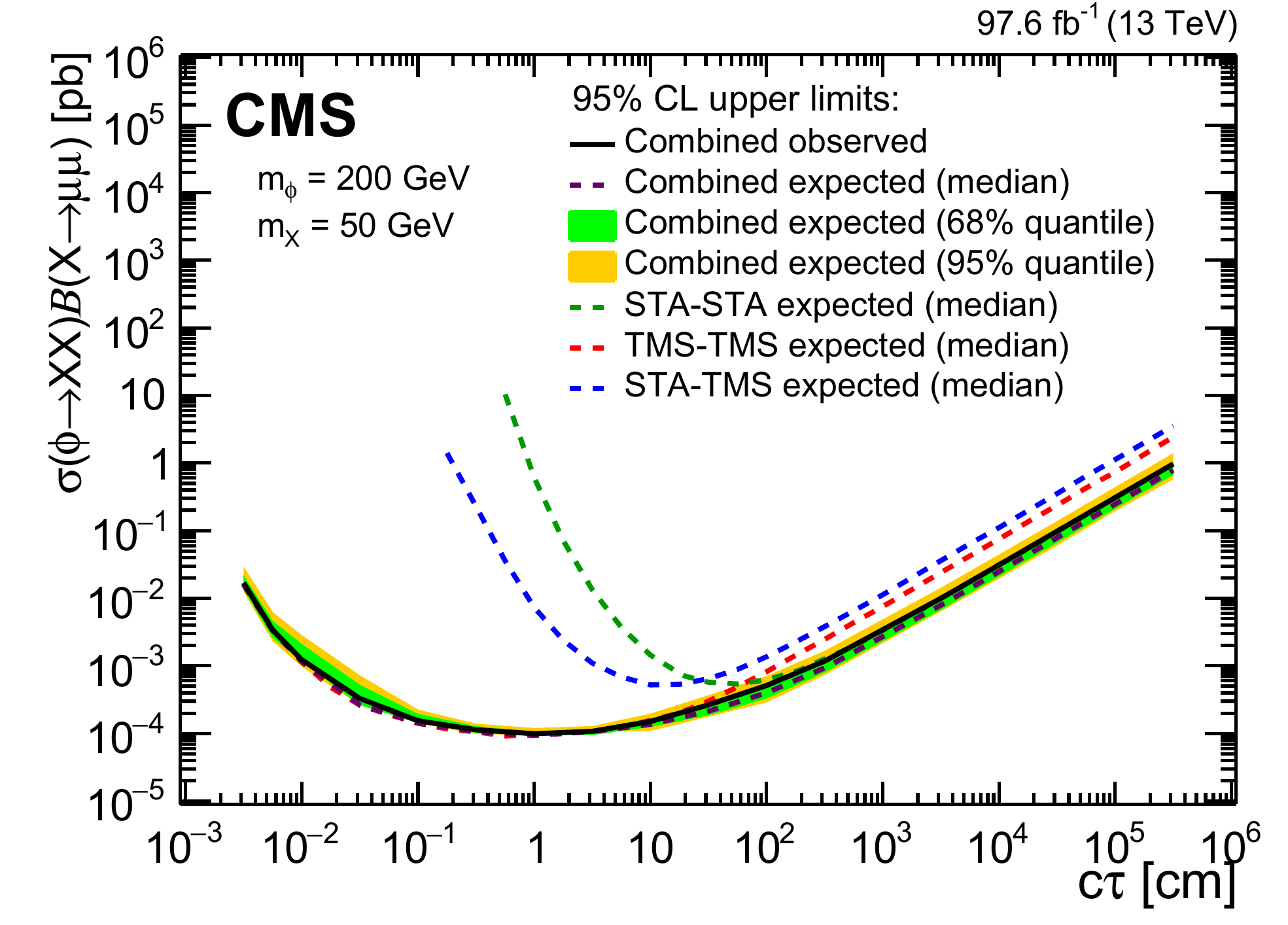}
  \caption{The 95\% \CL upper limits on $\sigma(\PBSMHiggs \to \PLLP
    \PLLP)\mathcal{B}(\PLLP\to\Pgm\Pgm)$ as a function of $\cTau(\PLLP)$
    in the heavy-scalar model, for $\mH=200\GeV$ and (left) $\mX=20\GeV$
    and (right) $\mX=50\GeV$.  The median expected limits
    obtained from the STA-STA, STA-TMS, and TMS-TMS dimuon categories
    are shown as dashed green, blue, and red curves, respectively; the
    combined median expected limits are shown as dashed black curves; and
    the combined observed limits are shown as solid black curves.
    The green and yellow bands correspond, respectively, to the
    68 and 95\% quantiles for the combined expected
    limits.  \label{fig:benchmark_limits_200}}
\end{figure}

\begin{figure}[htb]
  \centering
  \includegraphics[width=\DSquareWidth]{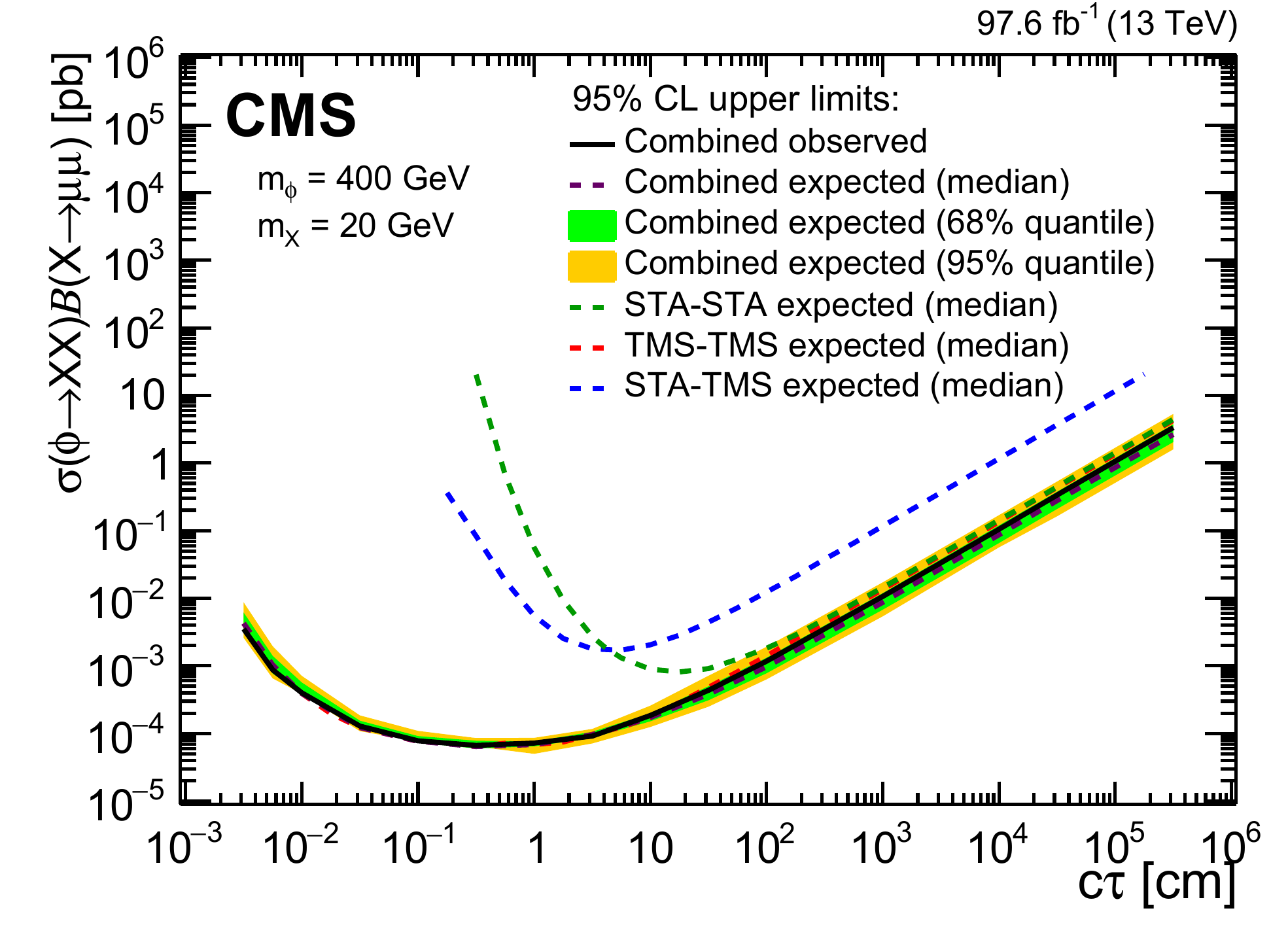}
  \hspace*{-1.3em}
  \includegraphics[width=\DSquareWidth]{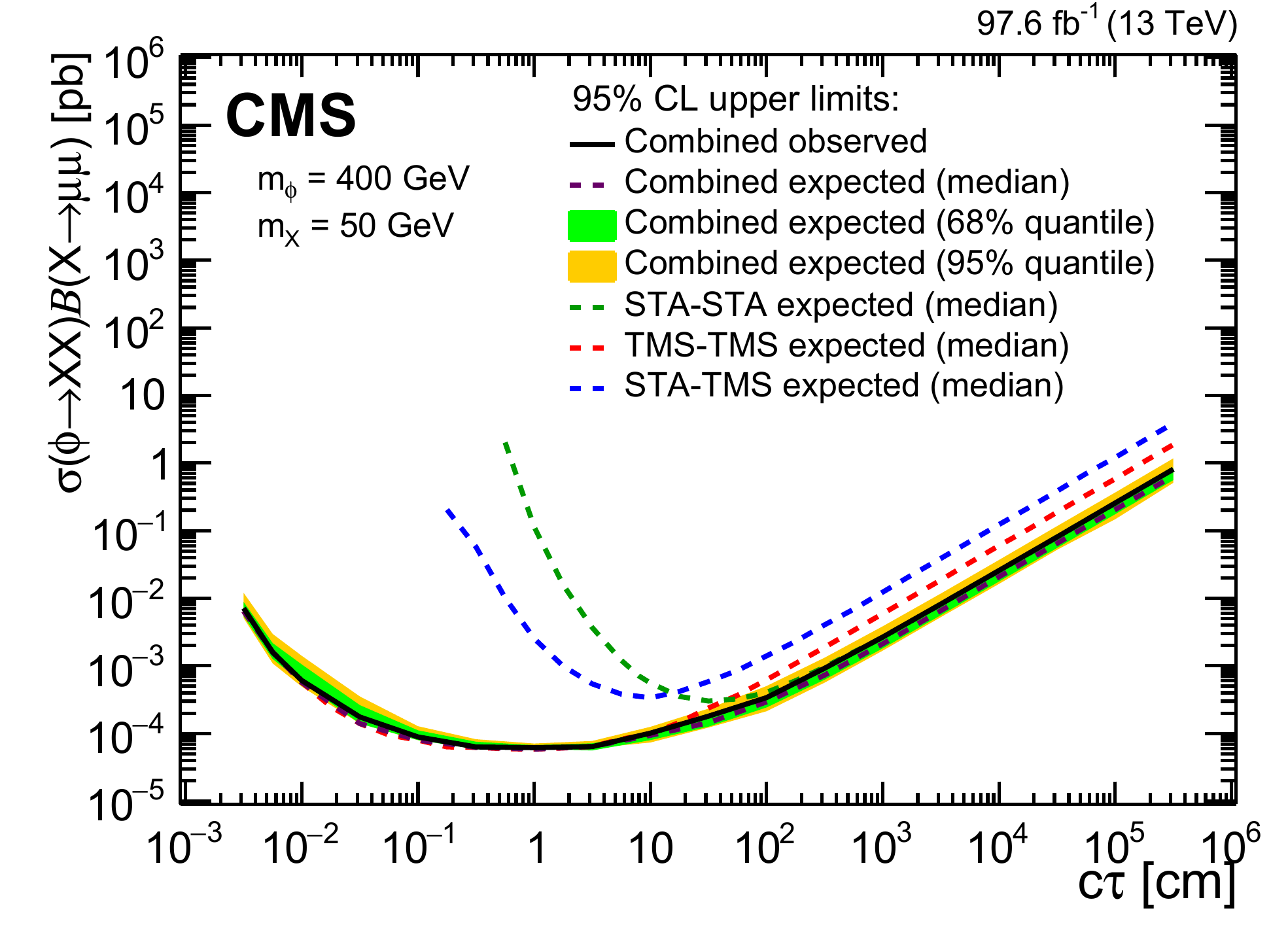}
  \includegraphics[width=\DSquareWidth]{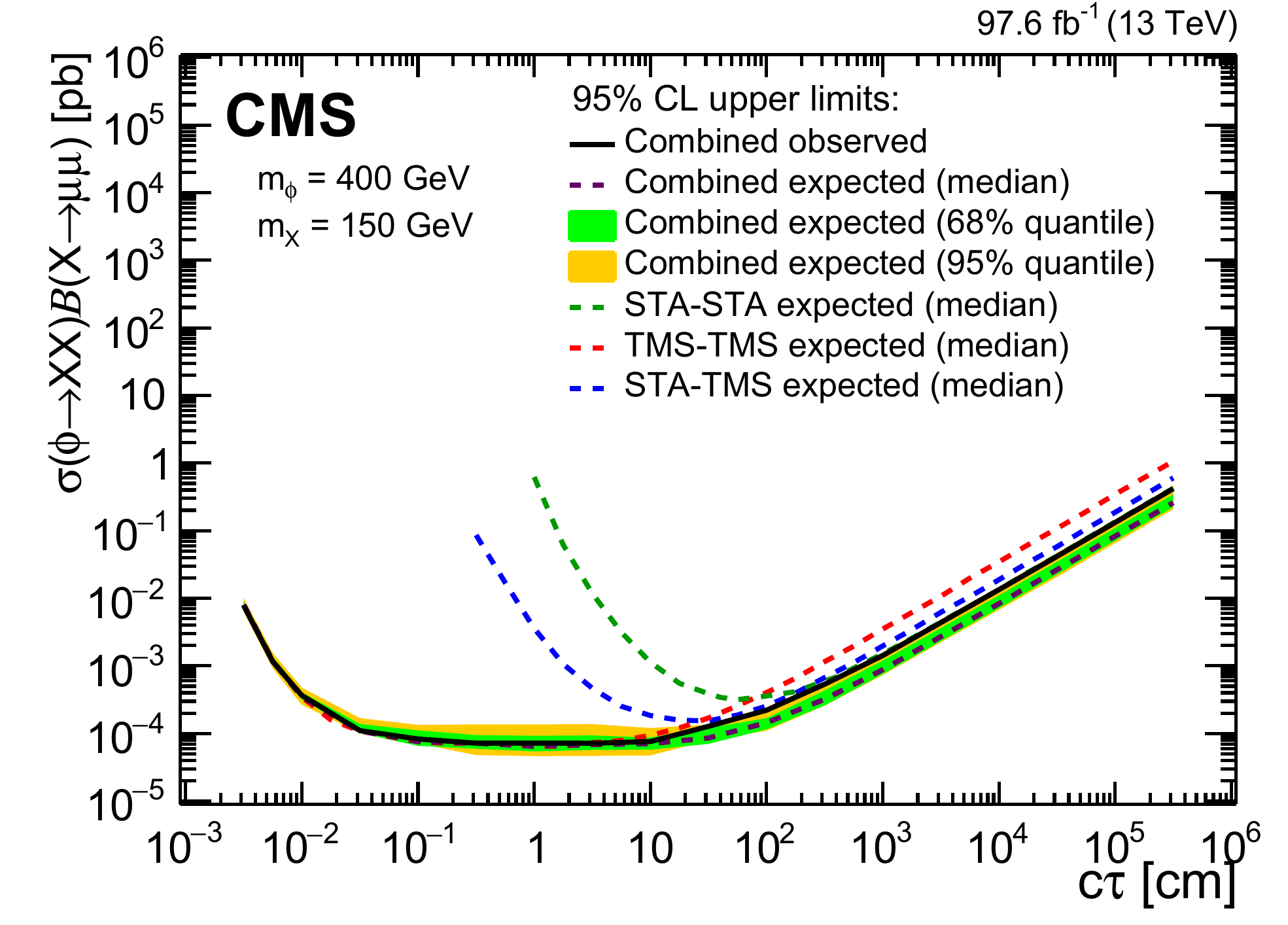}
    \caption{The 95\% \CL upper limits on $\sigma(\PBSMHiggs \to \PLLP
      \PLLP)\mathcal{B}(\PLLP\to\Pgm\Pgm)$ as a function of
      $\cTau(\PLLP)$ in the heavy-scalar model, for $\mH=400\GeV$ and
      (upper left) $\mX=20\GeV$, (upper right) $\mX=50\GeV$, and (lower)
      $\mX=150\GeV$.  The median expected limits obtained from
      the STA-STA, STA-TMS, and TMS-TMS dimuon categories are shown as
      dashed green, blue, and red curves, respectively; the combined
      median expected limits are shown as dashed black curves; and the
      combined observed limits are shown as solid black curves.  The
      green and yellow bands correspond, respectively, to the 68
      and 95\% quantiles for the combined expected
      limits.  \label{fig:benchmark_limits_400}}
\end{figure}

\begin{figure}[htbp]
  \centering
  \includegraphics[width=\DSquareWidth]{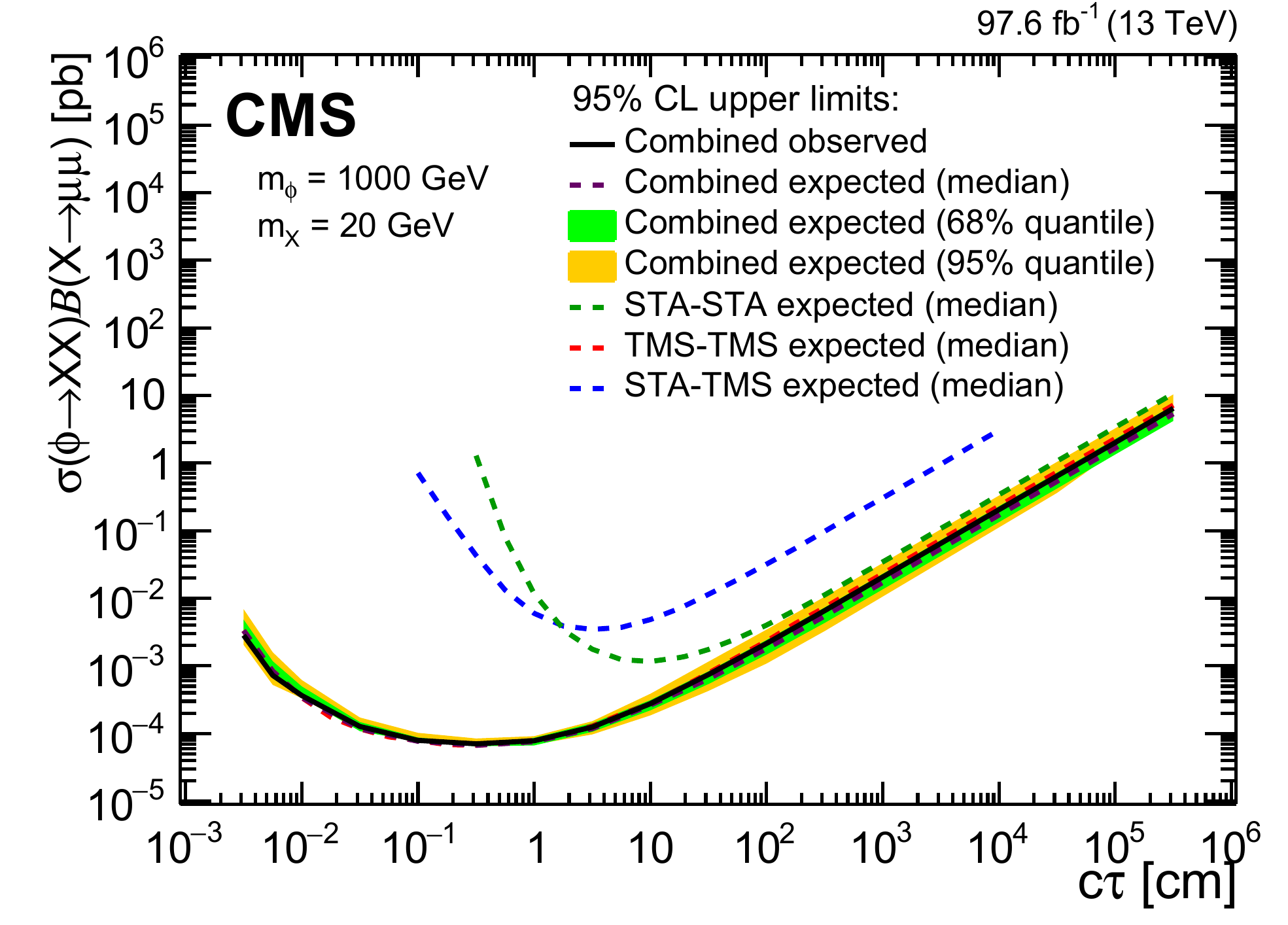}
  \hspace*{-1.3em}
  \includegraphics[width=\DSquareWidth]{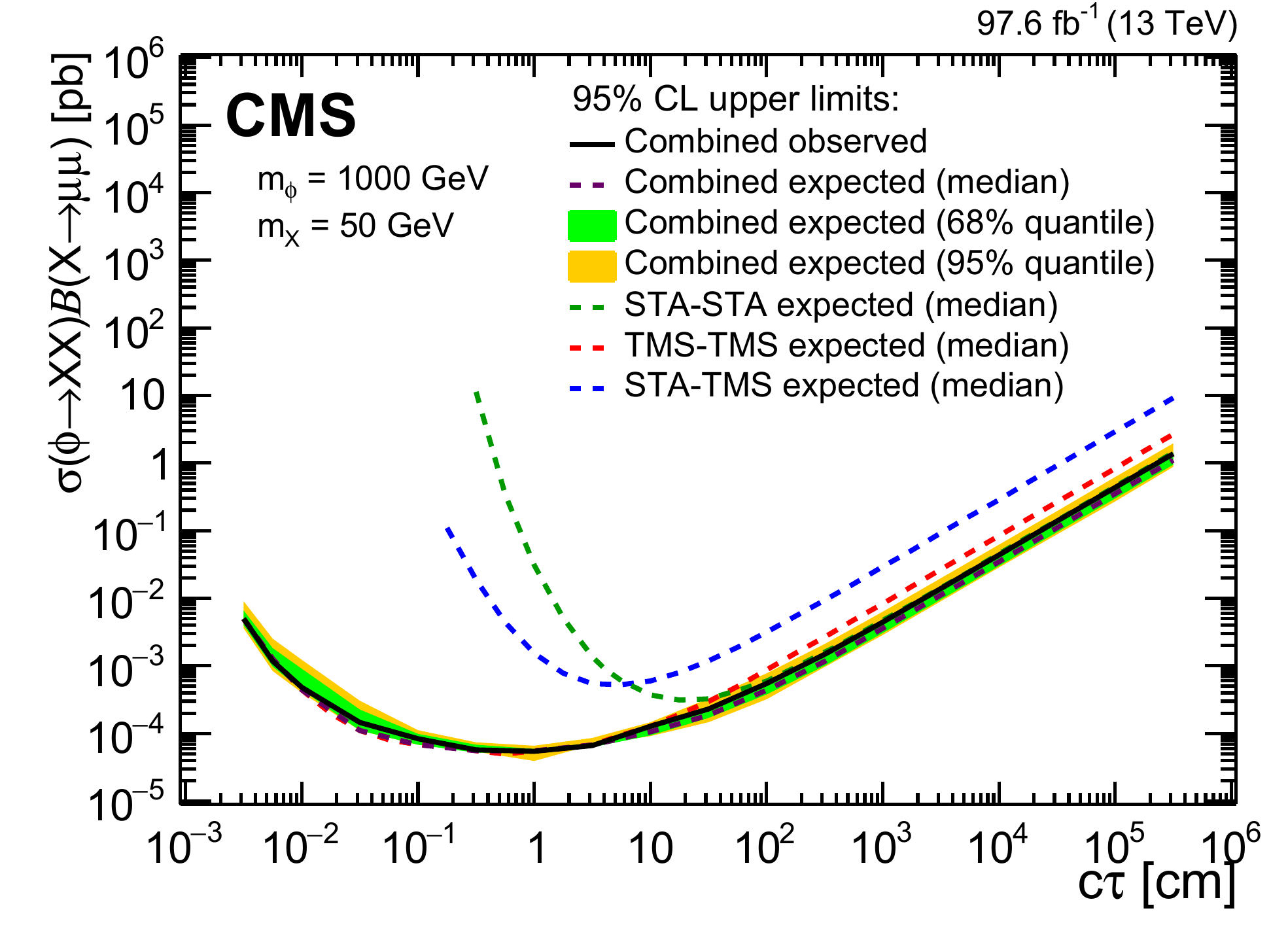}
  \includegraphics[width=\DSquareWidth]{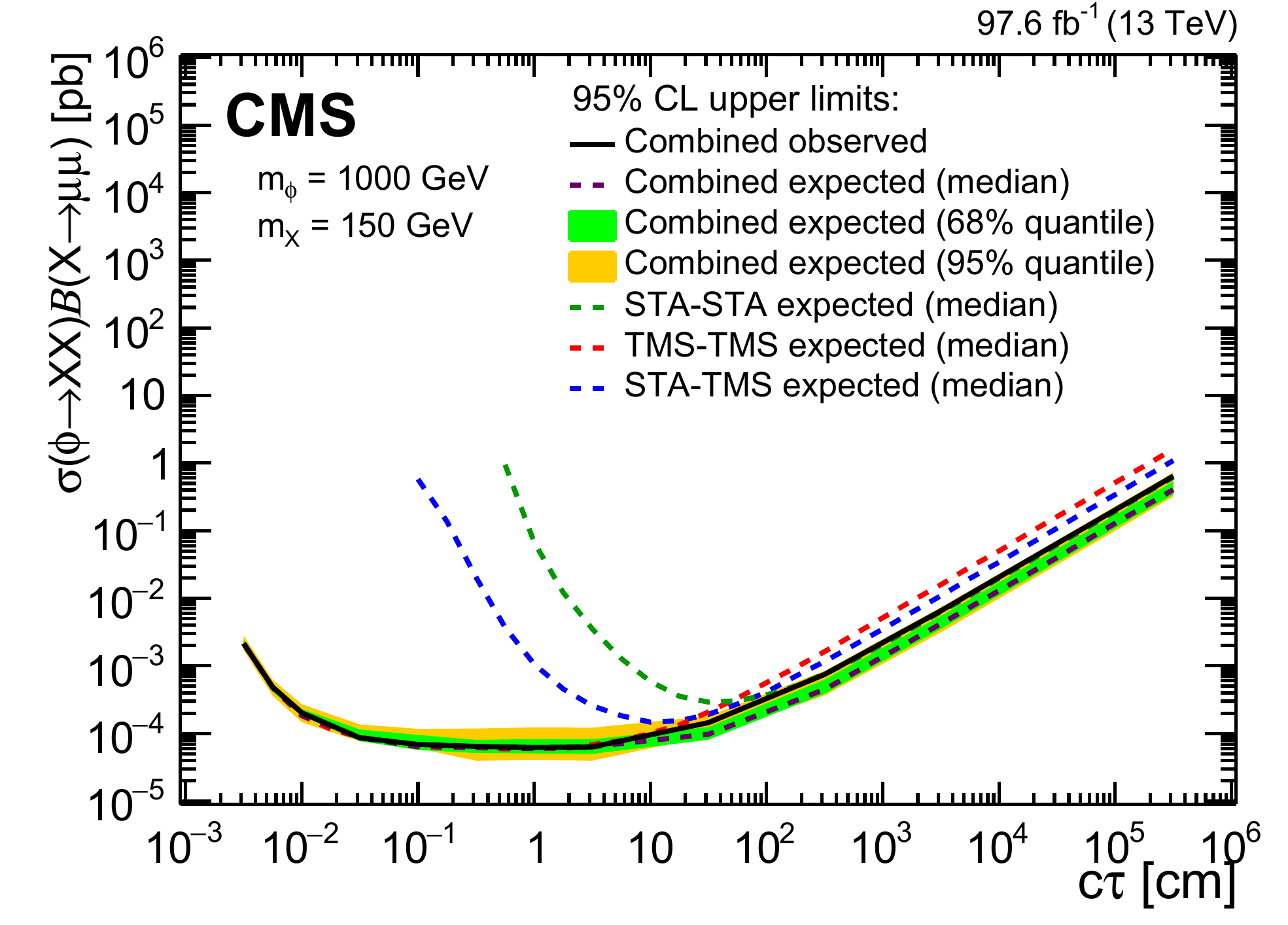}
  \hspace*{-1.3em}
  \includegraphics[width=\DSquareWidth]{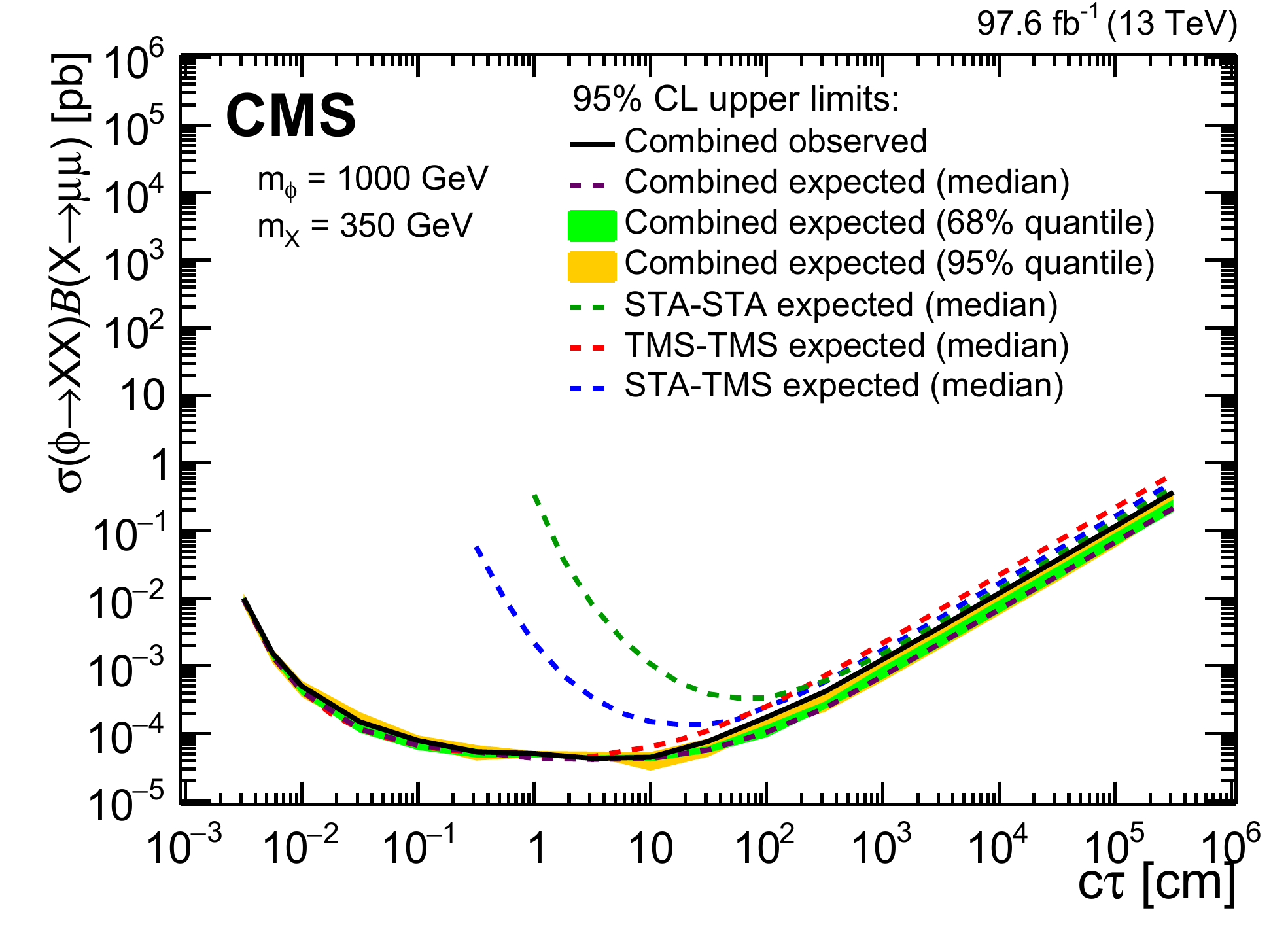}
    \caption{The 95\% \CL upper limits on $\sigma(\PBSMHiggs \to \PLLP
      \PLLP)\mathcal{B}(\PLLP\to\Pgm\Pgm)$ as a function of
      $\cTau(\PLLP)$ in the heavy-scalar model, for $\mH=1\TeV$ and
      (upper left) $\mX=20\GeV$, (upper right) $\mX=50\GeV$,
      (lower left) $\mX=150\GeV$, and (lower right) $\mX=350\GeV$.
      The median expected limits obtained from the STA-STA, STA-TMS,
      and TMS-TMS dimuon categories are shown as dashed green, blue,
      and red curves, respectively; the combined median expected
      limits are shown as dashed black curves; and the combined observed
      limits are shown as solid black curves.  The green and
      yellow bands correspond, respectively, to the 68 and 95\%
      quantiles for the combined expected
      limits.  \label{fig:benchmark_limits_1000}}
\end{figure}

The observed and expected 95\% \CL upper limits on the product
$\sigma(\PBSMHiggs \to \PLLP \PLLP) \mathcal{B}(\PLLP\to\Pgm\Pgm)$
as a function of $\cTau(\PLLP)$ in the simplified heavy-scalar model
for $\mH=125$, 200, 400, and 1000\GeV 
are shown in Figs.~\ref{fig:benchmark_limits_125},
\ref{fig:benchmark_limits_200},
\ref{fig:benchmark_limits_400}, and
\ref{fig:benchmark_limits_1000}, respectively.  Each figure
shows the limits obtained with the ensemble of 2016 and 2018 data in the
individual dimuon categories, as well as their combination.
The search is sensitive to a broad range of $\cTau$ from 30\micron to
more than 1\unit{km}.  As expected, the three dimuon categories reach
maximum sensitivity at different $\cTau$ values and all give relevant
contributions to the overall sensitivity of the search.  The limits
are most restrictive for $\cTau$ between 0.1\mm and 10--100\unit{m},
excluding $\sigma(\PBSMHiggs \to \PLLP
\PLLP)\mathcal{B}(\PLLP\to\Pgm\Pgm)$ smaller than 1\unit{fb}, and become
more stringent at high LLP masses, owing to a higher signal efficiency
and lower background.  The smallest $\sigma(\PBSMHiggs \to \PLLP
\PLLP)\mathcal{B}(\PLLP\to\Pgm\Pgm)$ value excluded is 0.05\unit{fb}, at
$\mX = 350\GeV$ and $\cTau$ between 0.3 and 30\cm.

\begin{figure}[htbp]
  \centering
  \includegraphics[width=\DSquareWidth]{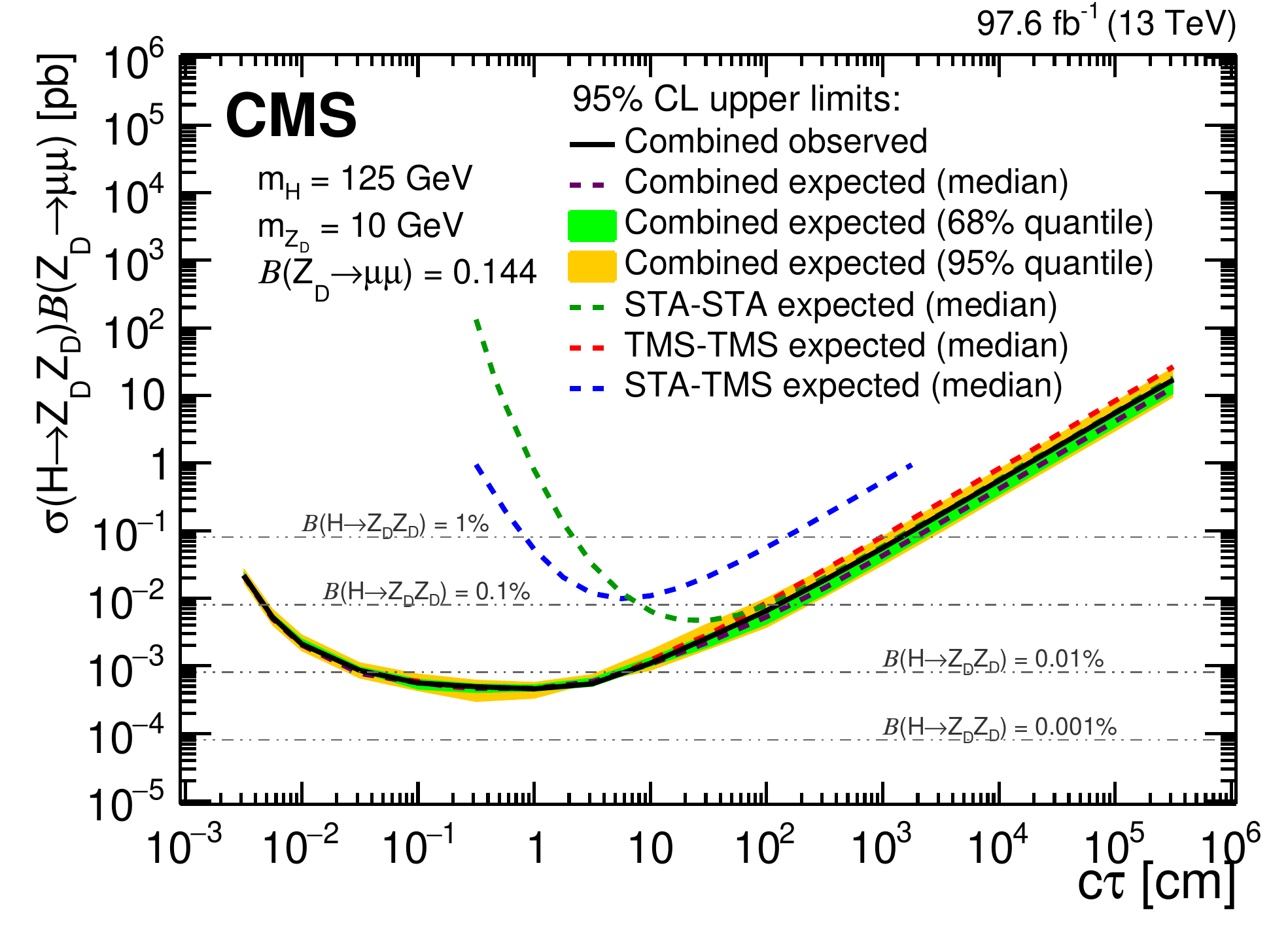}
  \hspace*{-1.3em}
  \includegraphics[width=\DSquareWidth]{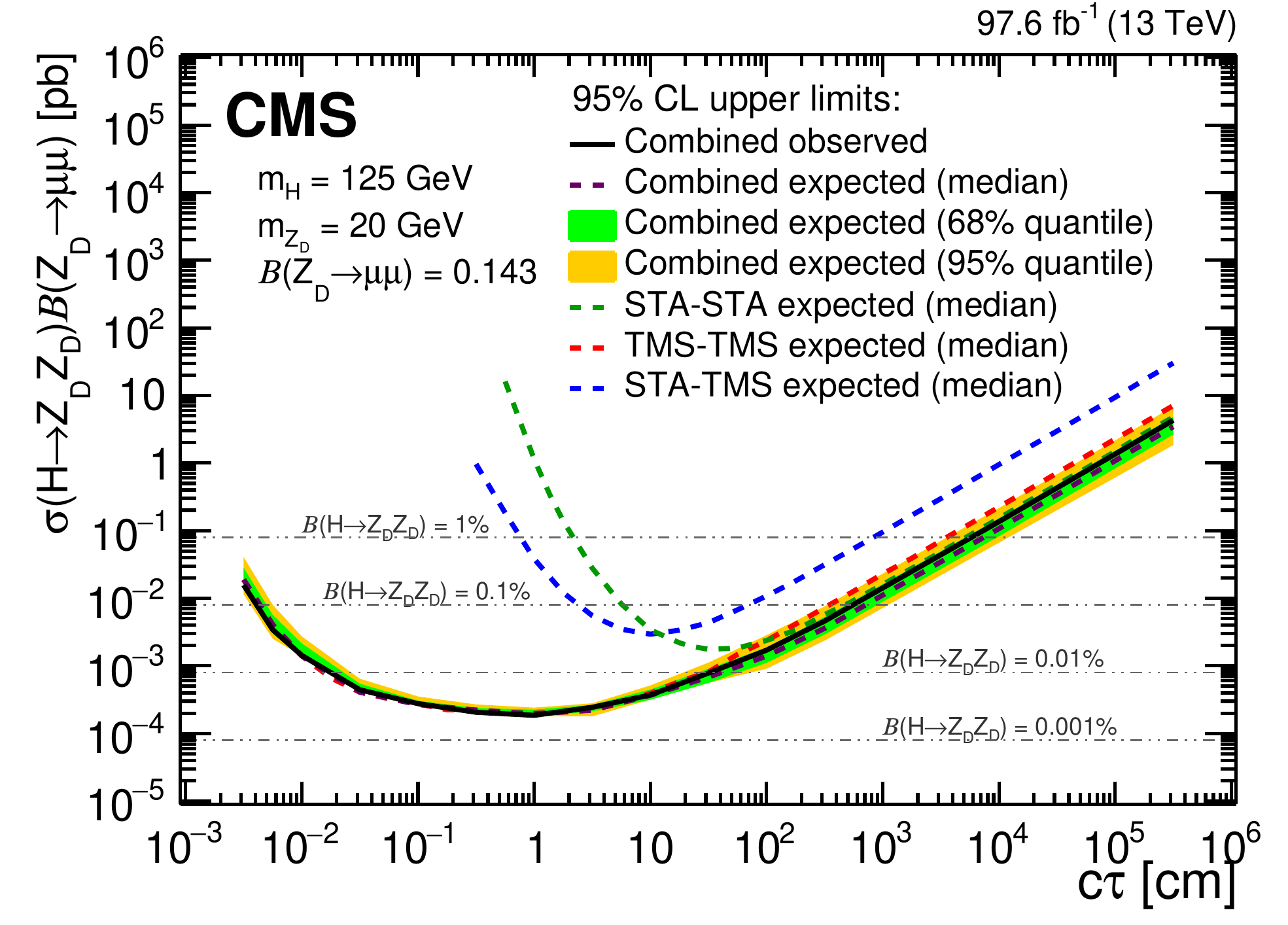}
  \includegraphics[width=\DSquareWidth]{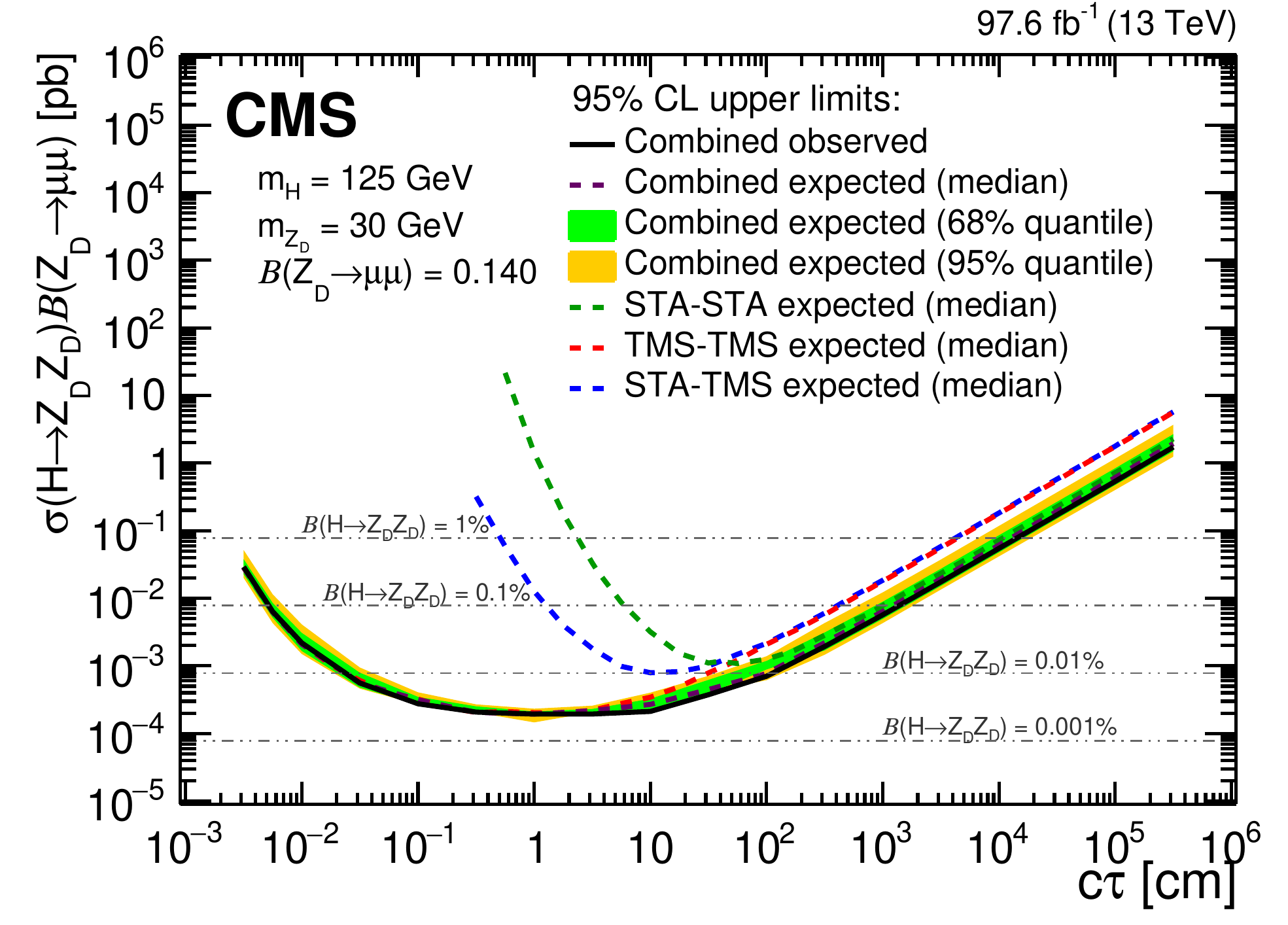}
  \hspace*{-1.3em}
  \includegraphics[width=\DSquareWidth]{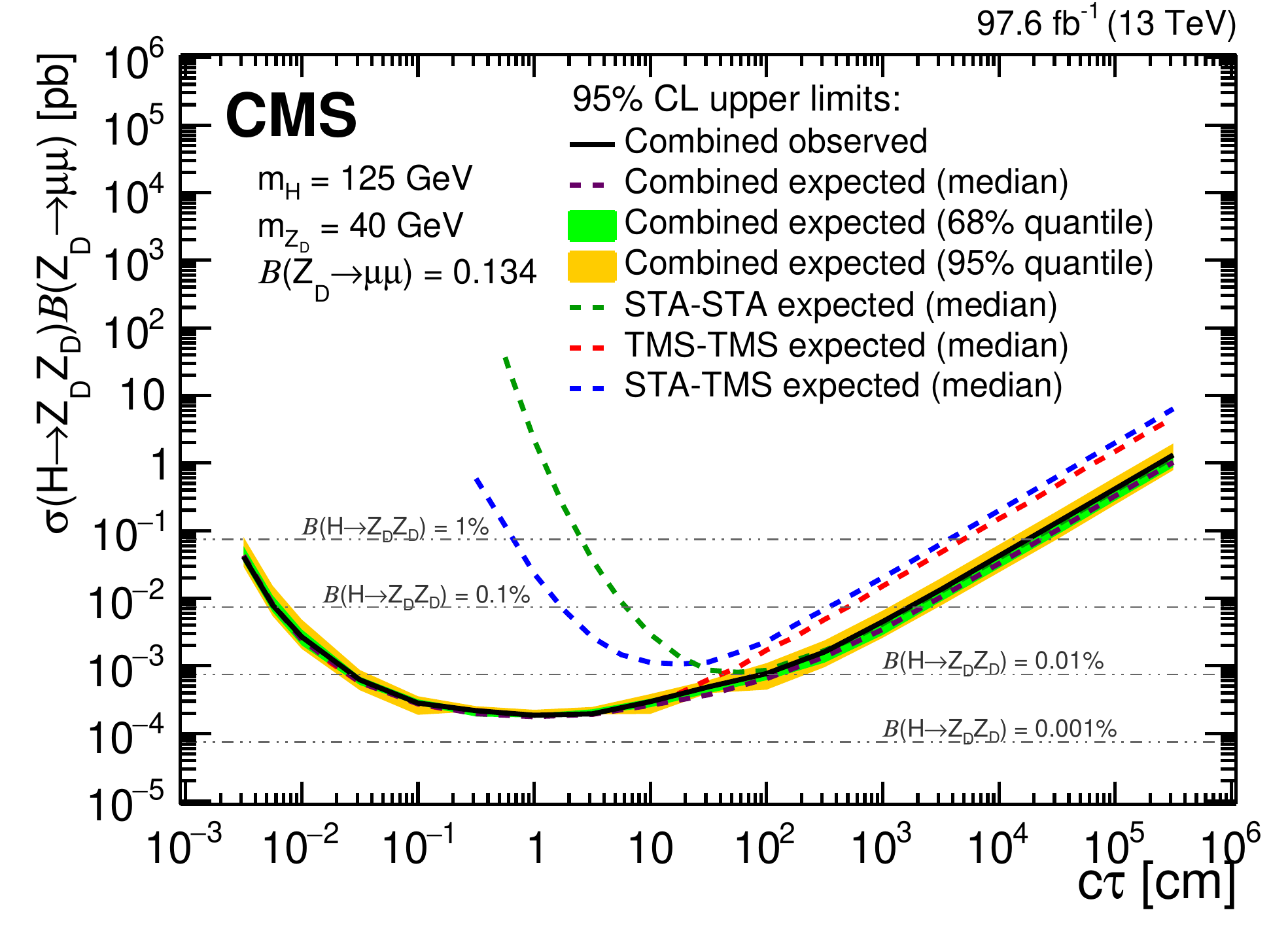}
  \includegraphics[width=\DSquareWidth]{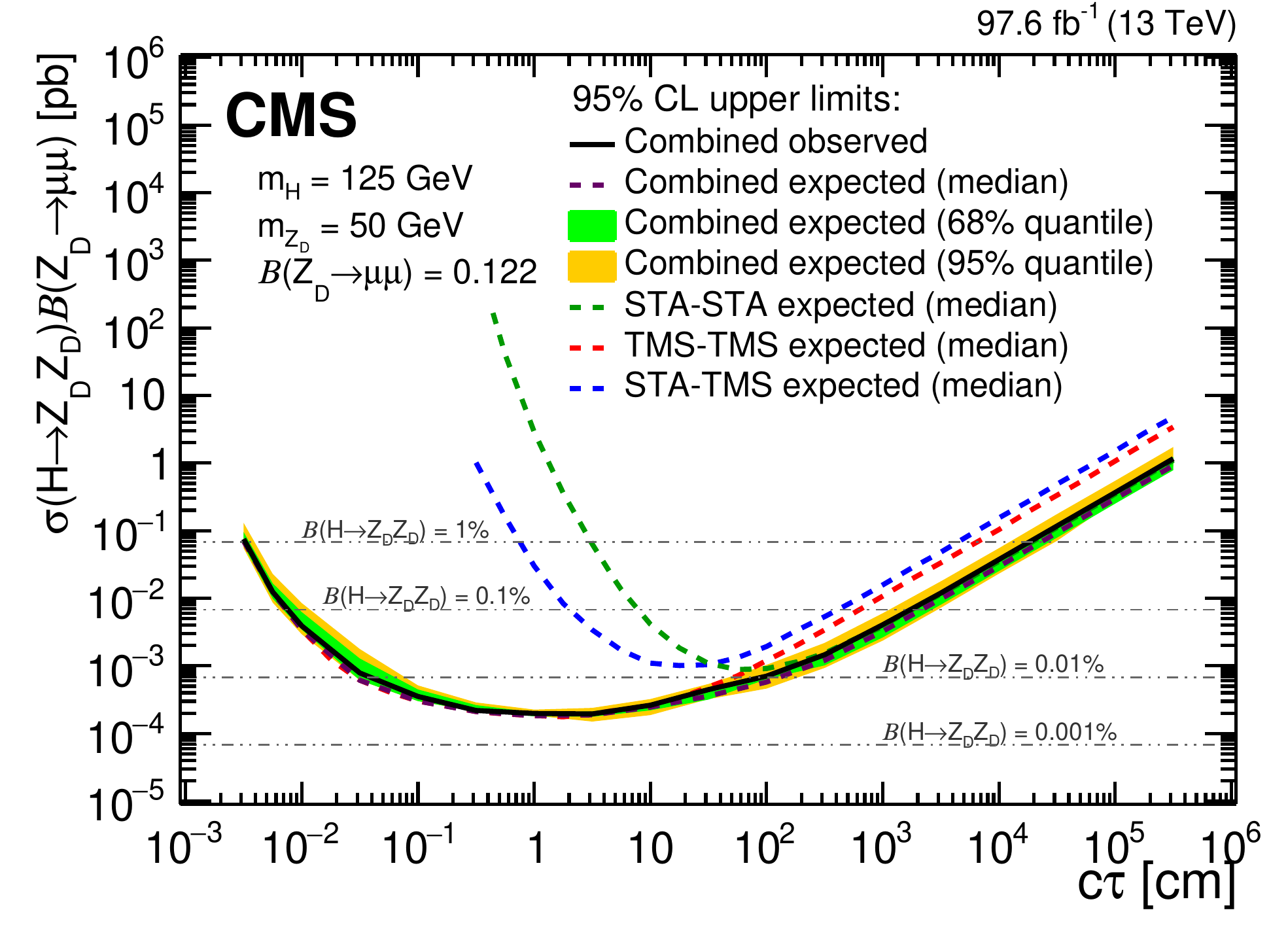}
  \hspace*{-1.3em}
  \includegraphics[width=\DSquareWidth]{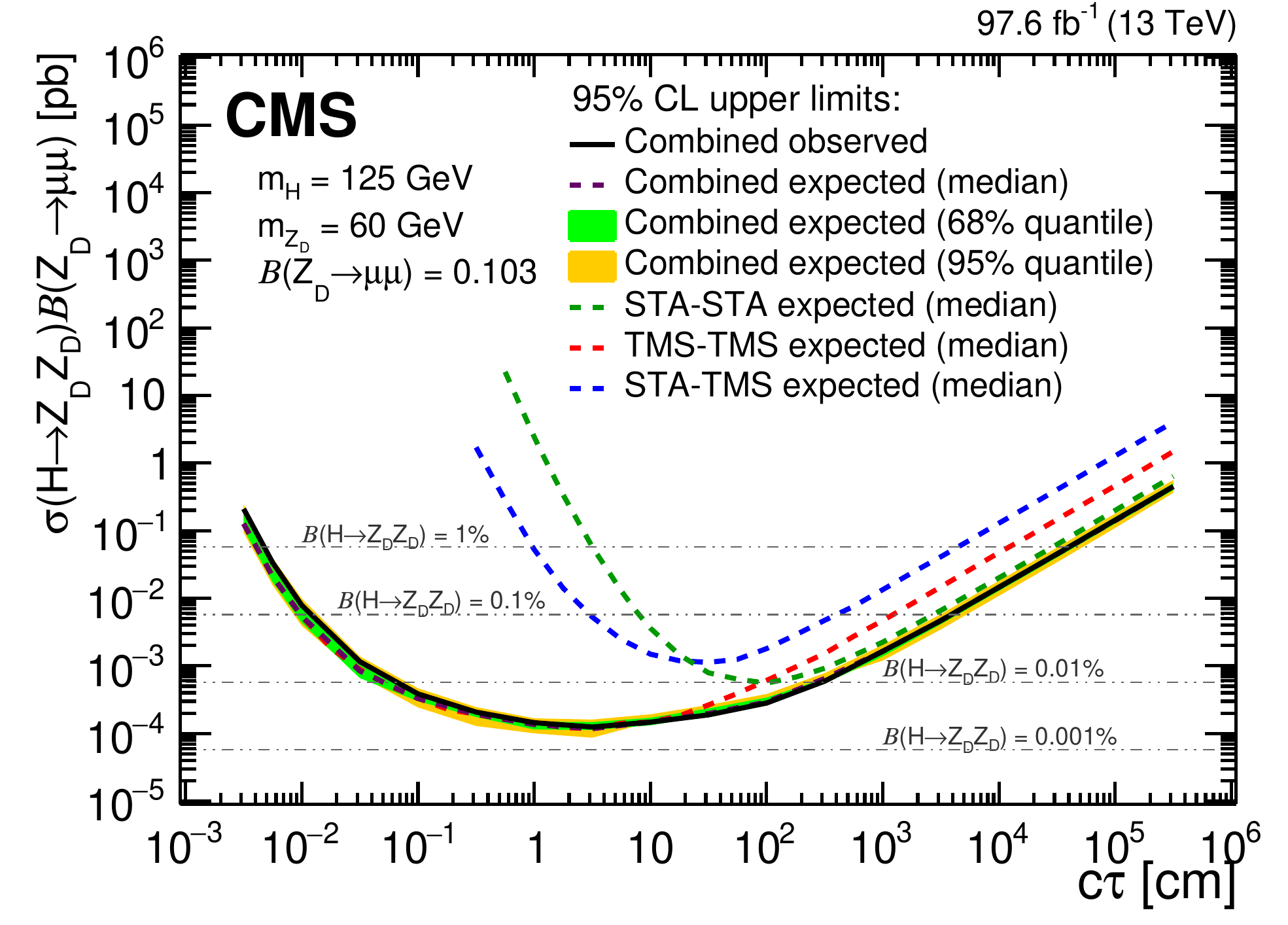}
  \caption{The 95\% \CL upper limits on $\sigma(\PSMHiggs \to \PZD
    \PZD)\mathcal{B}(\PZD\to\Pgm\Pgm)$ as a function of $\cTau(\PZD)$ in
    the HAHM model, for $\mZD$ ranging from 10\GeV (upper left) to
    60\GeV (lower right).  The median expected limits obtained from
    the STA-STA, STA-TMS, and TMS-TMS dimuon categories are shown as
    dashed green, blue, and red curves, respectively; the combined
    median expected limits are shown as dashed black curves; and the
    combined observed limits are shown as solid black curves.
    The green and yellow bands correspond, respectively, to the
    68 and 95\% quantiles for the combined expected limits.
    The horizontal lines in gray correspond to the theoretical
    predictions for values of $\mathcal{B}(\PSMHiggs \to \PZD \PZD)$
    indicated next to the lines.
    \label{fig:darkphoton_limits}}
\end{figure}

Figure~\ref{fig:darkphoton_limits} shows the observed and expected
95\% \CL upper limits obtained in the framework of the HAHM model
under the assumption of $\mH > \mh/2$.
The limits in the individual dimuon categories, as well as
their combination, are set on the product $\sigma(\PSMHiggs \to \PZD
\PZD) \mathcal{B}(\PZD\to\Pgm\Pgm)$ for different $\mZD$ as a
function of the mean proper decay length of $\PZD$.
The limits are compared to the theoretical predictions for a set of
representative $\mathcal{B}(\PSMHiggs \to \PZD \PZD)$ values ranging from
1\% to 0.001\%.  In the $\mZD$ range of 20--60\GeV,
the branching fraction $\mathcal{B}(\PSMHiggs \to \PZD
\PZD)$ of 1\% is excluded in the $\cTau(\PZD)$ range
from a few tens of $\mum$ to approximately 100\unit{m}, whereas $\mathcal{B}(\PSMHiggs
\to \PZD \PZD)$ as low as 0.01\% is excluded in the
range of 1\mm to 1\unit{m}.  These constraints on rare SM Higgs
boson decays are tighter than those derived from searches for
invisible Higgs boson decays~\cite{CMS:2022qva} and from indirect
constraints from measurements of the SM Higgs boson
couplings~\cite{Sirunyan:2018koj}.  At $\mZD> 20\GeV$, the limits
obtained are the best to date for all $\cTau(\PZD)$ values except
those between ${\approx}0.5$ and 500\mm (depending on $\mZD$), where
our search is complemented by the CMS search using data collected with
a dedicated high-rate data stream~\cite{CMS:2021sch}.

\begin{figure}[htbp]
  \centering
  \includegraphics[width=0.97\DSquareWidth]{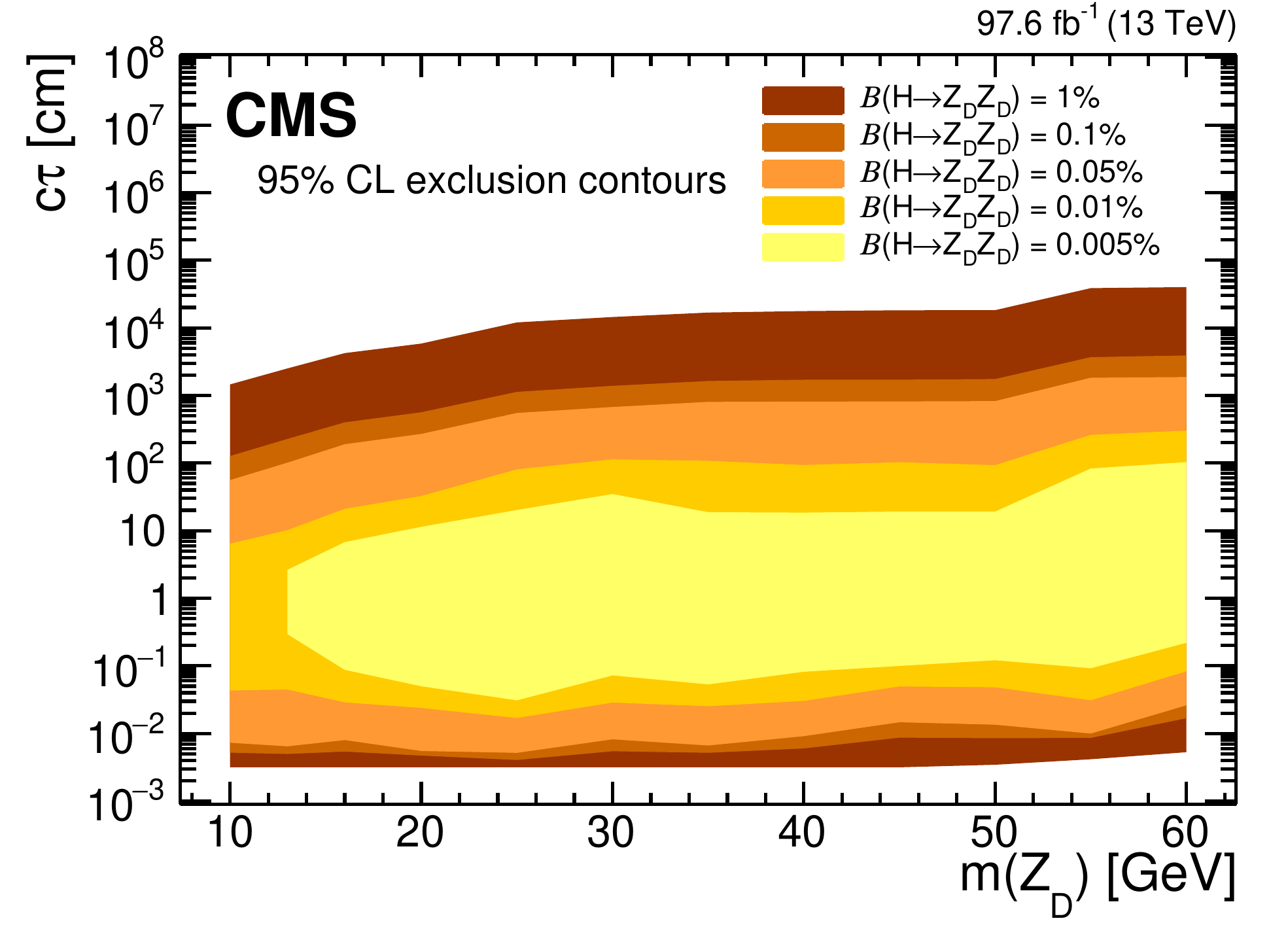}
  \includegraphics[width=0.97\DSquareWidth]{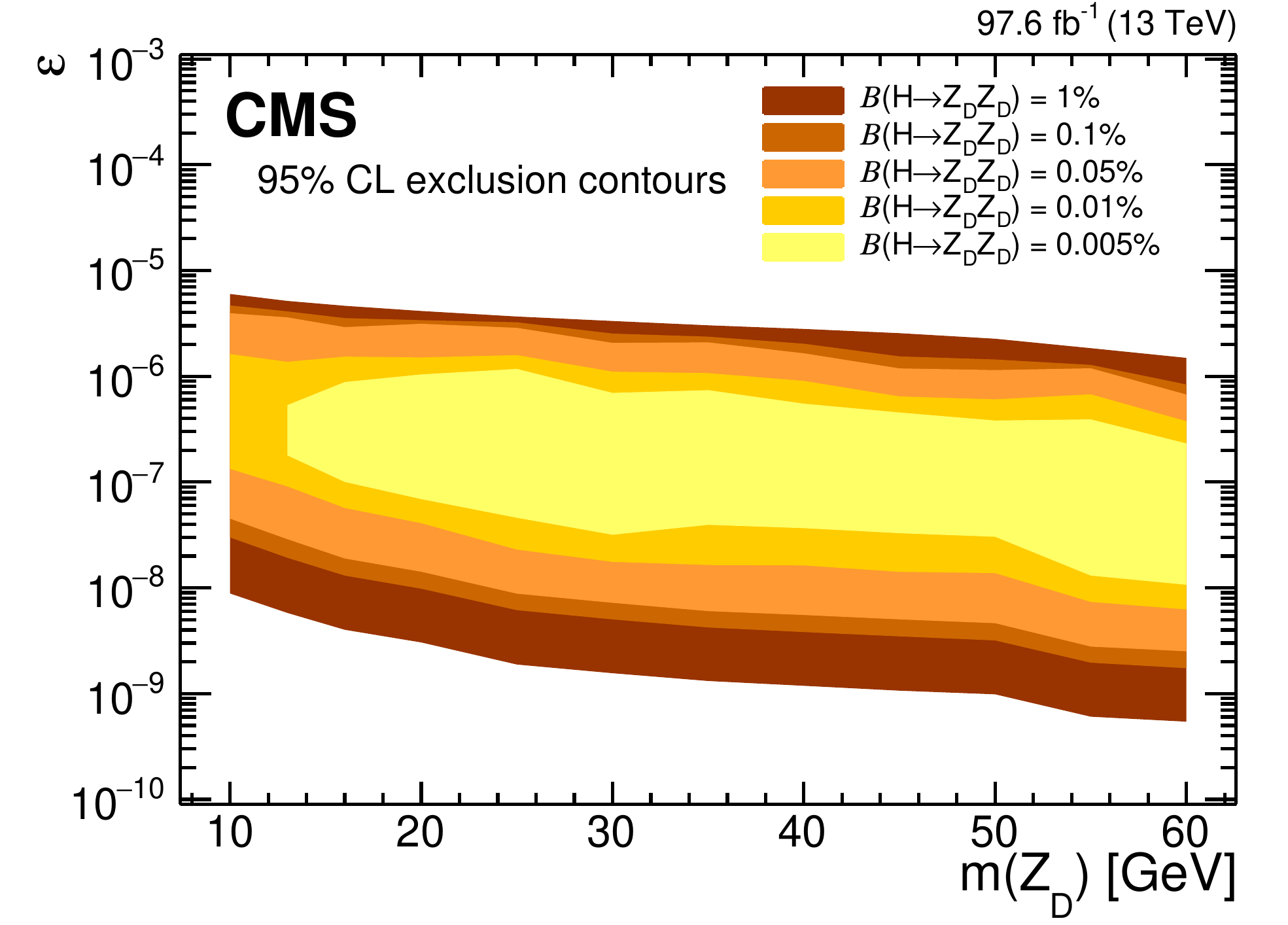}
  \caption{Observed 95\% \CL exclusion contours in the HAHM model, in
    the (left) ($\mZD$, $\cTau(\PZD)$) and (right) ($\mZD$, $\epsilon$)
    planes.  The contours correspond to several
    representative values of $\mathcal{B}(\PSMHiggs \to \PZD \PZD)$
    ranging from 0.005 to 1\%.  
    \label{fig:darkphoton_contours}}
\end{figure}

The observed 95\% \CL exclusion contours in the ($\mZD$, $\cTau(\PZD)$)
plane, determined from the observed 2016-18 upper limits on
$\cTau(\PZD)$ for several representative values of $\mathcal{B}(\PSMHiggs
\to \PZD \PZD)$, are shown in Fig.~\ref{fig:darkphoton_contours}
(left).  These results can be translated into limits on the
kinetic mixing $\epsilon$ following the relationship between the dark
photon mass, lifetime, and $\epsilon$.  The resulting 95\% \CL
exclusion contours in the ($\mZD$, $\epsilon$) plane are shown in
Fig.~\ref{fig:darkphoton_contours} (right).  Our analysis excludes a
wide range of $\epsilon$ values, between $9 \times 10^{-9}$ and $6
\times 10^{-6}$ at $\mZD = 10\GeV$ and between $5 \times 10^{-10}$ and
$1.5 \times 10^{-6}$ at $\mZD = 60\GeV$ for $\mathcal{B}(\PSMHiggs \to
\PZD \PZD) = 1\%$.

\section{Summary} \label{sec:summary}
Data collected by the CMS experiment in proton-proton collisions
at $\sqrt{s} = 13\TeV$ in 2016 and 2018 and corresponding to an
integrated luminosity of 97.6\fbinv have been used to conduct an
inclusive search for long-lived exotic neutral particles (LLPs)
decaying to a pair of oppositely charged muons.  The search is largely model-independent
and is sensitive to a broad range of LLP lifetimes and masses.  No significant
excess of events above the standard model background is observed.  The
results are interpreted as limits on the parameters of the
hidden Abelian Higgs model, in which the Higgs boson decays to a pair
of long-lived dark photons $\PZD$, and of a simplified model, in which
LLPs are produced in decays of an exotic heavy neutral scalar boson.
In the mass range $20<\mZD<60\GeV$, a
branching fraction of the Higgs boson to dark photons of 1\%
is excluded at 95\% confidence level 
in the range of proper decay length $\cTau(\PZD)$ from a few tens of $\mum$ to
${\approx}100\unit{m}$.  The results of this search significantly 
extend the previously excluded range of model parameters.
For the hidden Abelian Higgs model with $\mZD$ greater than 20\GeV and
less than half the mass of the Higgs boson, they provide the best
limits to date on the branching fraction of the Higgs boson to dark photons
for $\cTau(\PZD)$ (varying with $\mZD$) between 0.03 and
${\approx}0.5$\mm, and above ${\approx}0.5$\unit{m}. At exotic scalar boson
masses larger than the Higgs boson mass, our results represent the
best current constraints for all considered LLP masses and lifetimes.

\begin{acknowledgments}
We congratulate our colleagues in the CERN accelerator departments for the excellent performance of the LHC and thank the technical and administrative staffs at CERN and at other CMS institutes for their contributions to the success of the CMS effort. In addition, we gratefully acknowledge the computing centres and personnel of the Worldwide LHC Computing Grid and other centres for delivering so effectively the computing infrastructure essential to our analyses. Finally, we acknowledge the enduring support for the construction and operation of the LHC, the CMS detector, and the supporting computing infrastructure provided by the following funding agencies: BMBWF and FWF (Austria); FNRS and FWO (Belgium); CNPq, CAPES, FAPERJ, FAPERGS, and FAPESP (Brazil); MES and BNSF (Bulgaria); CERN; CAS, MoST, and NSFC (China); MINCIENCIAS (Colombia); MSES and CSF (Croatia); RIF (Cyprus); SENESCYT (Ecuador); MoER, ERC PUT and ERDF (Estonia); Academy of Finland, MEC, and HIP (Finland); CEA and CNRS/IN2P3 (France); BMBF, DFG, and HGF (Germany); GSRI (Greece); NKFIH (Hungary); DAE and DST (India); IPM (Iran); SFI (Ireland); INFN (Italy); MSIP and NRF (Republic of Korea); MES (Latvia); LAS (Lithuania); MOE and UM (Malaysia); BUAP, CINVESTAV, CONACYT, LNS, SEP, and UASLP-FAI (Mexico); MOS (Montenegro); MBIE (New Zealand); PAEC (Pakistan); MES and NSC (Poland); FCT (Portugal); J MESTD (Serbia); MCIN/AEI and PCTI (Spain); MOSTR (Sri Lanka); Swiss Funding Agencies (Switzerland); MST (Taipei); MHESI and NSTDA (Thailand); TUBITAK and TENMAK (Turkey); NASU (Ukraine); STFC (United Kingdom); DOE and NSF (USA).
  
\hyphenation{Rachada-pisek} Individuals have received support from the Marie-Curie programme and the European Research Council and Horizon 2020 Grant, contract Nos.\ 675440, 724704, 752730, 758316, 765710, 824093, 884104, and COST Action CA16108 (European Union); the Leventis Foundation; the Alfred P.\ Sloan Foundation; the Alexander von Humboldt Foundation; the Belgian Federal Science Policy Office; the Fonds pour la Formation \`a la Recherche dans l'Industrie et dans l'Agriculture (FRIA-Belgium); the Agentschap voor Innovatie door Wetenschap en Technologie (IWT-Belgium); the F.R.S.-FNRS and FWO (Belgium) under the ``Excellence of Science -- EOS" -- be.h project n.\ 30820817; the Beijing Municipal Science \& Technology Commission, No. Z191100007219010; the Ministry of Education, Youth and Sports (MEYS) of the Czech Republic; the Hellenic Foundation for Research and Innovation (HFRI), Project Number 2288 (Greece); the Deutsche Forschungsgemeinschaft (DFG), under Germany's Excellence Strategy -- EXC 2121 ``Quantum Universe" -- 390833306, and under project number 400140256 - GRK2497; the Hungarian Academy of Sciences, the New National Excellence Program - \'UNKP, the NKFIH research grants K 124845, K 124850, K 128713, K 128786, K 129058, K 131991, K 133046, K 138136, K 143460, K 143477, 2020-2.2.1-ED-2021-00181, and TKP2021-NKTA-64 (Hungary); the Council of Science and Industrial Research, India; the Latvian Council of Science; the Ministry of Education and Science, project no. 2022/WK/14, and the National Science Center, contracts Opus 2021/41/B/ST2/01369 and 2021/43/B/ST2/01552 (Poland); the Funda\c{c}\~ao para a Ci\^encia e a Tecnologia, grant CEECIND/01334/2018 (Portugal); the National Priorities Research Program by Qatar National Research Fund; MCIN/AEI/10.13039/501100011033, ERDF ``a way of making Europe", and the Programa Estatal de Fomento de la Investigaci{\'o}n Cient{\'i}fica y T{\'e}cnica de Excelencia Mar\'{\i}a de Maeztu, grant MDM-2017-0765 and Programa Severo Ochoa del Principado de Asturias (Spain); the Chulalongkorn Academic into Its 2nd Century Project Advancement Project, and the National Science, Research and Innovation Fund via the Program Management Unit for Human Resources \& Institutional Development, Research and Innovation, grant B05F650021 (Thailand); the Kavli Foundation; the Nvidia Corporation; the SuperMicro Corporation; the Welch Foundation, contract C-1845; and the Weston Havens Foundation (USA).
\end{acknowledgments}

\bibliography{auto_generated} 
\cleardoublepage \appendix\section{The CMS Collaboration \label{app:collab}}\begin{sloppypar}\hyphenpenalty=5000\widowpenalty=500\clubpenalty=5000
\cmsinstitute{Yerevan Physics Institute, Yerevan, Armenia}
{\tolerance=6000
A.~Tumasyan\cmsAuthorMark{1}\cmsorcid{0009-0000-0684-6742}
\par}
\cmsinstitute{Institut f\"{u}r Hochenergiephysik, Vienna, Austria}
{\tolerance=6000
W.~Adam\cmsorcid{0000-0001-9099-4341}, J.W.~Andrejkovic, T.~Bergauer\cmsorcid{0000-0002-5786-0293}, S.~Chatterjee\cmsorcid{0000-0003-2660-0349}, K.~Damanakis\cmsorcid{0000-0001-5389-2872}, M.~Dragicevic\cmsorcid{0000-0003-1967-6783}, A.~Escalante~Del~Valle\cmsorcid{0000-0002-9702-6359}, P.S.~Hussain\cmsorcid{0000-0002-4825-5278}, M.~Jeitler\cmsAuthorMark{2}\cmsorcid{0000-0002-5141-9560}, N.~Krammer\cmsorcid{0000-0002-0548-0985}, S.~Kulkarni\cmsAuthorMark{3}, L.~Lechner\cmsorcid{0000-0002-3065-1141}, D.~Liko\cmsorcid{0000-0002-3380-473X}, I.~Mikulec\cmsorcid{0000-0003-0385-2746}, P.~Paulitsch, F.M.~Pitters, J.~Schieck\cmsAuthorMark{2}\cmsorcid{0000-0002-1058-8093}, R.~Sch\"{o}fbeck\cmsorcid{0000-0002-2332-8784}, D.~Schwarz\cmsorcid{0000-0002-3821-7331}, M.~Sonawane\cmsorcid{0000-0003-0510-7010}, S.~Templ\cmsorcid{0000-0003-3137-5692}, W.~Waltenberger\cmsorcid{0000-0002-6215-7228}, C.-E.~Wulz\cmsAuthorMark{2}\cmsorcid{0000-0001-9226-5812}
% ...
\par}
\vskip\cmsinstskip
\dag:~Deceased\\
$^{1}$Also at Yerevan State University, Yerevan, Armenia\\
$^{2}$Also at TU Wien, Vienna, Austria\\
$^{3}$Now at Institute of Physics, University of Graz, Graz, Austria\\
% ...
\end{sloppypar}
%%% END EDITABLE REGION %%%
% skeleton_end
\end{document}